\documentclass[%
 reprint,
 bibnotes,
 amsmath,amssymb,
 aps,
 pra,
nofootinbib]{revtex4-2}

\usepackage{graphicx}
\usepackage{dcolumn}
\usepackage{bm}
\usepackage{float}
\usepackage{placeins}
\usepackage{xfrac}
\usepackage{footnote}
\usepackage{tabularx}

\begin{document}

\preprint{APS/123-QED}

\title{Using Cryogenic CMOS Control Electronics To Enable A Two--Qubit Cross--Resonance Gate}%

\author{Devin Underwood}\email{devin.underwood@ibm.com}
\author{Joseph A. Glick}\email{joseph.a.glick@ibm.com}
\author{Ken Inoue}
\author{David J. Frank}
\author{John Timmerwilke}
\author{Emily Pritchett}
\author{Sudipto Chakraborty}
\author{Kevin Tien}
\author{Mark Yeck}
\author{John  F. Bulzacchelli}
\author{Chris Baks}

\affiliation{IBM Quantum, T.J. Watson Research Center, 1101 Kitchawan Rd., Yorktown Heights, 10598, NY, USA}

\author{Pat Rosno}
\affiliation{IBM Quantum, 2800 37th St. NW, Rochester, MN, USA}

\author{Raphael Robertazzi}
\author{Matthew Beck}
\author{Rajiv V. Joshi}
\author{Dorothy Wisnieff}
\affiliation{IBM Quantum, T.J. Watson Research Center, 1101 Kitchawan Rd., Yorktown Heights, 10598, NY, USA}

\author{Daniel Ramirez}
\author{Jeff Ruedinger}
\affiliation{IBM Quantum, 2800 37th St. NW, Rochester, MN, USA}

\author{Scott Lekuch}
\author{Brian P. Gaucher}
\author{Daniel J. Friedman}
\affiliation{IBM Quantum, T.J. Watson Research Center, 1101 Kitchawan Rd., Yorktown Heights, 10598, NY, USA}

\date{\today}

\begin{abstract}
Qubit control electronics composed of CMOS circuits are of critical interest for next generation quantum computing systems. A CMOS--based application specific integrated circuit (ASIC) fabricated in 14nm FinFET technology was used to generate and sequence qubit control waveforms and demonstrate a two--qubit cross resonance gate between fixed frequency transmons. The controller was thermally anchored to the T = 4K stage of a dilution refrigerator and the measured power was 23 mW per qubit under active control. The chip generated single--side banded output frequencies between 4.5 and 5.5 GHz with a maximum power output of -18 dBm. Randomized benchmarking (RB) experiments revealed an average number of 1.71 instructions per Clifford (IPC) for single--qubit gates, and 17.51 IPC for two--qubit gates. A single--qubit error per gate of $\epsilon_{\text{1Q}}$=8e-4 and two--qubit error per gate of $\epsilon_\text{2Q}$=1.4e-2 is shown. A drive--induced Z--rotation is observed by way of a rotary echo experiment; this observation is consistent with expected qubit behavior given measured excess local oscillator (LO) leakage from the CMOS chip. The effect of spurious drive induced Z--errors is numerically evaluated with a two--qubit model Hamiltonian, and shown to be in good agreement with measured RB data. The modeling results suggest the Z--error varies linearly with pulse amplitude. 
\end{abstract}

\maketitle

\section{\label{sec:level1} Introduction}

Next generation quantum computers will undergo a paradigm shift whereupon multi--qubit devices will predominantly perform fault tolerant quantum circuits. This new era of quantum computing will require orders of magnitude more qubits than are currently being integrated in today's systems~\cite{Bravyi2022}. For large quantum computing systems comprised of solid state quantum processors (superconducting qubits or quantum dots), cryogenic control electronics is considered a key enabling technology ~\cite{Gambetta2017,VanMeter2013,Hornibrook2015}. There has been significant development in CMOS electronics for quantum dot processors ~\cite{xue2021,Pauka2021,Charbon2016,Reilly2019}; primarily due to the increased I/O demands, and the potential for integrating CMOS electronics with qubits at temperatures $>$ 100mK. More recently, cryogenic CMOS electronics for superconducting qubits have been developed and have been used for single qubit gate demonstrations ~\cite{Bardin2019,Frank2022,Chakraborty2022}.

Research in cryogenic CMOS control electronics has primarily focused on achieving the low--power analog requirements to operate within the thermal load limitations of a dilution refrigerator (DR). While power dissipation is an important specification for cryogenic control technologies, minimalistic controllers are not sufficient for practical qubit control. For a fully cryogenic control architecture to be useful, a classical processor capable of producing relevant pulse sequences will be required. This classical processor presents an additional source of power dissipation, and specialized digital architectures will be needed for cryogenic integration to achieve required performance within the limited available power budget.

Another important consideration for cryogenic control technologies is the maximum thermal load of the different stages in a dilution refrigerator. For thermalization of the CMOS--chip at the $T = 4$K plate, the power limitation of a cryogen--free DR is set by the second stage of a cryo--cooler, and the maximum thermal load will be determined by the number of cryo--coolers in the DR. Notably, the maximum load of cryo--coolers is temperature dependent; they yield more cooling power at higher temperatures~\cite{Choi2007}. If the second stage were allowed to operate at higher temperatures, then the DR would be able to support more cryogenic control channels. Assuming higher operating temperatures of the second stage do not impact the cooling of other temperature stages, then this strategy will make the integration of cryogenic control technologies more feasible. 

\begin{figure*}
    \centering
    \includegraphics[width=2.0\columnwidth]{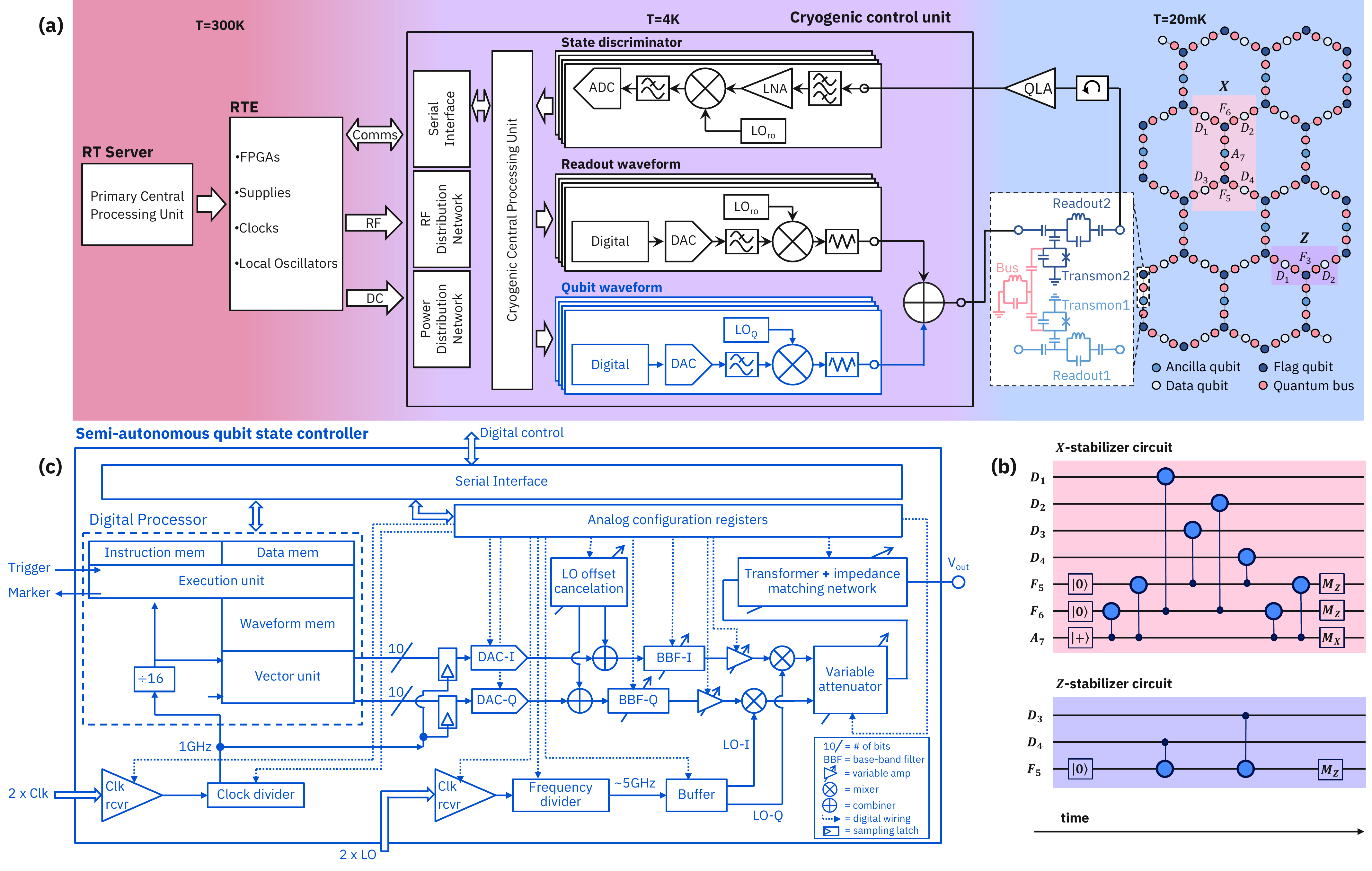}
    \caption{\textbf{(a)} A block diagram for a fault tolerant quantum computing architecture in which control signals for the quantum processor (QP) are digitally synthesized at the 4K stage of the dilution refrigerator. Here, the QP is composed of fixed frequency transmon qubits coupled together through a fixed frequency quantum bus and arranged in a heavy hexagonal lattice. The cryogenic control unit (CCU) is composed of a cryogenic central processing unit (CCPU), qubit waveform generators, readout waveform generators, and quantum state discriminators. A CCU containing these key elements would be capable of autonomous operation, yielding a shorter latency loop when running deterministic quantum circuits. A room temperature (RT) processor performs classical computations, orchestrates high level quantum operations, and interprets results of quantum algorithms \cite{Bravyi2022}. RT support electronics are necessary to power, clock, and program active cryogenic electronics. The support electronics interface with a RT server which performs classical computations necessary to run quantum algorithms. \textbf{(b)} The \textbf{$X$} and \textbf{$Z$} stabilizer circuits for performing error correcting protocols on a heavy hexagonal lattice. Stabilizers represent the primary protocol for logical qubit maintenance and the CCU oversees these protocols; which include monitoring physical qubits, decoding errors on physical qubits \cite{Das2021}, and generating conditional sequences of pulses. \textbf{(c)} An expanded block diagram of the qubit waveform generator (blue box in (a)), used for this manuscript. }
    \label{fig_architecture}
\end{figure*}

Cryo--controlled architectures such as the one shown in Fig.~\ref{fig_architecture}\textbf{(a)} may also yield systematic scaling advantages that would warrant using low power CMOS ASICs. Examples of such advantages include: lower communication latency~\cite{guo2022}, lower thermal noise floor~\cite{krinner2019}, reduced dispersion of base--band signals~\cite{rol2020}, reduction in RF loss due to signal delivery~\cite{simbierowicz2022}, wire count reduction from room temperature to the $T = 4$K stage, a reduction in power per channel, a reduction in cost per channel, and a reduction in total system size.

In this manuscript, we present measurement results on a transmon--based quantum processor (QP)~\cite{Koch2007,Chow2011} controlled via a custom cryo--CMOS application specific integrated circuit (ASIC). The ASIC was designed for cryogenic operation and is capable of generating pulse sequences for qubit characterization, verification, and validation (QCVV) experiments common for use with transmon qubits. Here the CMOS chip is a dual--channel semi--autonomous qubit state controller fabricated in 14nm FinFET technology~\cite{Frank2022,Chakraborty2022}. A single channel block diagram is shown in Fig.~\ref{fig_architecture}\textbf{(c)}. The on--chip processor facilitates autonomy through its ability to play pre--defined sequences of qubit control waveforms, a necessary requirement for both quantum error correction (QEC)~\cite{Shore1995,Bacon2006,Fowler2012,Cross2007} and quantum error mitigation (QEM) workloads~\cite{Kandala_2019,Havlicek_2019,Suzuki2022}. In the experiments described in this work, emphasis was placed on understanding the demands of the classical processor (CP), which represents a near--term development challenge for cryogenic control technologies. For example, present day QPs are primarily used for physics learning and qubit characterization experiments~\cite{blumekohout2017, cross2019, mckay2019}, which can be difficult to support in a low--power processor.

An important qubit gate characterization experiment that presents challenges to limited memory control hardware is randomized benchmarking (RB)~\cite{Magesan2012}. For RB experiments, a random sequence of Clifford gates followed by an inversion pulse are required to be stored in memory. Individual Cliffords correspond to waveforms with independent amplitudes, phases, and durations, while the sequence of Cliffords corresponds to a set of instructions. For complex experiments like RB, compressing large instruction sets into limited memory is challenging; however, not all experiments are as demanding. In a quantum computing system that primarily performs QEC protocols (Fig.~\ref{fig_architecture}\textbf{(a)(b)}), the set of required pulse sequences is simple and repetitive, especially when compared to those needed for QCVV experiments~\cite{eisert2020}. Understanding the complexity and characteristics of the pulse sequences demanded by these experiments is important for developing optimal qubit control technologies, and could lead to special purpose instruction set architectures (ISA).

Here we report the use of cryo--CMOS to generate qubit control waveforms for a suite of characterization experiments including: $T_1$, $T_2$, $T_2^*$, Carr–Purcell–Meiboom–Gill sequences (CPMG), rotary echo, Hamiltonian tomography, and RB of single--qubit and two--qubit gates. The classical processor was characterized during qubit measurements, highlighting the efficiency of the ISA. Transmon calibration routines were performed in order to realize the aforementioned characterization experiments. These measurements serve as a promising demonstration of cryo--CMOS based control technology, with single--qubit and two--qubit error per gate (EPG) observed to be $\epsilon_{\text{1Q}}= 8$e-4 and $\epsilon_\text{2Q}=1.4$e-2, respectively. We show through Lindblad master equation simulations that the observed error is set by control noise. The primary error source is a pulse--induced qubit Z--axis rotation that arises due to spurious spectral content observed in the CMOS controller output. Additionally, a Gaussian distributed pulse amplitude noise was observed on the RF pulses, but simulations showed that this noise did not significantly impact gate errors.

\section{Cryo-CMOS Control Electronics}\label{sec_architecture}

An ideal quantum control architecture will have the capability to autonomously generate waveform sequences conditioned on the measurement of physical qubits~\cite{Corcoles2021}. A control unit able to satisfy this requirement will be composed of circuit blocks for qubit waveform generation, entanglement waveform generation, readout waveform generation, and qubit state discrimination. Furthermore, the unit will require a central processor that manages these blocks and conditionally asserts logic for running the appropriate quantum circuits (Fig.~\ref{fig_architecture}\textbf{(a)}). CMOS is an ideal technology for realizing such a control unit because of existing industrial fabrication capabilities and ease of integration of the different circuit blocks required.

In a quantum computing system optimized to execute specific quantum algorithms, the fault tolerant operations are in principle deterministic based on the decoding of physical qubit measurements~\cite{Das2021,Overwater_2022,Chen_2022}, implying full autonomy is achievable. If the above mentioned capabilities are performed autonomously and within the dilution refrigerator, this approach will yield a reduced latency control configuration (Fig.~\ref{fig_architecture}\textbf{(a)}), thereby increasing the achievable number of circuit layer operations per second (CLOPS)~\cite{Wack2021}. The primary latency concerns addressed here are a reduction in the round--trip transient time and the response time for conditional waveform generation. 

The proposed autonomous control unit is composed of distinguishable circuit blocks, which can be developed either as stand--alone chips in a multi--chip configuration or as a single larger integrated circuit. The work detailed in this manuscript focuses on a demonstration with a stand--alone circuit block consisting of two distinct RF channels for qubit waveform generation. Here the same RF generator is used for both single qubit control and generating entanglement between qubit pairs; this approach leverages the inherent  wiring advantage associated with the cross--resonance based architecture \cite{Chow2011}. As shown in Fig.~\ref{fig_architecture}\textbf{(c)}, this semi--autonomous qubit state controller consists of an analog block with a low power digital to analog converter (DAC), a special purpose classical processor, and a serial interface for communicating with RT electronics~\cite{Frank2022,Chakraborty2022}.

\subsection{Analog Block}
The qubit state controller was designed and fabricated in 14 nm FinFET technology, a technology choice which was desirable due to its high switching efficiency, large transistor on/off ratio, and lower threshold voltages that lead to reduced power dissipation~\cite{Bhattacharya2014}. The choice of a highly--scaled transistor technology also reduces the cost of adding a high degree of digital programmability. For example, the qubit controller is capable of being configured to the following modes of conversion operation: double--sideband with suppressed carrier (DSB-SC), single--sideband direct conversion lower sideband (SSB-LSB), and single--sideband direct conversion upper sideband (SSB-USB). The reported experiments utilized SSB--LSB as a mode of operation; which provided additional filtering of the LO when it was placed high in frequency than the transmons.

For experimental versatility, the qubit state controller's analog control block was made configurable utilizing over 200 bits of digital control. The analog control block is composed of two 10--bit DACs (in--phase and quadrature, or I/Q), two baseband filters (I/Q), a complex mixer to provide, e.g., SSB--LSB output, and a tunable output stage. The complex mixer receives quadrature clock signals for upconversion of the DAC's baseband signal to the qubit's $\vert 0 \rangle$ to $\vert 1 \rangle$ transition frequency $\omega_{01}$. The SSB mode was chosen as the primary mode of operation in order to reduce circuit complexity while also reducing analog power consumption. In order to maximize dynamic range while simultaneously minimizing noise generation, a fully differential current mode design was implemented. Notable advantages of the design include: current reuse among multiple functional blocks, high bandwidth interfaces between circuit elements, convenient implementation of the variable gain stages (using current scaling and current steering \cite{OSullivan2004}), and low switching noise at the output~\cite{Frank2022,Chakraborty2022}. 

The differential wiring extends from the DAC output to the output stage of the chip. The output stage consists of a balun which converts the differential signal to single--ended; the balun resonance frequency is tunable to support the range of desired  SSB frequencies. Balun resonance and shape tuning are controlled using 4 bits of center frequency adjustment and 2 bits of quality factor adjustment. The output impedance is adjustable to match the fridge wiring that is connected to the balun output. A variable attenuator provides 20dB of programmable attenuation for noise reduction, plus an additional 25dB to be switched in for blanking the AWG during readout. To satisfy dynamic range requirements, two variable gain stages were used in the analog control path. The first gain stage was placed between the baseband filter (BBF) and the SSB upconverter while the second gain stage was placed at the output of the SSB up--converter. Both gain stages are unidirectional (to provide reverse isolation) and yield a total of 34 dB of gain control with an average step size close to 2 dB. The BBF bandwidth (as reflected in 3dB cutoff frequency) is configurable over a range of 100--800 MHz using  5 bandwidth control configuration bits. A variable attenuator 

The DAC can be programmed to produce an IF offset within $\sim$ 400 MHz of the LO, and the in--band spurious tones at the output were suppressed to a spurious--free dynamic range of $40$ dB out to 500 MHz~\cite{Frank2022,Chakraborty2022}. A microwave source with a frequency between 8 and 12~GHz is delivered from room temperature to the cryo-CMOS chip, on which LO signals in a range of 4--6~GHz are generated with a 2:1 frequency divider. Leveraging the analog control path's differential current mode architecture, programmable DC currents are added to the output currents of the DACs in order to compensate differential offsets, which helps reduce LO leakage in the RF output. 

\subsection{Digital Block} 
\label{digital_block}

The digital architecture features a processor implementing 32 bit fixed--point instructions for programming flexibility, including special instructions for waveform generation and phase rotations, and is designed to minimize power consumption for cryogenic operation. The processor's instruction set architecture (ISA) defines 32 general purpose instructions (8 branch/flow control, 10 data movement, and 14 arithmetic) to enable trigger--controlled loops, subroutines, and computation, as well as 5 special instructions for the generation of waveforms and digital output signals. The processor core uses three SRAM banks with 32 KB dedicated for instructions, 20 KB dedicated for waveforms, and 32 KB dedicated for data, respectively. To minimize power consumption, the processor has a fast clock domain operating at the sampling frequency $f_s$ of the DACs for providing waveform data to the DACs, and a slow clock domain operating at $f_\text{CLK} = f_s/16$ for program control. The microarchitecture implements instruction fetch, decode, branch resolution, and scalar arithmetic execution within one slow clock cycle. 

Waveform data is stored as an envelope modulated by intermediate frequency I/Q sinusoids with an initial phase of zero. This approach reduces waveform memory footprint and avoids the power overhead associated with the sine/cosine evaluations of a numerically controlled oscillator (NCO) \cite{Park2021AFI,Patra2020}. The Compute Waveform Coefficients (CWC) instruction prepares the I/Q coefficients used by the Play Waveform (PW) instruction to set the phase and amplitude of the output waveform. These coefficients are calculated relative to the frame phase which can be modified by special instructions such as Add Frame Phase (ADDFP) to effect a virtual Z--rotation of the qubit phase \cite{McKay_2017}. The coefficients are applied to the stored waveform data through 16--way single instruction, multi data vector arithmetic logic in the slow clock domain. One PW instruction can play up to 4096 waveform samples. The samples are serialized into the fast clock domain and sent to the I/Q DACs in the analog section. The waveform retrieval and processing functions progress independently of the program flow and control facilities of the processor.

\section{CMOS Processor for Qubit Control}

\begin{table}
\begin{center}
\begin{tabular}{ |p{1.5cm}||p{1.75cm}|p{1.8cm}|p{2.75cm}| }
 \hline
 \multicolumn{4}{|c|}{CMOS parameters} \\
 \hline
 Channel & LO(GHz) & IF(MHz) & Digital Clock(GHz) \\
 \hline
 $\text{CH}_{\text{CTRL}}$   & 5.6  & 261.77   & 2.25 \\
 \hline
 $\text{CH}_{\text{TRGT}}$   & 5.6  & 348.81   & 2.25 \\
 \hline
\end{tabular}

\begin{tabular}{ |p{1.5cm}||p{1.75cm}|p{1.8cm}|p{2.75cm}| }
 \hline
 \multicolumn{4}{|c|}{Qubit parameters} \\
 \hline
 Qubit& T1($\mu s$) & T2($\mu s$) & ZZ(kHz) \\
 \hline
 $Q_{CTRL}$   & $57.57\pm 4.46$  & $68.99 \pm 2.04$ & $103.0 \pm 67$\\
 \hline
 $Q_{TRGT}$   & $61.58\pm 5.99$  & $69.16 \pm 4.80$ & $103.0 \pm 67$ \\
 \hline
\end{tabular}
 \label{table_device_params}
\end{center}
\caption{Cryo-CMOS parameters used during qubit control experiments and average qubit parameters measured using the cryo-CMOS chip. The reported coherence measurements were interleaved between 2Q RB measurements shown in Fig.~\ref{fig_RB_vs_gate_length}\textbf{(f)}}
\end{table}

\begin{figure}
    \centering
    \includegraphics[width=1.0\columnwidth]{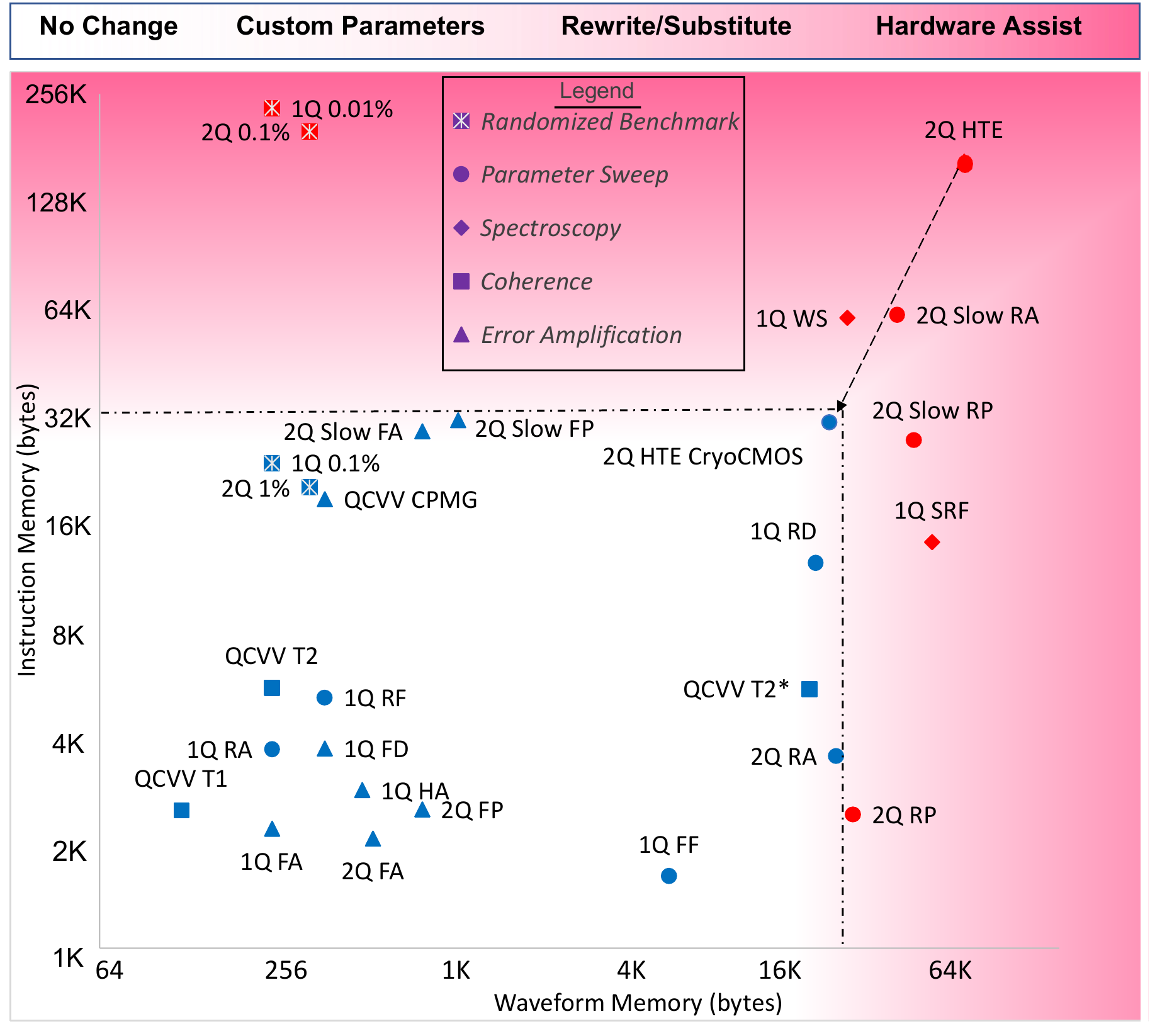}
    \caption{Instruction and waveform memory requirements for each of the calibration sequences, QCVV experiments, and RB experiments, using nominal pulse widths of 42.67 ns for single qubit gates, 71.1 ns for the shortest CR pulse width, and 711.1ns for longest CR pulse width (marked "Slow" in the data labels). The X and Y axes unit is memory size in bytes, in logarithmic scale. CryoCMOS memory limit of 32KB for instructions and 20KB for waveforms are denoted by the dotted square. ‘1Q’ in the data labels identifies single qubit calibrations and RB experiments, ‘2Q’ or '2Q Slow' identifies two qubit calibrations and RB experiments, and ‘QCVV’ identifies the characterization experiments (T1, T2, T2 star and CPMG). For calibrations, the last
    string identifies the type of calibration, encoded as [ R=Rough $\mid$ F=Fine ] [ A=Amplitude $\mid$ F=Frequency $\mid$ D=Drag $\mid$ P=Phase ]. For RBs, the last string specifies the errors per Clifford, and the corresponding marker identifies the memory size needed to reach P(1)=0.49 for the error rate. WS stands for wideband spectroscopy, SRF stands for super rough frequency calibration, and HTE for Hamiltonian Tomography with Echoed CR. Vast differences in the memory requirements are evident for these different experiments. For instance, CR amplitude, CR phase, spectroscopy and HTE require large waveform memory, while RB, CPMG, spectroscopy and HTE require very large instruction memory. The colors indicate the degree of difficulty to accommodate the CryoCMOS memory limitations. Experiments that fall in the ‘No Change’ zone (white) runs without any modifications. ‘Parameter Change’ requires reducing and/or carefully recalculating the experiment’s parameters such as number of steps and range (lower resolution / narrower range). ‘Rewrite/Substitute’ requires rewriting the experiment itself and/or substituting pulse types, to devise an equivalent experiment that fits better in memory by exploiting CryoCMOS core features. 2Q HTE is in this category. ‘Hardware Assist’ (purple) identifies experiments that may require changes to the current CryoCMOS hardware and/or instruction set. 1Q and 2Q RB with lower error rates fall in this category, requiring further innovation going forward.}
    \label{fig_memory-vs-experiments}
\end{figure}

Even with a specialized ISA, some experiments were challenging to accommodate with this low power processor, primarily due to its limited memory.  Note that all experiments performed with the cryoCMOS processor were originally developed using room temperature electronics that offer significantly more memory and higher performance as compared to the custom processor design. This discrepancy created issues for the porting of experiments to the low power processor. Details regarding issues encountered are illustrated in Fig.~\ref{fig_memory-vs-experiments} and are further discussed in section ~\ref{programming_cmos}. Here Fig.~\ref{fig_memory-vs-experiments} shows the memory usage of the processor for the different experiments performed. Many experiments required no change from routines developed with room--temperature electronics, but in some cases additional effort was required in order to fit pulse sequences into processor memory. The default approach to reducing memory demands was to decrease the point density by customizing parameters, while ensuring enough data points were collected to extract accurate fit results. In cases when the default approach was not sufficient, it was necessary to rewrite experiment routines or substitute new pulse types. One experiment for which new pulse definitions were required is Hamiltonian tomography~\cite{Sheldon2016} and details for how this experiment was made to work are reviewed in section~\ref{programming_cmos}.

With respect to how they were operated using room temperature control electronics, most experiments fell into the category of not requiring change or only needing custom parameters to achieve successful execution. RB is an example of an experiment made to work through parameter adjustment: in this case, the instruction memory was the limitation to be overcome. The strategy used to address this challenge was first to reduce the number of Clifford sequences, and then to use logarithmic spacing between the different sequence lengths, both of which helped to reduce the number of instructions required. However, as shown in Fig.~\ref{fig_RB1Q}\textbf{(a)} and Fig.~\ref{fig_RB2Q}\textbf{(a)}, the last data point in an RB experiment consumes the most memory, implying simple point reduction methods will not scale as error rates improve. Lower error rates will require more Clifford gates for the exponential decay to converge; which will require more instruction memory. As shown in Fig.~\ref{fig_memory-vs-experiments}, the instruction memory requirements for an RB experiment are predicted to increase from ~32KB to ~256KB when the 1Q error and 2Q error are reduced from 0.1\% to 0.01\% and 1\% to 0.1\%, respectively. 

To accommodate longer sequences, ASICs may require  more effective instruction memory, a more specialized ISA, or a custom compiler designed to specifically for the ISA~\cite{moro_2021}. Alternatively, new measurement techniques with lower memory overhead could also be adapted to extract error rates, e.g.: using simple characterization measurements along with error models to infer errors~\cite{Pedersen_2007,Omalley_2015,Willsch_2017,Tahereh_2022}, interpreting the error through an under-sampled RB experiment~\cite{Kelly_2014}, or performing experiments like quantum process tomography~\cite{Rodionov_2014,gaikwad_2022} that place lesser demands on memory. It is also worth noting that longer sequences may not be needed in future quantum computers. Experiments like RB are useful for performance evaluation during hardware and software development, but are not the intended use case for quantum computers. Experiments such as QEC~\cite{Shore1995,Bacon2006,Fowler2012,Cross2007} and QEM~\cite{Kandala_2019,Havlicek_2019,Suzuki2022}, will be the primary use cases for useful applications of quantum computers. 

\section{Memory reduction in a Low Power Processor}
\label{programming_cmos}

The on-chip processor supports 32KB of SRAM for instructions, and 20 KB of SRAM for waveforms. For simplicity these memory banks have a single purpose designation: waveform memory is used store waveform data, and instruction memory is used to store programs that play sequences of qubit control pulses. The resource bottlenecks were observed when waveforms became too long, and/or sequences became too long. When this happened memory reduction techniques were necessary. One example technique was to reduce waveform memory by partitioning waveforms. 

\begin{figure}[H]
    \centering
    \includegraphics[width=0.5\columnwidth]{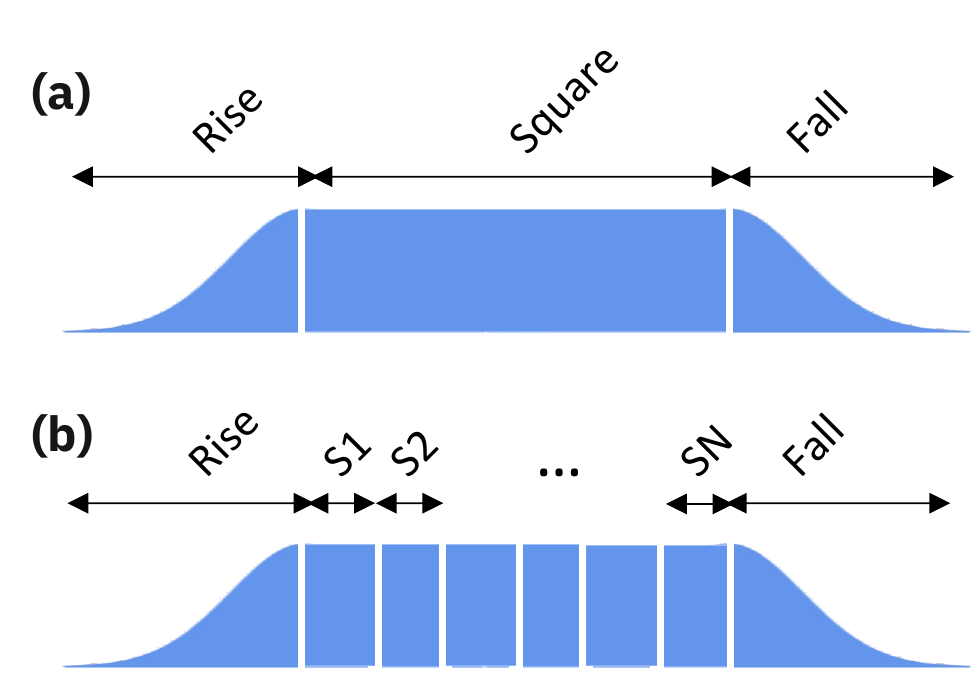}
    \caption{\textbf{(a)} The standard waveform used for a cross-resonance Hamiltonian tomography experiments. \textbf{(b)} Illustration of waveform partitioning used for memory reduction of pulses that were too long to be stored in waveform memory.} 
    \label{fig_waveform_partition}
\end{figure}

A key example is the Hamiltonian tomography experiment from Fig.~\ref{fig_calibrations}\textbf{(d)}, which required many long square--Gaussian waveforms of different lengths that did not fit in waveform memory. To overcome this problem, the flat section of the square--Gaussian waveform was partitioned into small equal--sized segments, and the waveform was constructed by stepping through instruction memory to add segments together, as shown in Fig.~\ref{fig_waveform_partition}. This technique made a fundamental physics experiment feasible with limited waveform memory, but at the cost of more instruction memory. The width of the segment was used as a sweep parameter in order to optimize the trade-space between waveform size, and the number of instructions.

\begin{figure}
    \centering
    \includegraphics[width=1.0\columnwidth]{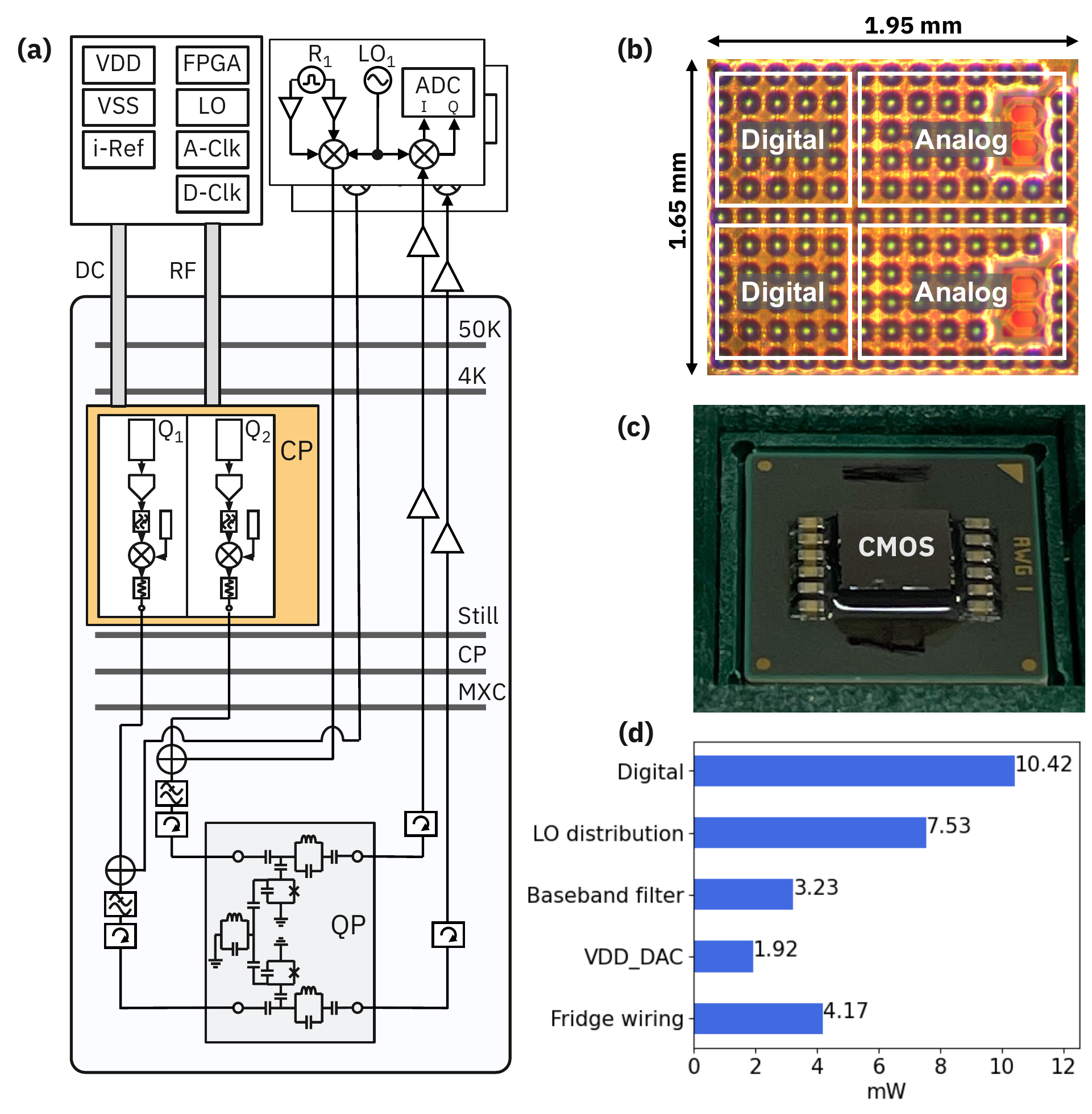}
    \caption{\textbf{(a)} Block diagram of the experimental setup. The cryoCMOS payload (CP) is mounted to the $T = 4$K plate of a dilution refrigerator and drives control signals down to a pair of transmons in the qubit payload (QP) on mixing chamber (MXC) plate at 10 mK to perform single--qubit and two--qubit cross--resonance gates. Between the CP and the QP there is 22 dB of cold attenuation, a Mini Circuits VLF 5500 low-pass filter, a ferrite isolator, and a directional coupler which combines the control and the readout signals together. The DC bias supplies, reference currents, clock frequencies, local oscillator (LO), readout electronics, and FPGA are all located at room temperature. \textbf{(b)} Micrograph of the CMOS chip highlighting the digital and analog sections of the two AWG channels. \textbf{(c)} The CMOS chip bonded to a laminate with NPO ceramic decoupling caps, that sits inside in a pogo--pin socket. The chip is thermally anchored to $T=4K$ through a Cu backing plate on the lid of the socket. \textbf{(d)} Power dissipation of the cryo--CMOS chip measured while under active control, for each sub--component of the chip, and the passive heat-load due to wiring from 50K to 4K. Including wiring, the total power dissipation per control channel is 27.27 mW. The reported powers are extracted from the on-chip supply voltage measured by a sense line, and the current sourced by the power supply. Power dissipation due to fridge wiring is calculated using cryogenic material models\footnote{Cryogenic properties of commonly used metals, https://trc.nist.gov/cryogenics/materials/materialproperties.htm}.}
    \label{fig_experimental_setup}
\end{figure}

This waveform memory reduction technique was also applied to other 2Q calibrations for consistency. In these cases the width of the smallest segment was chosen to align with the fastest 2Q CR pulse. To increase the pulse width, the number of instructions increased linearly with the number of segments. As was observed in Fig.~\ref{fig_memory-vs-experiments} for 2Q calibrations marked with the "Slow" prefix. In this specific case the smallest segment was chosen to be 71.1 ns and the slowest CR pulse used was 711.1 ns; resulting in $\approx10x$ increase in instruction memory usage. It is conceivable that the processor's arithmetic and branch facilities could be exploited to reduce the 10x increase, but this is not trivial because of phase alignment requirements between adjacent segments.
As mentioned in Section~\ref{digital_block}, the processor supports phase manipulation instructions so in theory this can be overcome, but in practice it is difficult to implement correctly and accurately.

Another challenge associated with this simplistic ISA model is resource starvation, because there are often no free clock cycles to allow execution of arithmetic or branch instructions. For example, in these experiments the shortest 1Q gate instrumented in Fig.~\ref{fig_RB_vs_gate_length}\textbf{(e)} is just 28.44 ns wide, which corresponds to two processor clock cycles. In order to generate the shortest pulse, two instructions are required: one to compute waveform coefficients (CWC), and another to play a waveform (PW). Shorter gates are desirable because they are shown to reduce errors, but a dependency exists between the processor clock frequency and the 1Q gate length. The processor clock frequency sets a lower bound on the 1Q gate speed. Increasing the processor clock frequency would speed up the execution of the CWC and PW instructions, but would result in more power dissipation.

The difficulty in optimizing instruction sequences on the ISA, is compounded by the layered structure of the existing software stack; in which high level pulse definitions are generated in a manner that supports a variety of room temperature control hardware. Since off-the-shelf AWGs do not provide quantum specific instructions, such as frame phase adjustments or parametric looping; the existing pulse generation software did not support the efficiency yielded by a quantum specific processor. Ideally a special purpose compiler would be used to take in high level pulse definitions and convert them into instructions for special purpose processors like the one presented in this manuscript \cite{moro_2021}.

\section{Experimental Setup}

\label{sup_sec_experimental_setup}
The experimental setup in a closed-cycle dilution refrigerator, detailed in Fig.~\ref{fig_experimental_setup}(a), consists primarily of a cryoCMOS AWG payload (labeled CP) mounted to the 4K plate, and a qubit payload (labeled QP) mounted to the mixing chamber (MXC) plate at 10 mK. The CP composed of an Au plated Cu machined mount that mechanically and thermally anchors a printed circuit board (PCB). The PCB houses the cryoCMOS AWG chip, decoupling capacitors, and routed traces that connect the various pins of the chip to the wiring connectors. The cryoCMOS AWG chip, shown in Fig.~\ref{fig_experimental_setup}(b), is bump bonded to a laminate with NPO ceramic decoupling capacitors and is inserted into a pogo-pin socket on the PCB, shown in Fig.~\ref{fig_experimental_setup}(c). The cryoCMOS chip is thermally anchored to the 4K stage via a Cu block on the lid of the socket and a Cu strap that is connected directly to the Cu back-plane of the mount. DC power supplies, reference currents, clock and local oscillator (LO) sources, and an FPGA all used to power and the control the chip are located at room temperature (RT) near the cryostat. The output of the two cryoCMOS AWG channels, set to the resonant frequencies of the qubits ($\approx$5 GHz), are sent via coaxial cables down to the MXC plate. At the MXC plate the signals pass through a total of -22 dB of cold attenuation, a Mini Circuits VLF 5500 low-pass filter, and a ferrite cryogenic isolator before connecting to the qubits. 

The QP consists of a pair of transmon qubits that are connected by an LC cancellation bus that helps reduce the effect of constant ZZ-type errors. The transmons are operated and measured in transmission, each with their own designated Purcell filters and readout resonators. The readout pulses, supplied by RT mixer-based signal generators, are set to the frequency of the respective qubit readout resonators ($\approx$7 GHz) and driven into the fridge. The readout control signals are combined at the MXC plate with the cryoCMOS AWG signals into a single line using a directional coupler. After passing through the QP the output signals from each of the two qubits are sent through another cryogenic isolator, a HEMT amplifier, and a RT amplifier, before being sent to an ADC to be digitized. The maximum power level of the control signals from the cryoCMOS AWG that can be delivered to the qubits (after passing through the -22 dB of cold attenuation between them) is $\approx$ -40 dBm. The net gain in the readout path is $\approx$16 dB, defined to be the total amplification of the HEMT and RT amplifiers, minus the losses associated with the coaxial cables. The readout chain did not include use of a quantum limited amplifier (such as a Josephson-based traveling-wave parametric amplifier), but if desired one could be added to improve the signal-to-noise ratio of the output signal. Even without the use of such amplifiers, state discrimination measurements showed that fidelities of $\approx$97$\%$ are achieved. 

Proper synchronization, timing, and triggering between the cryoCMOS AWGs and the readout control electronics is critical to performing all of the qubit calibrations and measurements. All the supplies (Clock, LO, and those for the readout electronics) are connected to a common 10 MHz clock reference. When an experiment is to be performed, the waveform data and the instruction sequences are first loaded onto the cryoCMOS AWG via the FGPA and the serial communication interface, then a signal is relayed back to the FPGA to confirm that the program was successfully loaded onto memory. At the beginning of each cryoCMOS AWG program there is a WAIT instruction. A trigger signal, marking the start of the experiment, is sent from the readout electronics at RT to the cryoCMOS AWG, satisfying the WAIT condition and commencing the programmed pulse sequence. At the end of a sequence, when a readout pulse is being played, the cryoCMOS AWG uses its on-chip programmable attenuation to blank the output of the AWG signal by $\approx$45 dB. By adjusting the pulse timing within the sequence itself and by using a buffer delay on the RT readout electronics, it was ensured that the readout pulses begin soon after the control pulses end, $\approx$10 ns after the blanker feature on the cryoCMOS AWG is engaged. The coordination and timing between the control pulses, the blanking window, and the readout pulses was confirmed using a high-speed oscilloscope. If necessary for a given experiment, the cryoCMOS AWGs can output their own marker pulses to trigger one another, or to indicate that a sequence is complete. At the end of a sequence the processor then loops back to the beginning of the program and waits for another trigger. Through this procedure the cryoCMOS AWG can initialize the qubits into any desired Clifford state, kick off a pulse sequence, such as for calibration, QEC, or to perform a specific qubit experiment, and ensure that at the end they are followed up by well-aligned readout pulses to complete the measurement.

\section{Qubit Calibrations} \label{methods}

\begin{figure*}
    \centering
    \includegraphics[width=1.6\columnwidth]{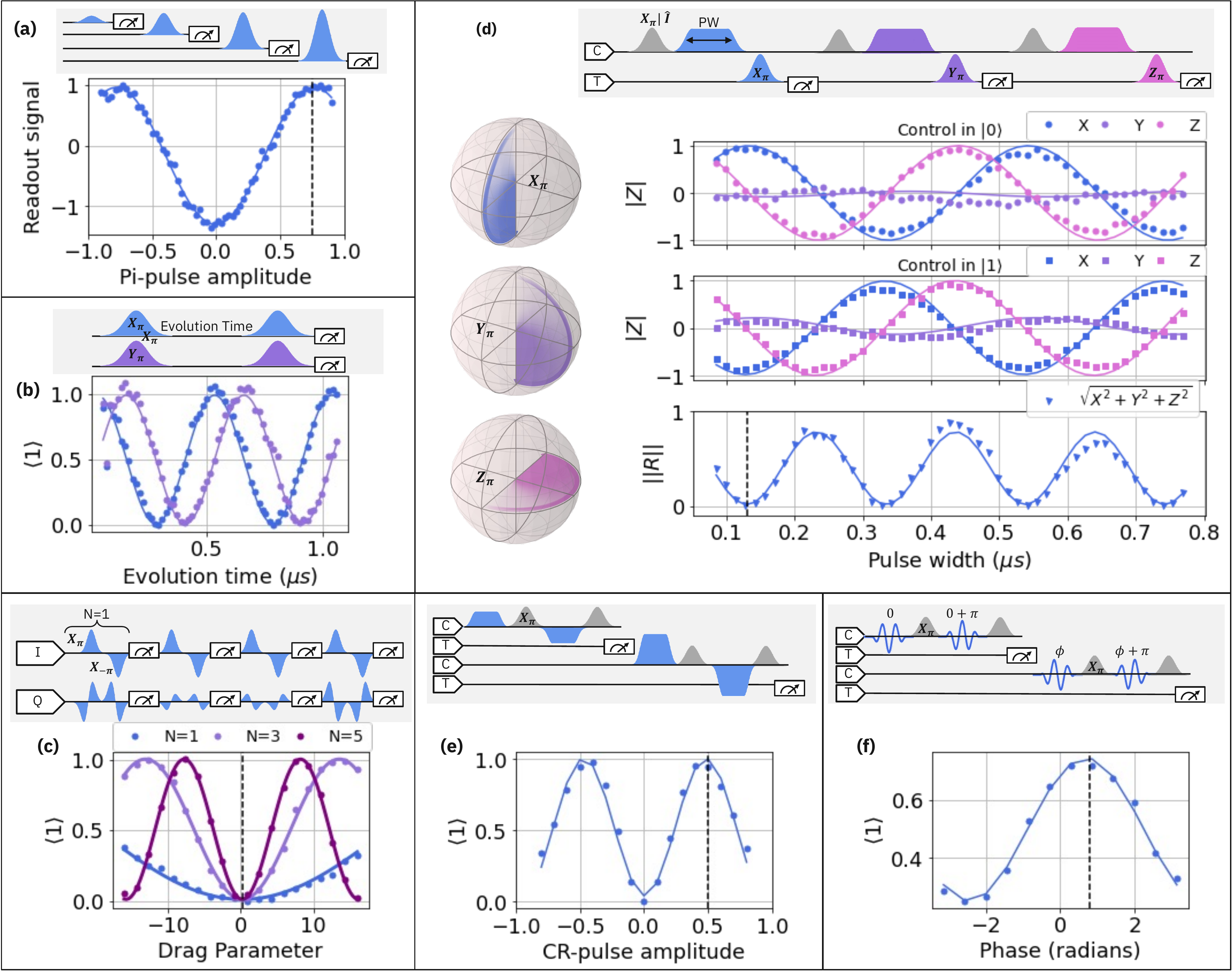}
    \caption{Calibration data for single--qubit (1Q) and two--qubit (2Q) waveforms and the corresponding pulse sequences. Where relevant, dashed vertical lines indicate optimal values extracted from calibration. \textbf{(a)} 1Q Rabi measurement to tune the $\pi$--pulse amplitude, optimally at the maxima in the curve. \textbf{(b)} 1Q Ramsey measurement to tune the qubit frequency. Playing either a X($\pi/2$) (blue) or Y($\pi/2$) (purple) pulse as the second pulse in the sequence yields two curves with a relative phase of $\pi /2$. The period of the curve(s) determines how offset the driven frequency is from the qubit's transition frequency.\textbf{(c)} Derivative Removal by Adiabatic Gate (DRAG) calibration to add a derivative of Gaussian quadrature component to the pulse shape. The $\pi$--pulses are repeated $N$ times within a pulse sequence for different sizes of the DRAG parameter. Each sequence yields a curve with a different period. The optimal DRAG parameter is the collective minima of the different curves. \textbf{(d)} Hamiltonian tomography measured as a function of the cross resonance (CR) pulse width. The $|Z|$ state of the target qubit is measured after projecting into $\langle X \rangle$, $\langle Y \rangle$, $\langle Z \rangle$. These measurements are performed with the control qubit prepared in either the $\vert 0 \rangle$, $\vert 1 \rangle$ state. The oscillations on the target are fit to a Hamiltonian, which can be used to extract device parameters, provide information about the optimal pulse width and the extent of errors such as IY. By computing the Bloch vector $\vert \vec{R} \vert$, one can extract the optimal CR pulse length at the curve's first minima. The qubit-2-qubit coupling was extracted to be $J=2.7$ MHz. \textbf{(e)} CR amplitude calibration, optimally at the first maxima. \textbf{(f)} CR phase calibration, optimally at the first maxima.}
    \label{fig_calibrations}
\end{figure*}

Performing calibration routines with room temperature control electronics are common practice for transmon--based devices, and detailed discussions can be found in references ~\cite{qiskit_text,Patterson2019}. Executing these routines using a novel, low power cryo-CMOS ASIC represents an important demonstration of functionality. As shown in Fig.~\ref{fig_calibrations}, experiments to optimize the amplitude, frequency, phase, and width of the control pulses were performed, from which a set of optimized parameters were found and stored in waveform memory. The successful execution of these calibrations is necessary to facilitate the high fidelity two--qubit echoed CR gate demonstrated in this manuscript.

\subsection{Single Qubit Waveforms}
 The DACs produce in--phase and quadrature signals of the form $V_I(t) = \Omega(t)\cos\left(\omega_{SSB} t -\phi\right)$ and $V_Q(t) = \Omega(t)\sin\left(\omega_{SSB} t -\phi\right)$, respectively, where $\omega_{SSB}$ is the single side band frequency and $\phi$ is the phase. For single--qubit gates, the signals are shaped with a Gaussian envelope of the form, 

\begin{equation}
    \Omega_{G}(t))=
        \begin{cases}
            \Omega_0 \frac{e^{\sfrac{t^2}{2\sigma^2}} - e^{\sfrac{t_g^2}{2\sigma^2}}} {1-e^{\sfrac{t_g^2}{2\sigma^2}}} ,& \; t\le t_g\\
            0 &, \; \text{else} 
        \end{cases}
\end{equation}

where  $\Omega_0$ is the amplitude, the width of the pulse is defined by $t_g$, and $\sigma$ is the standard deviation. The functional form of $\Omega_G$ is chosen to enforce that the pulse start and end with zero amplitude~\cite{Gambetta2011,McKay_2017}. The pulse amplitudes for X($\pi$) and X($\sfrac{\pi}{2}$) were calibrated by driving the qubit at the $\vert 0 \rangle$ to $\vert 1 \rangle $ transition frequency: During the Gaussian pulse, the higher energy levels of the transmon experience a Stark shift, resulting in a drive--induced phase shift about the Z--axis of the Bloch sphere. This phase shift is amplitude dependent, and is mitigated with a derivative removal via adiabatic gate (DRAG) calibration~\cite{Motzoi2009,Gambetta2011,Chow2010,McKay_2017}. The DRAG pulse is implemented by applying $\Omega_{G}(t)$ and the derivative $\beta\dot{\Omega}_{G}(t)$ to the in--phase and quadrature channels, respectively. As shown in Fig.~\ref{fig_calibrations}\textbf{(c)}, the DRAG pulse is repeated N times while sweeping the DRAG scale parameter $\beta$; the collective minima correspond to the optimal parameter value. Subsequent single qubit calibrations consisted of error amplification measurements for fine tuning all pulse parameters~\cite{Sheldon2016_v2,Vitanov_2020}

\subsection{Two Qubit Waveforms}
In order to drive the cross--resonance interaction, a square--Gaussian waveform is applied to the control qubit at the target qubit's resonant frequency~\cite{Malekakhlagh_2022}. The pulse shape is defined by a Gaussian rise and fall of length $\tau_r$ and standard deviation $\sigma_r$, has a flat--top of length $\tau_p - 2\tau_r$ and amplitude $\Omega_0$, and a total pulse length of $\tau_p$. The expression describing the pulse shape is given by

\begin{equation}
\Omega_{GS}(t)=
    \begin{cases}
        \Omega_0\frac{e^{\frac{(t-\tau_r)^2}{2\sigma^2_r}}-e^{\frac{\tau_r^2}{2\sigma^2_r}}}{1-e^{\frac{\tau_r^2}{2\sigma^2_r}}} &, \; 0<t<\tau_r\\
        \Omega_0 &, \; \tau_r<t<\tau_p-\tau_r \\
        \Omega_0\frac{e^{\frac{[t-(\tau_p-\tau_r)]^2}{2\sigma^2_r}}-e^{\frac{\tau_r^2}{2\sigma^2_r}}}{1-e^{\frac{\tau_r^2}{2\sigma^2_r}}} &, \; \tau_p-\tau_r<t<\tau_p
    \end{cases}
\end{equation}

The interactions between control and target qubits can be driven by sweeping the pulse width for a fixed amplitude, or conversely by sweeping the amplitude for a fixed pulse width. In Fig.~\ref{fig_calibrations}\textbf{(d)}, full--state Hamiltonian tomography is performed on the target qubit. This calibration is a method to identify the coherent error terms relevant to the microwave drive~\cite{Sheldon2016}. 

The calibration is performed by sweeping the width of the CR waveform applied to the control, then applying a $X_{\pi}$, $Y_{\pi}$, or $Z_{\pi}$ pulse in order to project the state of the target onto the $X$, $Y$, or $Z$ axes of the Bloch sphere. This process is carried out for the control in $\vert 0 \rangle$ and $\vert 1 \rangle $. The data is then fit to a block--diagonal Hamiltonian and the six interaction terms $IX$, $IY$, $IZ$, $ZX$, $ZY$ and $ZZ$ are parameterized. The quantity $\|R\|=\sqrt{(\langle X_0 \rangle + \langle X_1 \rangle)^2 + (\langle Y_0 \rangle + \langle Y_1 \rangle)^2 + (\langle Z_0 \rangle + \langle Z_1 \rangle)^2}$ in Fig.~\ref{fig_calibrations}\textbf{(d)} is the two--norm distance between Bloch vectors of the target for the two states of the control. When $\|R\|=0$, the two qubits are maximally entangled, and indicates an optimal pulse width $\tau_p$.

\section{Benchmarking a Cross-Resonance Gate}
The cryo--CMOS chip was specially designed to generate waveforms for transmon qubits in a cross-resonance (CR) based architecture, as shown in Fig.~\ref{fig_architecture}(a). A detailed description of the experimental setup with the CMOS chip thermalized to the $T = 4$K stage in a dilution refrigerator is provided in Section~\ref{sup_sec_experimental_setup}. In a cross--resonance based qubit device, entanglement between neighboring qubits is generated via the cross-resonance (CR) interaction~\cite{Chow2011,Sheldon2016,Sheldon2016_v2,Magesan_2020,Malekakhlagh_2020,Sundaresan2020,Kandala2021}. The CR interaction gives rise to a $ZX$ term in the Hamiltonian, which describes a state--dependent Rabi oscillation on the target qubit which in turn depends on the state of the control qubit. The entanglement is generated by applying an RF drive to the control qubit at the target qubit's $\vert 0 \rangle$ to $\vert 1 \rangle $ transition frequency. The always on coupling in CR devices gives rise to parasitic terms in the Hamiltonian such as $ZZ$, $ZI$, and $IZ$. These undesirable terms can be mitigated in hardware through cancellation buses~\cite{Kandala2021}, and digitally through echoed gate sequences~\cite{Sheldon2016,Sundaresan2020}. Here an echoed gate was realized on a device with a cancellation bus. This gate sequence used Gaussian waveforms for single--qubit rotations and square--Gaussian waveforms for generating entanglement.


\begin{figure}
    \centering
    \includegraphics[width=0.8\columnwidth]{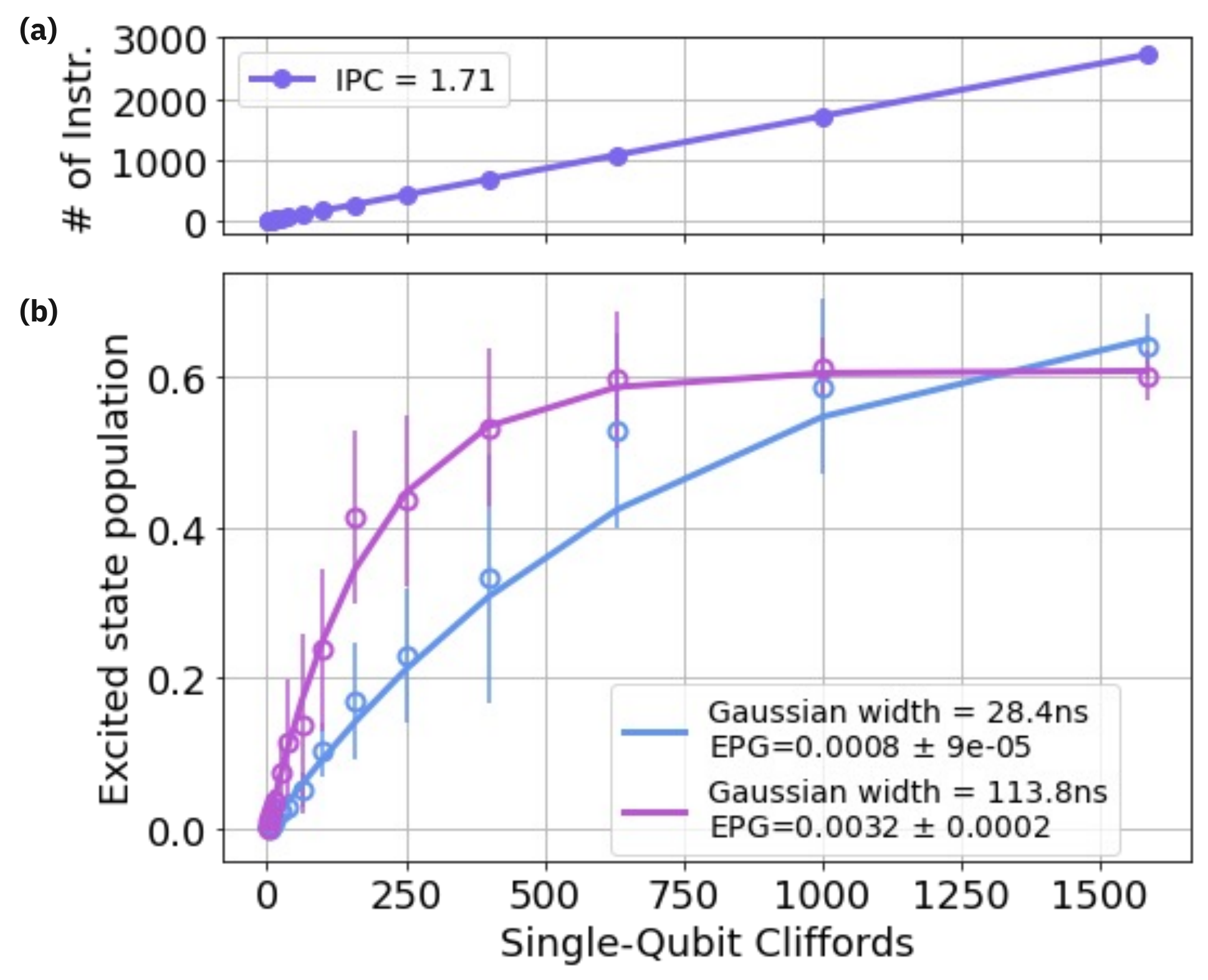}
    \caption{RB data for single--qubit experiments. \textbf{(a)} The number of instructions increase linearly with Clifford count, with $36.94\%$ of the instructions coming from the last RB data point. A single-qubit RB experiment requires 1.71 instructions per Clifford, which is extracted from the slope of the plotted line. \textbf{(b)} Single--qubit RB data for Gaussian widths of 28.4ns and 113.8ns, respectively. The longer gate has more error $\epsilon_{1Q} = 0.0032$, but converges more quickly, requiring fewer instructions. The shorter gate has less error $\epsilon_{1Q} = 0.0008$, but requires longer Clifford sequences to measure and thus more instructions. The observed decay does not converge to 0.5, indicating leakage outside of the computational basis \cite{McKay_2017}. For example, leakage into the $|2\rangle$ state gives rise to IQ counts that are different from $|0\rangle$ and $|1\rangle$. This yields a measurement result of the form $V_0\cdot p_0 + V_1\cdot p_1 + V_2\cdot p_2$, which converges above 0.5, without proper binning of the higher excited states. We believe this effect to be caused by spurious spectral content such as LO leakage.}
    \label{fig_RB1Q}
\end{figure}

\begin{figure}
    \centering
    \includegraphics[width=0.8\columnwidth]{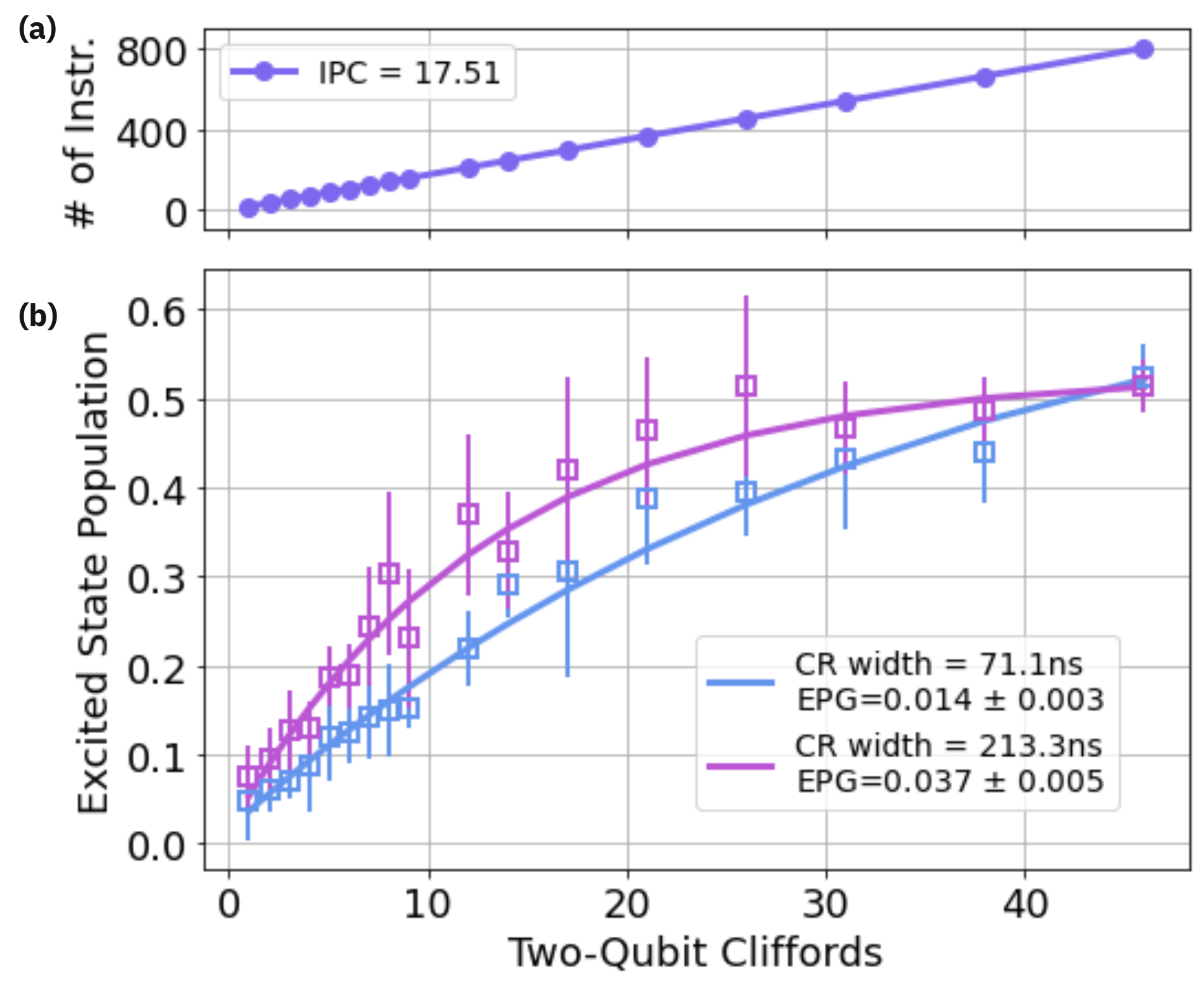}
    \caption{RB data for two--qubit experiments. \textbf{(a)} Two--qubit RB is more demanding on the processor requiring 17.51 instructions per Clifford, with $18.41\%$ of the instructions coming from the last data point. For both RB experiments, a small fraction of the total number of instructions come from pre--sequence calibration, buffering, and pulse idle times, and contribute to IPC. \textbf{(b)} Two--qubit RB measurements for CR pulse widths of 71.1 ns and 213.3 ns, respectively. The shorter gate length have less error $\epsilon_{2Q}=0.014$ compared to $\epsilon_{2Q}=0.037$, but require more computational overhead in order to measure with precision using traditional RB. Error bars for RB experiments are averaged over ten rounds of the same random seed.}
    \label{fig_RB2Q}
\end{figure}

\subsection{Single-Qubit Randomized Benchmarking}
\begin{figure*}
    \centering
    \includegraphics[width=1.8\columnwidth]{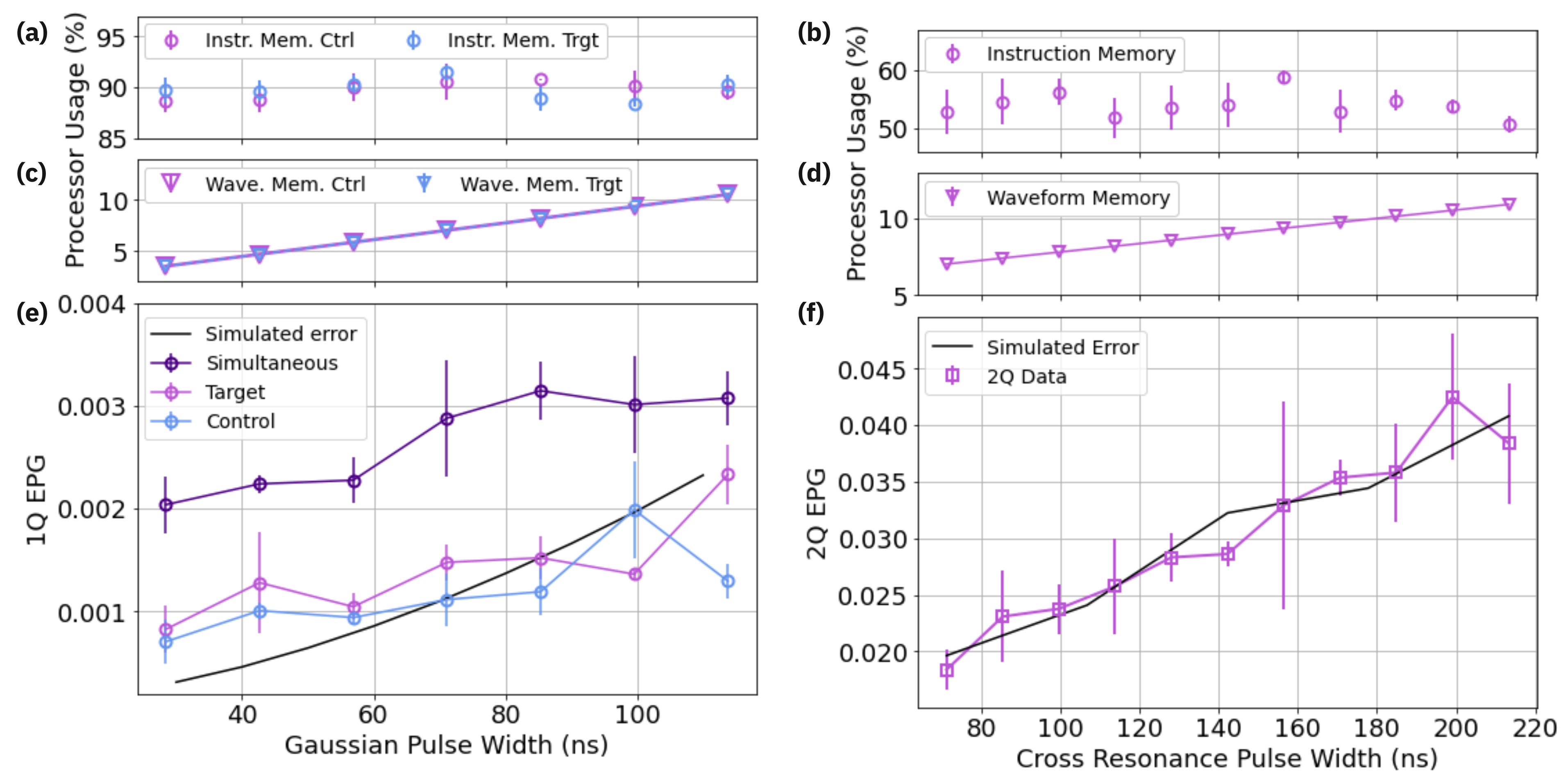}
    \caption{Two--qubit RB data for different single--qubit and two--qubit gate lengths. All data and error bars are averaged over three different RB experiments performed at each gate length. \textbf{(a--b)} The random Clifford gate sequences are stored in instruction memory, and the memory demands are plotted as a percentage of fullness. Instruction memory usage fluctuates because random sequences may have slightly different lengths. \textbf{(c--d)} Pulse definitions for the 1Q and 2Q gates are stored in waveform memory, which is shown to increase linearly with gate length and does not increase the number of instructions. \textbf{(e)} Single--qubit RB measured on each qubit individually and both qubits simultaneously, while sweeping the Gaussian pulse width. The individual RB error is modeled with a Hamiltonian simulation assuming 83.1 kHz of Z--rotation; consistent with oscillations observed in the rotary echo experiment. For simultaneous RB, each qubit is measured and the average error is reported. The simultaneous error is believed to be due to an increase in quantum cross--talk from $ZZ$, and classical cross--talk from the CMOS chip.  \textbf{(f)} Two--qubit RB of an echoed cross--resonance gate as a function of the width of the cross--resonance pulse. Faster gates are observed to have reduced error, consistent with a reduction in decoherence errors; however, simulations reveal qubit coherence is not the leading source of error. Using a parameterized two-qubit Hamiltonian, the additional error was modeled by assuming an amplitude dependent Z--error on the target qubit, and a constant Z--error on the control qubit.}
    \label{fig_RB_vs_gate_length}
\end{figure*}

Single qubit RB experiments were performed both individually and simultaneously on the control and target qubits and characterized as a function of the gate length $t_g$. For each $t_g$, the waveforms were calibrated and measurements of $T_1$ and $T_2$ were interleaved between RB experiments. The results for individual RB experiments are displayed in Fig.~\ref{fig_RB_vs_gate_length}\textbf{(e)}. The data show an error reduction as a function of gate length $t_g$. However, the observed errors are not solely explained by decoherence. For example, the errors do not track with the first order single qubit error model $\epsilon_{1Q}$ ~\cite{Pritchett_2023,Tahereh_2022}. This discrepancy implies that the control electronics are contributing to the excess error measured in the RB experiments. Potential control--related error sources include spectral content as shown in Fig.~\ref{fig_Z_error_meas} and pulse amplitude noise detailed in the section ~\ref{sec_amp_noise}. The spurious spectral content and quasi-static amplitude noise give rise to over/under rotation errors that vary quadratically with noise source amplitude.

During simultaneous operation, each control channel plays unique random Clifford sequences, which gives rise to coherent quantum cross--talk errors of the form $\epsilon^{ZZ}_{1Q}=\frac{1}{6}\left(2\pi ZZ \ t_g \right)^2$ that contribute to the total observed error~\cite{Omalley_2015}. The increase in simultaneous error observed in Fig.~\ref{fig_RB_vs_gate_length}\textbf{(e)} is not explained by error analysis that assumes only coherent quantum cross--talk, a result which implies an external noise source. It is suspected that classical cross-talk on the CMOS chip arises during simultaneous waveform generation, which gives rise to the observed error.

\subsection{Two-Qubit Randomized Benchmarking}
Two--qubit gates are more difficult in practice than single qubit gates. When compared to single qubit gates, two--qubit gates are longer and there are more ways in which errors can arise; furthermore, the coupled qubits are more sensitive than single qubits to error sources. For example, in the decoherence error model $\epsilon_{2Q}$ ~\cite{Pritchett_2023,Tahereh_2022}, the error coefficients are larger and there are more terms that factor into the error (see section ~\ref{sec_error_modeling}). In addition to decoherence, there are other potential sources of error, including phase noise~\cite{Ball2016}, amplitude noise, pulse--induced decay~\cite{Yan2013,Bylander2011}, spurious spectral tones, coherent quantum cross-talk~\cite{Ku2020}, microwave cross-talk~\cite{Ding2020,Wanga2022}, and leakage outside of the computational basis~\cite{McKay_2017,Werninghaus2021}. Additionally, the two--qubit waveform generation is more complex, requiring on average 17.51 instructions per Clifford (IPC) (Fig.~\ref{fig_RB2Q}), compared to 1.71 IPC for single qubit gates (Fig.~\ref{fig_RB1Q}). 

The echoed two--qubit cross--resonance gate was calibrated and benchmarked for different two--qubit gate lengths, as shown in Fig.~\ref{fig_RB_vs_gate_length}\textbf{(d)}. The pulses driving single--qubit rotations were held constant, and measurements of $T_1$ and $T_2$ were interleaved with RB experiments. The error rates were simulated using a two-qubit model Hamiltonian parameterized with experimental data, and the measured RB data does not track with decoherence or coherent quantum cross-talk. 

The simulations suggest the error is due to an always--on Z--rotation combined with an amplitude dependent Z-error on the target qubit. The simulated Z-error is consistent with rotary echo experiments performed on the target qubit that are observed to have an average Z-rotation of 83.1kHz, and modeled using Linblad Master equations (Fig.~\ref{fig_Z_error_meas}\textbf{(a)}). The target Z-rotation is further consistent with spectrum analyzer measurements of the CMOS chip output that reveal excess LO leakage when measured $T=5$K in a closed-cycle \textsuperscript{4}He cryostat. The presence of off--resonance spectral content will give rise to an AC Stark effect~\cite{Schuster2005} which shifts the qubit frequency, resulting in a Z--rotation. As shown in Fig.~\ref{fig_Z_error_meas}\textbf{(c)} the spurious content was observed to be channel dependent, with the LO leakage on the control qubit's channel being the worst of the two. This random variation is in excess of what was expected in the chip's design phase, so the designed tuning range did not cover the distribution observed in hardware samples; the data shows an example where there was sufficient range for one channel and insufficient for the second.

\begin{figure*}
    \centering
    \includegraphics[width=1.75\columnwidth]{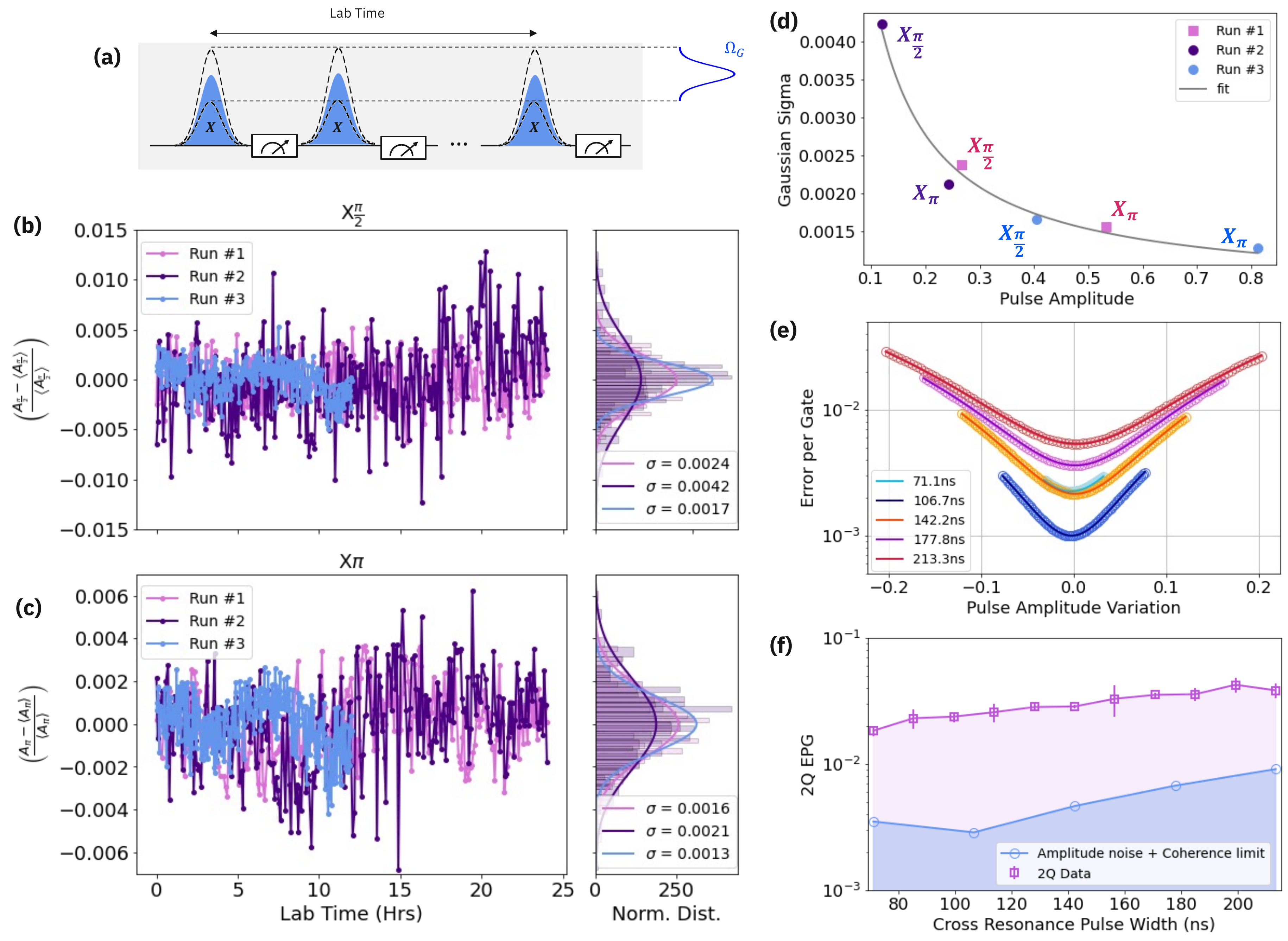}
    \caption{\textbf{(a)} Measurement sequence performed to extract quasi--static amplitude noise. The pulse amplitude calibrations were repeated consecutively, and the $X_{\frac{\pi}{2}}$ and $X_{\pi}$ calibrations were interleaved. The $t_g$ was the same for $X_{\frac{\pi}{2}}$ and $X_{\pi}$, and the amplitude was varied. For each calibration, the amplitude coefficient was observed to fluctuate, consistent with a normal noise distribution. \textbf{(b)},\textbf{(c)} The percent difference of the pulse amplitude coefficients plotted for $X_{\frac{\pi}{2}}$ and $X_{\pi}$, respectively. The time series of coefficients binned data, fit to a normal distribution, and are projected to the panel to the right. The noise was measured for three different experimental configurations: Run 1 was the standard experiment, Run 2 had additional filtering on the RT supply lines and a shorter $t_g$, and Run 3 had increased on-chip attenuation and a longer $t_g$. The pulse amplitude varied with $t_g$. \textbf{(d)} The width of the normal distributions are plotted as a function of the amplitude coefficient. The relationship between the quasi-static noise and pulse amplitude is shown to follow a $1/x$ dependence of the form $f(x) = 0.00041/x+0.0007$. This result indicates the SNR improves for larger DAC amplitudes. \textbf{(e)} Simulated gate error as a fuction of the quasi-static amplitude noise $\sigma$, and for different CR pulse widths. The error is fitted to a quadratic heuristic model of the form $c_{0} + c_{1} \Delta_{amp} + c_{2}\Delta_{amp}^2$. Coefficients are listed in \ref{table_heuristics}. \textbf{(f)} Measured 2Q gate error as a function of CR pulse width, along with the simulated error due to observed amplitude noise. Modeling implies amplitude noise is not leading source of error.}
    \label{fig_amp_noise_experiment}
\end{figure*}

\section{Amplitude Noise}
\label{sec_amp_noise}

In Fig.~\ref{fig_amp_noise_experiment}, a measurement was performed to observe quasi-static amplitude noise. The measurement sequence consisted of repeated amplitude calibrations for $X_{\pi}$ followed by $X_{\pi/2}$. For each calibration, the DAC amplitude coefficient was stored, and the percent difference was plotted as a function of lab time. The data were binned and then fit to a normal distribution. Three different measurements were performed each with slightly different experimental configurations: Run \verb|#|1 was performed with the nominal device configuration and gate length, Run \verb|#|2 was performed with a shorter gate length after adding additional low pass filtering to the supply voltages, and for Run \verb|#|3 the on--chip attenuation was maxed out requiring a long, low amplitude Gaussian pulse to generate a pi--rotation.

For each experimental run, the amplitude required to drive a $X_{\pi}$ and a $X_{\pi/2}$ changed due to different amounts of in-line attenuation. The relative fluctuations in the pulse amplitude were evaluated and shown to increase for larger pulse amplitudes. In Fig.~\ref{fig_amp_noise_experiment}, the distribution width $\sigma$ was plotted vs the pulse amplitude, which yielded a 1/x dependency. Since the different experimental configurations did not deviate from the 1/x dependency, these results imply that the source of the noise is on--chip, and could be related to an increase in low frequency noise at cryogenic temperatures \cite{Oka_2020}.

Error analysis was performed using the 1/x fit of the noise as an input into the error model. We find that the amplitude noise is larger than decoherence errors, but is not significant enough to explain the observed 2Q gate error. For small low frequency fluctuations the amplitude noise will result in either an over or under rotation during the gate.  The error per gate from an over/under rotation can be fit to the form $c_{0} + c_{1} \Delta_{amp} + c_{2}\Delta_{amp}^2$.  The error due to quasi-static amplitude noise was simulated using modeling techniques described in Section~\ref{sec_error_modeling}. The simulated error was fit to the quadratic error model, and the coefficients for different sample lengths as shown in Table \ref{table_heuristics}. Resulting in a simple heuristic error model.
\begin{table*}[!htbp]
\begin{center}
\begin{tabular}{||c c c c c c c||} 
 \hline
 coefficient & 71 ns & 106.7 ns & 142.2 ns & 177.8 ns & 213.3 ns & Mean \\ [0.5ex] 
 \hline\hline
 $c_{2}$ & 0.593127 & 0.353895 & 0.471876 & 0.523495 & 0.548932 & 0.498265\\ 
 \hline
 $c_{1}$ & 0.003120 & 0.001647 & -0.001617 & -0.002103 & -0.004213 & -0.000633 \\
 \hline
 $c_{0}$ & 0.002252 & 0.001001 & 0.002150 & 0.003619 & 0.005373 & 0.002879 \\ [1ex] 
 \hline
\end{tabular}
\end{center}
\caption{Fit parameters for the heuristic gate error model that assumes quasi-static amplitude noise. The fitting routines were applied to numerical results from Hamiltonian simulations using measured device parameters. The quadratic behavior is consistent with "theta squared" errors that arise from over/under angle rotations.}
\label{table_heuristics}
\end{table*}

\section{Rotary Echo Experiment}

\begin{figure}
    \centering
    \includegraphics[width=0.8\columnwidth]{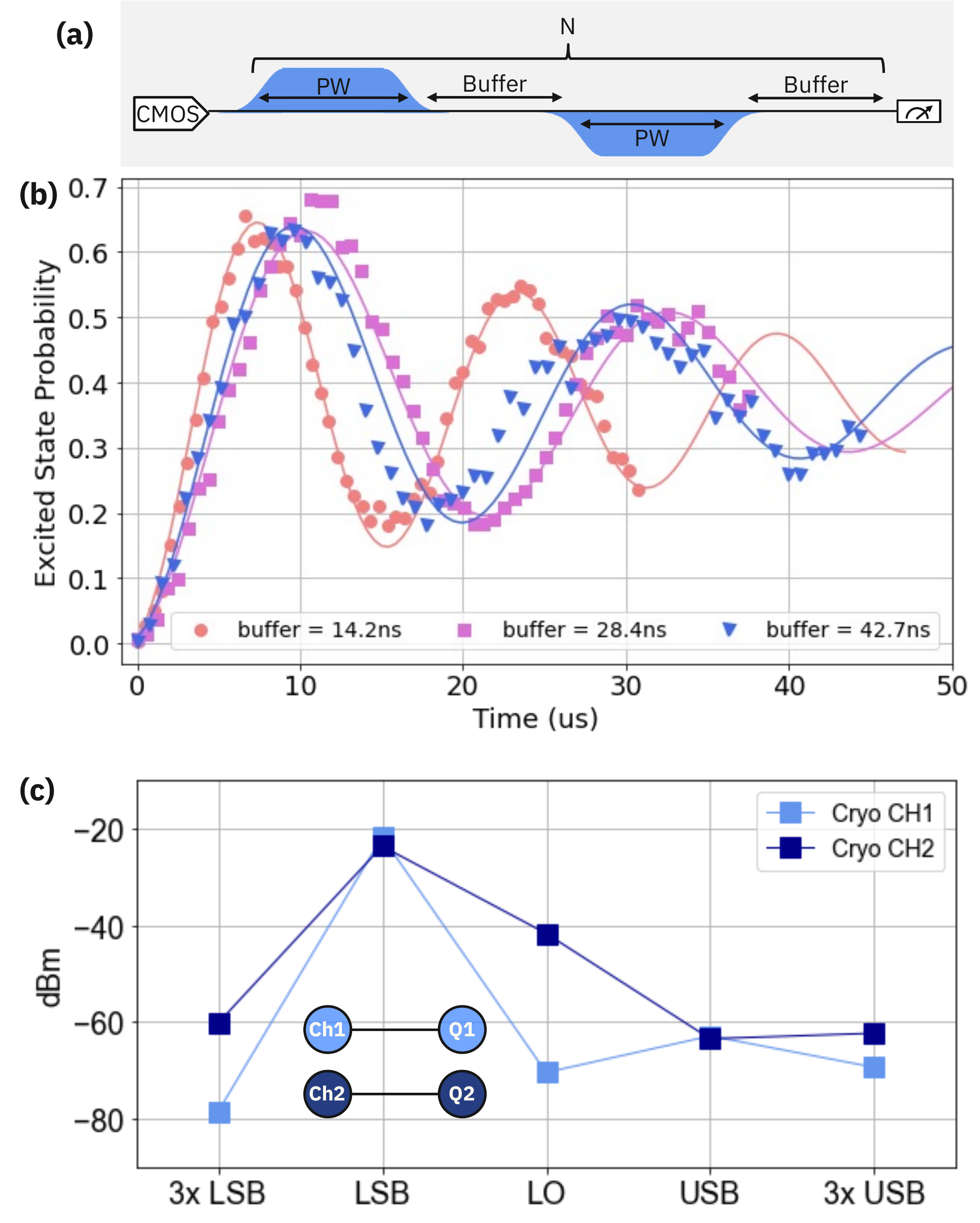}
    \caption{\textbf{(a)} A pulse sequence for a rotary echo experiment. \textbf{(b)} Rotary echo measurement on the control qubit shows oscillatory behavior rather than the expected exponential decay. Data is fit to a master equation simulation that includes: $T1$, $T2$, and a constant $Z$. \textbf{(c)} Measured amplitudes of the most prominent spurious peaks observed in a spectrum analyzer. The LO was set to 5 GHz, with a side band frequency of 250 MHz. The cryogenic measurements were performed at 5K in a \textsuperscript{4}He closed-cycle cryostat prior to loading into a dilution refrigerator for qubit testing. The same bias conditions for the CMOS chip were used for qubit measurements. Channel 1 was connected to the control qubit, and channel 2 was connected to the target qubit. Here channel 2 is observed to have more spurious content, which is consistent with observed behavior on the control qubit. The additional spurious content Stark shifts the qubit, which shifts the energy levels and gives rise to an always on $Z$ error on the target qubit.}
    \label{fig_Z_error_meas}
\end{figure}

 A rotary echo experiment~\cite{Gustavsson2012} was performed to measure driven decay, as shown in Fig.~\ref{fig_Z_error_meas}\textbf{(b)}. The qubit was pulsed in the $+X$ direction, followed by an idle gate, and then pulsed in the $-X$ direction. The pulse width was 200ns and the sequence was repeated for $N=31$ times. Perfectly symmetric and noiseless pulses will yield an exponential decay consistent with the qubit lifetime. Here a time dependent oscillation is observed and it is believed to be due to an AC Stark shift that arises during the pulse. The oscillations are fitted using a two level qubit model where an additional Z--rotation occurs during the $+X$ and $-X$ rotations.  This Z rotation arises from spurious peaks which stark shift the qubit while it is being driven. As shown in Fig.~\ref{fig_Z_error_meas}\textbf{(c)}
 
 An average Z--rotation strength of 83.1 KHz was extracted from fitting to the data with the three different buffer lengths: 14.2ns, 28.4ns, and 42.7ns.  The simulation consisted of solving the Lindblad master equation for a single qubit with a time dependent X--pulse with an additional time dependent Z--pulse.  All pulses were square--Gaussian~\cite{Malekakhlagh_2022} shaped with a 16 ns sigma.  The X rotation rate, Z rotation rate, T1, and T2 were all input variables used during the fit and most variation was tied to changes in the Z rotation strength.

\begin{table}
\begin{center}
\begin{tabular}{||c c c c c||} 
 \hline
 Buffer (ns) & X--rate (MHz) & Z--rate (KHz) & T1 (us) & T2 (us) \\ [0.5ex] 
 \hline\hline
 14.2 & 1.897 & 94.3 & 82.2 & 33.5 \\ 
 \hline
 28.4 & 1.927 & 72.5 & 82.2 & 32.6 \\
 \hline
 42.7 & 1.927 & 82.6 & 79.9 & 33.1 \\ [1ex] 
 \hline
\end{tabular}
\end{center}
\caption{Fit parameters extracted from the master equation simulation. A Nelder-Mead optimization routine was performed on the data-set from Fig.\ref{fig_Z_error_meas} to determine best fit. The Z-rate was a the free parameter in each fit. The X-rate, T1, and T2 were allowed small bounds in order to achieve an optimal fit. The optimal T1, T2 fit parameters are different than observed values reported in Table \ref{table_device_params}. The deviations are attributed to TLS fluctuations between the time the rotary echo data was collected \cite{carroll_2022,Thorbeck2023}.}
\label{table_rotary_sims}
\end{table}

\section{Modeling Cross-Resonance Gate Errors}
\label{sec_error_modeling}

To model the effect of spurious Z-errors on our qubits, we numerically calibrate the cross resonance (CR) gate using a two-qubit model Hamiltonian
\begin{eqnarray}\label{CRHam}
H=\sum_{c\in\{a,b\}}\left[\omega_{c}c^\dagger c + \frac{\delta_c}{2}c^\dagger c (c^\dagger c -\mathbb{I})\right] \nonumber 
\\ + J(a+a^\dagger)(b+b^\dagger) \nonumber 
\\+ \Omega_x(t)\cos(\omega_b t) (a+a^\dagger)
\end{eqnarray}
which describes two transmon qubits with lowering operators $a$ and $b$ within a Duffing model with anharmonicity $\delta_{a/b}$ and coupling strength $J$.  To generate the cross-resonance entangling interaction we drive the control qubit ($a+a^\dagger$) with the target qubit's frequency ($\omega_b$).  $\Omega_x(t)$ describes the pulse envelope which for CR gates we use the Gaussian square shape $\Omega_{\rm GS}(t)$ described in Eq. (2).   Our pulse is allowed to rise (and fall) over twice $\sigma_{\rm GS}$, the Gaussian width, before (and after) a square pulse of duration $\tau_{\rm p}-4\sigma_{\rm GS}$.  
For each pulse width considered, we have calibrate the amplitude of the pulse in (\ref{CRHam}) to minimize the total error of a $U_{\rm ideal}=e^{-i\pi ZX/4}$ rotation (the native entangler produced by the CR drive).  We do this by performing time-domain simulations of the Hamitonian in (\ref{CRHam}) to estimate $U_{\rm sim}$, then minimizing the two-qubit gate error as defined by

\begin{equation}
{\rm error}_{2Q}=1-
\left( \left\vert{\rm Tr}[U_{\rm sim}^\dagger\cdot U_{\rm ideal}]\right\vert^2/4+1\right)/5.
\end{equation}

\begin{figure}
    \centering
    \includegraphics[width=0.8\columnwidth]{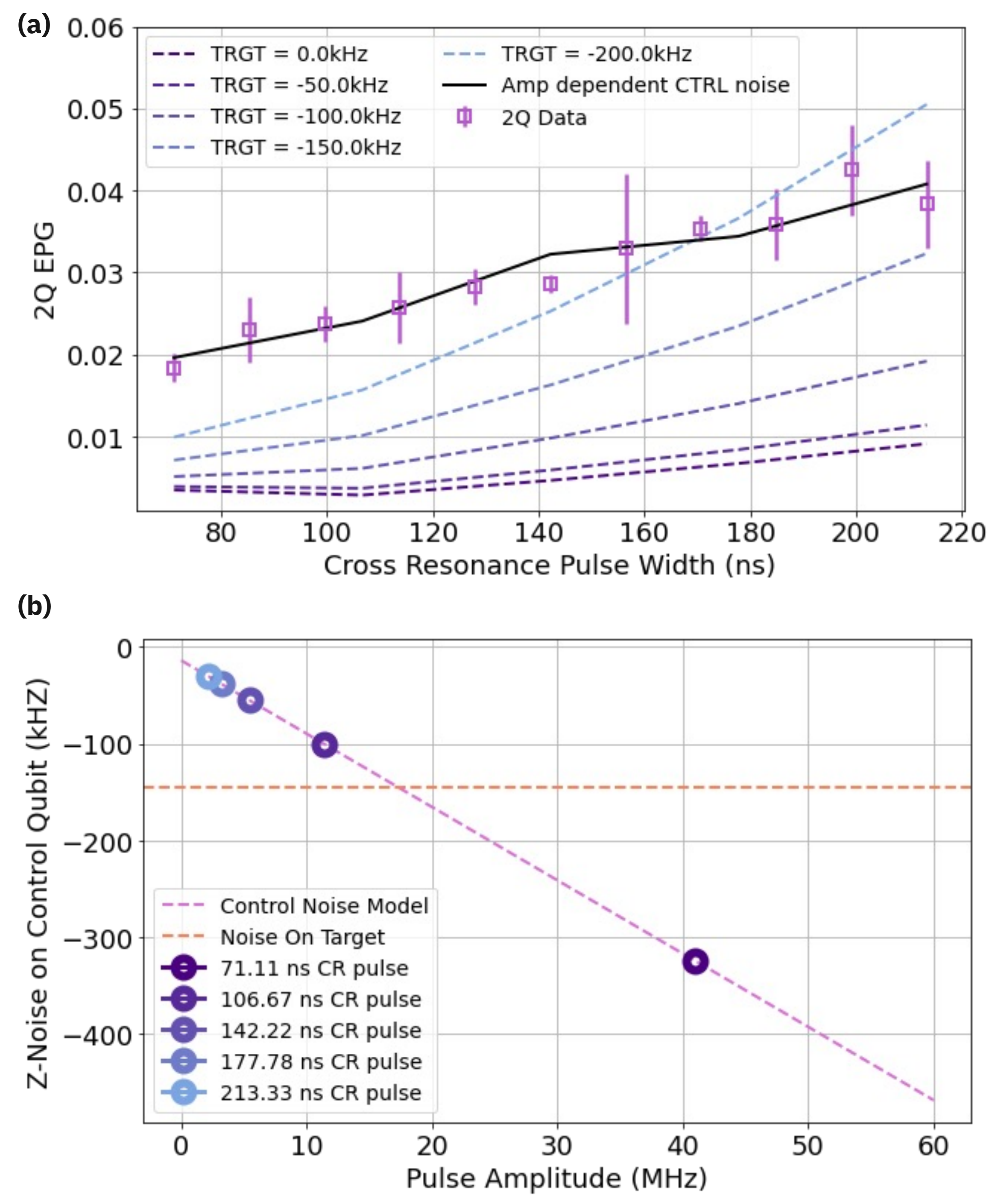}
    \caption{\textbf{(a)} Measured and simulated two--qubit EPG as a function of the CR pulse width. Simulated error is shown for different amounts of Z-noise on target qubit, while the Z--noise on the control qubit is held constant at the measured $85$kHz. The difference in slopes between measured error and simulated error implies a constant Z--noise model is not sufficient for describing the observed errors. A best fit is obtained by introducing an amplitude dependent Z--noise on the control qubit, with the Z--noise on the target qubit fixed to $145kHz$. We note that the amplitude dependent Z--noise on the control and the constant Z--noise on the target are not explained by the experimental data that was collected.
    \textbf{(b)} The simulated Z--noise as a function of the CR pulse amplitude. The observed EPG was best captured by assuming Z--noise that varies linearly with pulse amplitude.}
    \label{fig_z_error_simulation}
\end{figure}

\section{Outlook}
The primary concern with realizing cryo-CMOS electronics for controlling qubits is the cooling limitation set by the dilution refrigerator. The CMOS ASIC used in these experiments is comprised of two low power RF analog front ends that dissipate ~8.92mW of power per channel, and a quantum specific processor that dissipates ~10.42mW per channel (see Fig. \ref{fig_experimental_setup}). Assuming cooling powers in a standard dilution refrigerator, it is believed that up to 100 cryo-CMOS channels could be integrated into a quantum computing system. The outlook becomes more optimistic when taking into account circuit innovations and advances in cooling infrastructure. 

A key insight from this work is that the processor requirements for common calibrations and quantum information experiments will need to increase as the performance of the quantum processor improves, but increasing processor capabilities will lead to more power dissipation from the CMOS chip. To mitigate the demand for more processing power new instruction set architectures maybe required, or new low overhead experiments for calibrating and characterizing qubits will be needed. 

Operating CMOS control electronics within the dilution refrigerator has the potential for improved gate performance due to lower noise, reduced loss, and less dispersion; however, the technical nuances of cryogenic operation and system integration make it challenging to achieve this potential. For example, most foundry circuit models are not reliable below 50K, surface mount components do not meet spec at cryogenic temperatures, and the electrical path length between the support electronics and ASICs is long and lossy when compared to CMOS based servers. Consequently, this technology will take time to achieve it's full potential.

This manuscript describes the first demonstration of 2-qubit randomized benchmarking with a cryogenic CMOS controller, with an observed error per gate of $\epsilon_\text{2Q}$=1.4e-2. The leading source of error is shown to arise from the electronics; however, an advantage of custom CMOS electronics is that new circuits can be designed to mitigate error sources after they have been identified. The primary engineering challenge arises from being able to distinguish errors arising from devices physics versus errors arising from control electronics. Identifying error sources is a non-trivial task, but these efforts are becoming simpler due to innovative approaches being developed in the field of QCVV.

\section{Conclusions}
We have developed a low--power CMOS ASIC designed to operate at $T = 4$K that is able to generate sequences of RF waveforms for controlling, calibrating, and benchmarking a universal set of quantum gates between a pair of transmon qubits. The cryogenic control electronics were used to demonstrate high fidelity two--qubit cross-resonance gates. A two--qubit Hamiltonian model provides insight into the behavior of spurious Z--errors, which indicate the control electronics noise has an amplitude dependence. Modeling and analysis suggests that the observed drive-depended Z--rotation during rotary echo experiments and of LO leakage in the ASIC's output are connected, implying that spurious content from the CMOS--chip is the primary source of gate error.

The CMOS processor was characterized across a wide variety of qubit experiments demonstrating its viability for providing control pulses to next--generation quantum computers. Furthermore, these results highlight challenges with low-power cryogenic control electronics related to the instruction and memory requirements for standard qubit experiments. These results underscore the need for further innovation of digital architectures as gate error rates approach fault--tolerant thresholds.

\begin{acknowledgments}
We thank Oliver Dial, Muir Kumph, David McKay, and George Zettles for technical advice and support. We acknowledge Donald Bethune for his development of the cryogenic infrastructure, Thomas Fox for his contribution to the instruction set architecture, and thank Jiri Stehlik, David Zajak, and Seth Merkle for helpful technical discussions.
\end{acknowledgments}

\bibliography{bibliography}

\begin{thebibliography}{71}%
\makeatletter
\providecommand \@ifxundefined [1]{%
 \@ifx{#1\undefined}
}%
\providecommand \@ifnum [1]{%
 \ifnum #1\expandafter \@firstoftwo
 \else \expandafter \@secondoftwo
 \fi
}%
\providecommand \@ifx [1]{%
 \ifx #1\expandafter \@firstoftwo
 \else \expandafter \@secondoftwo
 \fi
}%
\providecommand \natexlab [1]{#1}%
\providecommand \enquote  [1]{``#1''}%
\providecommand \bibnamefont  [1]{#1}%
\providecommand \bibfnamefont [1]{#1}%
\providecommand \citenamefont [1]{#1}%
\providecommand \href@noop [0]{\@secondoftwo}%
\providecommand \href [0]{\begingroup \@sanitize@url \@href}%
\providecommand \@href[1]{\@@startlink{#1}\@@href}%
\providecommand \@@href[1]{\endgroup#1\@@endlink}%
\providecommand \@sanitize@url [0]{\catcode `\\12\catcode `\$12\catcode
  `\&12\catcode `\#12\catcode `\^12\catcode `\_12\catcode `\%12\relax}%
\providecommand \@@startlink[1]{}%
\providecommand \@@endlink[0]{}%
\providecommand \url  [0]{\begingroup\@sanitize@url \@url }%
\providecommand \@url [1]{\endgroup\@href {#1}{\urlprefix }}%
\providecommand \urlprefix  [0]{URL }%
\providecommand \Eprint [0]{\href }%
\providecommand \doibase [0]{https://doi.org/}%
\providecommand \selectlanguage [0]{\@gobble}%
\providecommand \bibinfo  [0]{\@secondoftwo}%
\providecommand \bibfield  [0]{\@secondoftwo}%
\providecommand \translation [1]{[#1]}%
\providecommand \BibitemOpen [0]{}%
\providecommand \bibitemStop [0]{}%
\providecommand \bibitemNoStop [0]{.\EOS\space}%
\providecommand \EOS [0]{\spacefactor3000\relax}%
\providecommand \BibitemShut  [1]{\csname bibitem#1\endcsname}%
\let\auto@bib@innerbib\@empty
\bibitem [{\citenamefont {Bravyi}\ \emph {et~al.}(2022)\citenamefont {Bravyi},
  \citenamefont {Dial}, \citenamefont {Gambetta}, \citenamefont {Gil},\ and\
  \citenamefont {Nazario}}]{Bravyi2022}%
  \BibitemOpen
  \bibfield  {author} {\bibinfo {author} {\bibfnamefont {S.}~\bibnamefont
  {Bravyi}}, \bibinfo {author} {\bibfnamefont {O.}~\bibnamefont {Dial}},
  \bibinfo {author} {\bibfnamefont {J.~M.}\ \bibnamefont {Gambetta}}, \bibinfo
  {author} {\bibfnamefont {D.}~\bibnamefont {Gil}},\ and\ \bibinfo {author}
  {\bibfnamefont {Z.}~\bibnamefont {Nazario}},\ }\bibfield  {title} {\bibinfo
  {title} {The future of quantum computing with superconducting qubits},\
  }\href {https://doi.org/10.1063/5.0082975} {\bibfield  {journal} {\bibinfo
  {journal} {Journal of Applied Physics}\ }\textbf {\bibinfo {volume} {132}},\
  \bibinfo {pages} {160902} (\bibinfo {year} {2022})}\BibitemShut {NoStop}%
\bibitem [{\citenamefont {Gambetta}\ \emph {et~al.}()\citenamefont {Gambetta},
  \citenamefont {Chow},\ and\ \citenamefont {Steffen}}]{Gambetta2017}%
  \BibitemOpen
  \bibfield  {author} {\bibinfo {author} {\bibfnamefont {J.~M.}\ \bibnamefont
  {Gambetta}}, \bibinfo {author} {\bibfnamefont {J.~M.}\ \bibnamefont {Chow}},\
  and\ \bibinfo {author} {\bibfnamefont {M.}~\bibnamefont {Steffen}},\
  }\bibfield  {title} {\bibinfo {title} {Building logical qubits in a
  superconducting quantum computing system},\ }\href
  {https://doi.org/10.1038/s41534-016-0004-0} {\ \textbf {\bibinfo {volume}
  {3}},\ \bibinfo {pages} {2}}\BibitemShut {NoStop}%
\bibitem [{\citenamefont {Van~Meter}\ and\ \citenamefont
  {Horsman}(2013)}]{VanMeter2013}%
  \BibitemOpen
  \bibfield  {author} {\bibinfo {author} {\bibfnamefont {R.}~\bibnamefont
  {Van~Meter}}\ and\ \bibinfo {author} {\bibfnamefont {D.}~\bibnamefont
  {Horsman}},\ }\bibfield  {title} {\bibinfo {title} {A blueprint for building
  a quantum computer}\ }\textbf {\bibinfo {volume} {56}},\ \href
  {https://doi.org/10.1145/2494568} {10.1145/2494568} (\bibinfo {year}
  {2013})\BibitemShut {NoStop}%
\bibitem [{\citenamefont {Hornibrook}\ \emph {et~al.}(2015)\citenamefont
  {Hornibrook}, \citenamefont {Colless}, \citenamefont {Conway~Lamb},
  \citenamefont {Pauka}, \citenamefont {Lu}, \citenamefont {Gossard},
  \citenamefont {Watson}, \citenamefont {Gardner}, \citenamefont {Fallahi},
  \citenamefont {Manfra},\ and\ \citenamefont {Reilly}}]{Hornibrook2015}%
  \BibitemOpen
  \bibfield  {author} {\bibinfo {author} {\bibfnamefont {J.~M.}\ \bibnamefont
  {Hornibrook}}, \bibinfo {author} {\bibfnamefont {J.~I.}\ \bibnamefont
  {Colless}}, \bibinfo {author} {\bibfnamefont {I.~D.}\ \bibnamefont
  {Conway~Lamb}}, \bibinfo {author} {\bibfnamefont {S.~J.}\ \bibnamefont
  {Pauka}}, \bibinfo {author} {\bibfnamefont {H.}~\bibnamefont {Lu}}, \bibinfo
  {author} {\bibfnamefont {A.~C.}\ \bibnamefont {Gossard}}, \bibinfo {author}
  {\bibfnamefont {J.~D.}\ \bibnamefont {Watson}}, \bibinfo {author}
  {\bibfnamefont {G.~C.}\ \bibnamefont {Gardner}}, \bibinfo {author}
  {\bibfnamefont {S.}~\bibnamefont {Fallahi}}, \bibinfo {author} {\bibfnamefont
  {M.~J.}\ \bibnamefont {Manfra}},\ and\ \bibinfo {author} {\bibfnamefont
  {D.~J.}\ \bibnamefont {Reilly}},\ }\bibfield  {title} {\bibinfo {title}
  {Cryogenic control architecture for large-scale quantum computing},\ }\href
  {https://doi.org/10.1103/PhysRevApplied.3.024010} {\bibfield  {journal}
  {\bibinfo  {journal} {Phys. Rev. Applied}\ }\textbf {\bibinfo {volume} {3}},\
  \bibinfo {pages} {024010} (\bibinfo {year} {2015})}\BibitemShut {NoStop}%
\bibitem [{\citenamefont {Xue}\ \emph {et~al.}()\citenamefont {Xue},
  \citenamefont {Patra}, \citenamefont {van Dijk}, \citenamefont {Samkharadze},
  \citenamefont {Subramanian}, \citenamefont {Corna}, \citenamefont
  {Paquelet~Wuetz}, \citenamefont {Jeon}, \citenamefont {Sheikh}, \citenamefont
  {Juarez-Hernandez}, \citenamefont {Esparza}, \citenamefont {Rampurawala},
  \citenamefont {Carlton}, \citenamefont {Ravikumar}, \citenamefont {Nieva},
  \citenamefont {Kim}, \citenamefont {Lee}, \citenamefont {Sammak},
  \citenamefont {Scappucci}, \citenamefont {Veldhorst}, \citenamefont
  {Sebastiano}, \citenamefont {Babaie}, \citenamefont {Pellerano},
  \citenamefont {Charbon},\ and\ \citenamefont {Vandersypen}}]{xue2021}%
  \BibitemOpen
  \bibfield  {author} {\bibinfo {author} {\bibfnamefont {X.}~\bibnamefont
  {Xue}}, \bibinfo {author} {\bibfnamefont {B.}~\bibnamefont {Patra}}, \bibinfo
  {author} {\bibfnamefont {J.~P.~G.}\ \bibnamefont {van Dijk}}, \bibinfo
  {author} {\bibfnamefont {N.}~\bibnamefont {Samkharadze}}, \bibinfo {author}
  {\bibfnamefont {S.}~\bibnamefont {Subramanian}}, \bibinfo {author}
  {\bibfnamefont {A.}~\bibnamefont {Corna}}, \bibinfo {author} {\bibfnamefont
  {B.}~\bibnamefont {Paquelet~Wuetz}}, \bibinfo {author} {\bibfnamefont
  {C.}~\bibnamefont {Jeon}}, \bibinfo {author} {\bibfnamefont {F.}~\bibnamefont
  {Sheikh}}, \bibinfo {author} {\bibfnamefont {E.}~\bibnamefont
  {Juarez-Hernandez}}, \bibinfo {author} {\bibfnamefont {B.~P.}\ \bibnamefont
  {Esparza}}, \bibinfo {author} {\bibfnamefont {H.}~\bibnamefont
  {Rampurawala}}, \bibinfo {author} {\bibfnamefont {B.}~\bibnamefont
  {Carlton}}, \bibinfo {author} {\bibfnamefont {S.}~\bibnamefont {Ravikumar}},
  \bibinfo {author} {\bibfnamefont {C.}~\bibnamefont {Nieva}}, \bibinfo
  {author} {\bibfnamefont {S.}~\bibnamefont {Kim}}, \bibinfo {author}
  {\bibfnamefont {H.-J.}\ \bibnamefont {Lee}}, \bibinfo {author} {\bibfnamefont
  {A.}~\bibnamefont {Sammak}}, \bibinfo {author} {\bibfnamefont
  {G.}~\bibnamefont {Scappucci}}, \bibinfo {author} {\bibfnamefont
  {M.}~\bibnamefont {Veldhorst}}, \bibinfo {author} {\bibfnamefont
  {F.}~\bibnamefont {Sebastiano}}, \bibinfo {author} {\bibfnamefont
  {M.}~\bibnamefont {Babaie}}, \bibinfo {author} {\bibfnamefont
  {S.}~\bibnamefont {Pellerano}}, \bibinfo {author} {\bibfnamefont
  {E.}~\bibnamefont {Charbon}},\ and\ \bibinfo {author} {\bibfnamefont
  {L.~M.~K.}\ \bibnamefont {Vandersypen}},\ }\bibfield  {title} {\bibinfo
  {title} {{CMOS}-based cryogenic control of silicon quantum circuits},\ }\href
  {https://doi.org/10.1038/s41586-021-03469-4} {\ \textbf {\bibinfo {volume}
  {593}},\ \bibinfo {pages} {205}}\BibitemShut {NoStop}%
\bibitem [{\citenamefont {Pauka}\ \emph {et~al.}()\citenamefont {Pauka},
  \citenamefont {Das}, \citenamefont {Kalra}, \citenamefont {Moini},
  \citenamefont {Yang}, \citenamefont {Trainer}, \citenamefont {Bousquet},
  \citenamefont {Cantaloube}, \citenamefont {Dick}, \citenamefont {Gardner},
  \citenamefont {Manfra},\ and\ \citenamefont {Reilly}}]{Pauka2021}%
  \BibitemOpen
  \bibfield  {author} {\bibinfo {author} {\bibfnamefont {S.~J.}\ \bibnamefont
  {Pauka}}, \bibinfo {author} {\bibfnamefont {K.}~\bibnamefont {Das}}, \bibinfo
  {author} {\bibfnamefont {R.}~\bibnamefont {Kalra}}, \bibinfo {author}
  {\bibfnamefont {A.}~\bibnamefont {Moini}}, \bibinfo {author} {\bibfnamefont
  {Y.}~\bibnamefont {Yang}}, \bibinfo {author} {\bibfnamefont {M.}~\bibnamefont
  {Trainer}}, \bibinfo {author} {\bibfnamefont {A.}~\bibnamefont {Bousquet}},
  \bibinfo {author} {\bibfnamefont {C.}~\bibnamefont {Cantaloube}}, \bibinfo
  {author} {\bibfnamefont {N.}~\bibnamefont {Dick}}, \bibinfo {author}
  {\bibfnamefont {G.~C.}\ \bibnamefont {Gardner}}, \bibinfo {author}
  {\bibfnamefont {M.~J.}\ \bibnamefont {Manfra}},\ and\ \bibinfo {author}
  {\bibfnamefont {D.~J.}\ \bibnamefont {Reilly}},\ }\bibfield  {title}
  {\bibinfo {title} {A cryogenic {CMOS} chip for generating control signals for
  multiple qubits},\ }\href {https://doi.org/10.1038/s41928-020-00528-y} {\
  \textbf {\bibinfo {volume} {4}},\ \bibinfo {pages} {64}}\BibitemShut
  {NoStop}%
\bibitem [{\citenamefont {Charbon}\ \emph {et~al.}(2016)\citenamefont
  {Charbon}, \citenamefont {Sebastiano}, \citenamefont {Vladimirescu},
  \citenamefont {Homulle}, \citenamefont {Visser}, \citenamefont {Song},\ and\
  \citenamefont {Incandela}}]{Charbon2016}%
  \BibitemOpen
  \bibfield  {author} {\bibinfo {author} {\bibfnamefont {E.}~\bibnamefont
  {Charbon}}, \bibinfo {author} {\bibfnamefont {F.}~\bibnamefont {Sebastiano}},
  \bibinfo {author} {\bibfnamefont {A.}~\bibnamefont {Vladimirescu}}, \bibinfo
  {author} {\bibfnamefont {H.}~\bibnamefont {Homulle}}, \bibinfo {author}
  {\bibfnamefont {S.}~\bibnamefont {Visser}}, \bibinfo {author} {\bibfnamefont
  {L.}~\bibnamefont {Song}},\ and\ \bibinfo {author} {\bibfnamefont {R.~M.}\
  \bibnamefont {Incandela}},\ }\bibfield  {title} {\bibinfo {title} {Cryo-cmos
  for quantum computing},\ }in\ \href
  {https://doi.org/10.1109/IEDM.2016.7838410} {\emph {\bibinfo {booktitle}
  {2016 IEEE International Electron Devices Meeting (IEDM)}}}\ (\bibinfo {year}
  {2016})\ pp.\ \bibinfo {pages} {13.5.1--13.5.4}\BibitemShut {NoStop}%
\bibitem [{\citenamefont {Reilly}(2019)}]{Reilly2019}%
  \BibitemOpen
  \bibfield  {author} {\bibinfo {author} {\bibfnamefont {D.~J.}\ \bibnamefont
  {Reilly}},\ }\bibfield  {title} {\bibinfo {title} {Challenges in scaling-up
  the control interface of a quantum computer},\ }in\ \href
  {https://doi.org/10.1109/IEDM19573.2019.8993497} {\emph {\bibinfo {booktitle}
  {2019 IEEE International Electron Devices Meeting (IEDM)}}}\ (\bibinfo {year}
  {2019})\ pp.\ \bibinfo {pages} {31.7.1--31.7.6}\BibitemShut {NoStop}%
\bibitem [{\citenamefont {Bardin}\ \emph {et~al.}(2019)\citenamefont {Bardin},
  \citenamefont {Jeffrey}, \citenamefont {Lucero}, \citenamefont {Huang},
  \citenamefont {Naaman}, \citenamefont {Barends}, \citenamefont {White},
  \citenamefont {Giustina}, \citenamefont {Sank}, \citenamefont {Roushan},
  \citenamefont {Arya}, \citenamefont {Chiaro}, \citenamefont {Kelly},
  \citenamefont {Chen}, \citenamefont {Burkett}, \citenamefont {Chen},
  \citenamefont {Dunsworth}, \citenamefont {Fowler}, \citenamefont {Foxen},
  \citenamefont {Gidney}, \citenamefont {Graff}, \citenamefont {Klimov},
  \citenamefont {Mutus}, \citenamefont {McEwen}, \citenamefont {Megrant},
  \citenamefont {Neeley}, \citenamefont {Neill}, \citenamefont {Quintana},
  \citenamefont {Vainsencher}, \citenamefont {Neven},\ and\ \citenamefont
  {Martinis}}]{Bardin2019}%
  \BibitemOpen
  \bibfield  {author} {\bibinfo {author} {\bibfnamefont {J.~C.}\ \bibnamefont
  {Bardin}}, \bibinfo {author} {\bibfnamefont {E.}~\bibnamefont {Jeffrey}},
  \bibinfo {author} {\bibfnamefont {E.}~\bibnamefont {Lucero}}, \bibinfo
  {author} {\bibfnamefont {T.}~\bibnamefont {Huang}}, \bibinfo {author}
  {\bibfnamefont {O.}~\bibnamefont {Naaman}}, \bibinfo {author} {\bibfnamefont
  {R.}~\bibnamefont {Barends}}, \bibinfo {author} {\bibfnamefont
  {T.}~\bibnamefont {White}}, \bibinfo {author} {\bibfnamefont
  {M.}~\bibnamefont {Giustina}}, \bibinfo {author} {\bibfnamefont
  {D.}~\bibnamefont {Sank}}, \bibinfo {author} {\bibfnamefont {P.}~\bibnamefont
  {Roushan}}, \bibinfo {author} {\bibfnamefont {K.}~\bibnamefont {Arya}},
  \bibinfo {author} {\bibfnamefont {B.}~\bibnamefont {Chiaro}}, \bibinfo
  {author} {\bibfnamefont {J.}~\bibnamefont {Kelly}}, \bibinfo {author}
  {\bibfnamefont {J.}~\bibnamefont {Chen}}, \bibinfo {author} {\bibfnamefont
  {B.}~\bibnamefont {Burkett}}, \bibinfo {author} {\bibfnamefont
  {Y.}~\bibnamefont {Chen}}, \bibinfo {author} {\bibfnamefont {A.}~\bibnamefont
  {Dunsworth}}, \bibinfo {author} {\bibfnamefont {A.}~\bibnamefont {Fowler}},
  \bibinfo {author} {\bibfnamefont {B.}~\bibnamefont {Foxen}}, \bibinfo
  {author} {\bibfnamefont {C.}~\bibnamefont {Gidney}}, \bibinfo {author}
  {\bibfnamefont {R.}~\bibnamefont {Graff}}, \bibinfo {author} {\bibfnamefont
  {P.}~\bibnamefont {Klimov}}, \bibinfo {author} {\bibfnamefont
  {J.}~\bibnamefont {Mutus}}, \bibinfo {author} {\bibfnamefont
  {M.}~\bibnamefont {McEwen}}, \bibinfo {author} {\bibfnamefont
  {A.}~\bibnamefont {Megrant}}, \bibinfo {author} {\bibfnamefont
  {M.}~\bibnamefont {Neeley}}, \bibinfo {author} {\bibfnamefont
  {C.}~\bibnamefont {Neill}}, \bibinfo {author} {\bibfnamefont
  {C.}~\bibnamefont {Quintana}}, \bibinfo {author} {\bibfnamefont
  {A.}~\bibnamefont {Vainsencher}}, \bibinfo {author} {\bibfnamefont
  {H.}~\bibnamefont {Neven}},\ and\ \bibinfo {author} {\bibfnamefont
  {J.}~\bibnamefont {Martinis}},\ }\bibfield  {title} {\bibinfo {title} {29.1 a
  28nm bulk-cmos 4-to-8ghz ¡2mw cryogenic pulse modulator for scalable quantum
  computing},\ }in\ \href {https://doi.org/10.1109/ISSCC.2019.8662480} {\emph
  {\bibinfo {booktitle} {2019 IEEE International Solid- State Circuits
  Conference - (ISSCC)}}}\ (\bibinfo {year} {2019})\ pp.\ \bibinfo {pages}
  {456--458}\BibitemShut {NoStop}%
\bibitem [{\citenamefont {Frank}\ \emph {et~al.}(2022)\citenamefont {Frank},
  \citenamefont {Chakraborty}, \citenamefont {Tien}, \citenamefont {Rosno},
  \citenamefont {Fox}, \citenamefont {Yeck}, \citenamefont {Glick},
  \citenamefont {Robertazzi}, \citenamefont {Richetta}, \citenamefont
  {Bulzacchelli}, \citenamefont {Ramirez}, \citenamefont {Yilma}, \citenamefont
  {Davies}, \citenamefont {Joshi}, \citenamefont {Chambers}, \citenamefont
  {Lekuch}, \citenamefont {Inoue}, \citenamefont {Underwood}, \citenamefont
  {Wisnieff}, \citenamefont {Baks}, \citenamefont {Bethune}, \citenamefont
  {Timmerwilke}, \citenamefont {Johnson}, \citenamefont {Gaucher},\ and\
  \citenamefont {Friedman}}]{Frank2022}%
  \BibitemOpen
  \bibfield  {author} {\bibinfo {author} {\bibfnamefont {D.~J.}\ \bibnamefont
  {Frank}}, \bibinfo {author} {\bibfnamefont {S.}~\bibnamefont {Chakraborty}},
  \bibinfo {author} {\bibfnamefont {K.}~\bibnamefont {Tien}}, \bibinfo {author}
  {\bibfnamefont {P.}~\bibnamefont {Rosno}}, \bibinfo {author} {\bibfnamefont
  {T.}~\bibnamefont {Fox}}, \bibinfo {author} {\bibfnamefont {M.}~\bibnamefont
  {Yeck}}, \bibinfo {author} {\bibfnamefont {J.~A.}\ \bibnamefont {Glick}},
  \bibinfo {author} {\bibfnamefont {R.}~\bibnamefont {Robertazzi}}, \bibinfo
  {author} {\bibfnamefont {R.}~\bibnamefont {Richetta}}, \bibinfo {author}
  {\bibfnamefont {J.~F.}\ \bibnamefont {Bulzacchelli}}, \bibinfo {author}
  {\bibfnamefont {D.}~\bibnamefont {Ramirez}}, \bibinfo {author} {\bibfnamefont
  {D.}~\bibnamefont {Yilma}}, \bibinfo {author} {\bibfnamefont
  {A.}~\bibnamefont {Davies}}, \bibinfo {author} {\bibfnamefont {R.~V.}\
  \bibnamefont {Joshi}}, \bibinfo {author} {\bibfnamefont {S.~D.}\ \bibnamefont
  {Chambers}}, \bibinfo {author} {\bibfnamefont {S.}~\bibnamefont {Lekuch}},
  \bibinfo {author} {\bibfnamefont {K.}~\bibnamefont {Inoue}}, \bibinfo
  {author} {\bibfnamefont {D.}~\bibnamefont {Underwood}}, \bibinfo {author}
  {\bibfnamefont {D.}~\bibnamefont {Wisnieff}}, \bibinfo {author}
  {\bibfnamefont {C.}~\bibnamefont {Baks}}, \bibinfo {author} {\bibfnamefont
  {D.}~\bibnamefont {Bethune}}, \bibinfo {author} {\bibfnamefont
  {J.}~\bibnamefont {Timmerwilke}}, \bibinfo {author} {\bibfnamefont {B.~R.}\
  \bibnamefont {Johnson}}, \bibinfo {author} {\bibfnamefont {B.~P.}\
  \bibnamefont {Gaucher}},\ and\ \bibinfo {author} {\bibfnamefont {D.~J.}\
  \bibnamefont {Friedman}},\ }\bibfield  {title} {\bibinfo {title} {A
  cryo-{CMOS} low-power semi-autonomous qubit state controller in 14nm {FinFET}
  technology},\ }in\ \href {https://doi.org/10.1109/ISSCC42614.2022.9731538}
  {\emph {\bibinfo {booktitle} {2022 IEEE International Solid- State Circuits
  Conference (ISSCC)}}},\ Vol.~\bibinfo {volume} {65}\ (\bibinfo {year}
  {2022})\ pp.\ \bibinfo {pages} {360--362}\BibitemShut {NoStop}%
\bibitem [{\citenamefont {Chakraborty}\ \emph {et~al.}(2022)\citenamefont
  {Chakraborty}, \citenamefont {Frank}, \citenamefont {Tien}, \citenamefont
  {Rosno}, \citenamefont {Yeck}, \citenamefont {Glick}, \citenamefont
  {Robertazzi}, \citenamefont {Richetta}, \citenamefont {Bulzacchelli},
  \citenamefont {Underwood}, \citenamefont {Ramirez}, \citenamefont {Yilma},
  \citenamefont {Davies}, \citenamefont {Joshi}, \citenamefont {Chambers},
  \citenamefont {Lekuch}, \citenamefont {Inoue}, \citenamefont {Wisnieff},
  \citenamefont {Baks}, \citenamefont {Bethune}, \citenamefont {Timmerwilke},
  \citenamefont {Fox}, \citenamefont {Song}, \citenamefont {Johnson},
  \citenamefont {Gaucher},\ and\ \citenamefont {Friedman}}]{Chakraborty2022}%
  \BibitemOpen
  \bibfield  {author} {\bibinfo {author} {\bibfnamefont {S.}~\bibnamefont
  {Chakraborty}}, \bibinfo {author} {\bibfnamefont {D.~J.}\ \bibnamefont
  {Frank}}, \bibinfo {author} {\bibfnamefont {K.}~\bibnamefont {Tien}},
  \bibinfo {author} {\bibfnamefont {P.}~\bibnamefont {Rosno}}, \bibinfo
  {author} {\bibfnamefont {M.}~\bibnamefont {Yeck}}, \bibinfo {author}
  {\bibfnamefont {J.~A.}\ \bibnamefont {Glick}}, \bibinfo {author}
  {\bibfnamefont {R.}~\bibnamefont {Robertazzi}}, \bibinfo {author}
  {\bibfnamefont {R.}~\bibnamefont {Richetta}}, \bibinfo {author}
  {\bibfnamefont {J.~F.}\ \bibnamefont {Bulzacchelli}}, \bibinfo {author}
  {\bibfnamefont {D.}~\bibnamefont {Underwood}}, \bibinfo {author}
  {\bibfnamefont {D.}~\bibnamefont {Ramirez}}, \bibinfo {author} {\bibfnamefont
  {D.}~\bibnamefont {Yilma}}, \bibinfo {author} {\bibfnamefont
  {A.}~\bibnamefont {Davies}}, \bibinfo {author} {\bibfnamefont {R.~V.}\
  \bibnamefont {Joshi}}, \bibinfo {author} {\bibfnamefont {S.~D.}\ \bibnamefont
  {Chambers}}, \bibinfo {author} {\bibfnamefont {S.}~\bibnamefont {Lekuch}},
  \bibinfo {author} {\bibfnamefont {K.}~\bibnamefont {Inoue}}, \bibinfo
  {author} {\bibfnamefont {D.}~\bibnamefont {Wisnieff}}, \bibinfo {author}
  {\bibfnamefont {C.~W.}\ \bibnamefont {Baks}}, \bibinfo {author}
  {\bibfnamefont {D.~S.}\ \bibnamefont {Bethune}}, \bibinfo {author}
  {\bibfnamefont {J.}~\bibnamefont {Timmerwilke}}, \bibinfo {author}
  {\bibfnamefont {T.}~\bibnamefont {Fox}}, \bibinfo {author} {\bibfnamefont
  {P.}~\bibnamefont {Song}}, \bibinfo {author} {\bibfnamefont {B.~R.}\
  \bibnamefont {Johnson}}, \bibinfo {author} {\bibfnamefont {B.~P.}\
  \bibnamefont {Gaucher}},\ and\ \bibinfo {author} {\bibfnamefont {D.~J.}\
  \bibnamefont {Friedman}},\ }\bibfield  {title} {\bibinfo {title} {A
  cryo-{CMOS} low-power semi-autonomous transmon qubit state controller in
  14-nm {FinFET} technology},\ }\href
  {https://doi.org/10.1109/JSSC.2022.3201775} {\bibfield  {journal} {\bibinfo
  {journal} {IEEE Journal of Solid-State Circuits}\ }\textbf {\bibinfo {volume}
  {57}},\ \bibinfo {pages} {3258} (\bibinfo {year} {2022})}\BibitemShut
  {NoStop}%
\bibitem [{\citenamefont {Choi}\ \emph {et~al.}(2006)\citenamefont {Choi},
  \citenamefont {Painter}, \citenamefont {Kim}, \citenamefont {Lee},
  \citenamefont {Yang}, \citenamefont {Weijers}, \citenamefont {Miller},\ and\
  \citenamefont {Miller}}]{Choi2007}%
  \BibitemOpen
  \bibfield  {author} {\bibinfo {author} {\bibfnamefont {Y.}~\bibnamefont
  {Choi}}, \bibinfo {author} {\bibfnamefont {T.}~\bibnamefont {Painter}},
  \bibinfo {author} {\bibfnamefont {D.}~\bibnamefont {Kim}}, \bibinfo {author}
  {\bibfnamefont {B.}~\bibnamefont {Lee}}, \bibinfo {author} {\bibfnamefont
  {H.}~\bibnamefont {Yang}}, \bibinfo {author} {\bibfnamefont {H.}~\bibnamefont
  {Weijers}}, \bibinfo {author} {\bibfnamefont {G.}~\bibnamefont {Miller}},\
  and\ \bibinfo {author} {\bibfnamefont {J.}~\bibnamefont {Miller}},\
  }\bibfield  {title} {\bibinfo {title} {Helium-liquefaction by cryocooler for
  high-field magnet cooling},\ }in\ \href@noop {} {\emph {\bibinfo {booktitle}
  {2007 International Cryocooler Conference}}}\ (\bibinfo {year} {2006})\ pp.\
  \bibinfo {pages} {655--661}\BibitemShut {NoStop}%
\bibitem [{\citenamefont {Das}\ \emph {et~al.}(2021)\citenamefont {Das},
  \citenamefont {Locharla},\ and\ \citenamefont {Jones}}]{Das2021}%
  \BibitemOpen
  \bibfield  {author} {\bibinfo {author} {\bibfnamefont {P.}~\bibnamefont
  {Das}}, \bibinfo {author} {\bibfnamefont {A.}~\bibnamefont {Locharla}},\ and\
  \bibinfo {author} {\bibfnamefont {C.}~\bibnamefont {Jones}},\ }\bibfield
  {title} {\bibinfo {title} {Lilliput: A lightweight low-latency lookup-table
  based decoder for near-term quantum error correction}\ }\href
  {https://doi.org/10.48550/ARXIV.2108.06569} {10.48550/ARXIV.2108.06569}
  (\bibinfo {year} {2021})\BibitemShut {NoStop}%
\bibitem [{\citenamefont {Guo}\ \emph {et~al.}()\citenamefont {Guo},
  \citenamefont {Lin}, \citenamefont {Han}, \citenamefont {Li}, \citenamefont
  {Sun}, \citenamefont {Liang}, \citenamefont {Li}, \citenamefont {Li},
  \citenamefont {Gong}, \citenamefont {Xu}, \citenamefont {Liao},\ and\
  \citenamefont {Peng}}]{guo2022}%
  \BibitemOpen
  \bibfield  {author} {\bibinfo {author} {\bibfnamefont {C.}~\bibnamefont
  {Guo}}, \bibinfo {author} {\bibfnamefont {J.}~\bibnamefont {Lin}}, \bibinfo
  {author} {\bibfnamefont {L.-C.}\ \bibnamefont {Han}}, \bibinfo {author}
  {\bibfnamefont {N.}~\bibnamefont {Li}}, \bibinfo {author} {\bibfnamefont
  {L.-H.}\ \bibnamefont {Sun}}, \bibinfo {author} {\bibfnamefont {F.-T.}\
  \bibnamefont {Liang}}, \bibinfo {author} {\bibfnamefont {D.-D.}\ \bibnamefont
  {Li}}, \bibinfo {author} {\bibfnamefont {Y.-H.}\ \bibnamefont {Li}}, \bibinfo
  {author} {\bibfnamefont {M.}~\bibnamefont {Gong}}, \bibinfo {author}
  {\bibfnamefont {Y.}~\bibnamefont {Xu}}, \bibinfo {author} {\bibfnamefont
  {S.-K.}\ \bibnamefont {Liao}},\ and\ \bibinfo {author} {\bibfnamefont
  {C.-Z.}\ \bibnamefont {Peng}},\ }\bibfield  {title} {\bibinfo {title}
  {Low-latency readout electronics for dynamic superconducting quantum
  computing},\ }\href {https://doi.org/10.1063/5.0088879} {\ \textbf {\bibinfo
  {volume} {12}},\ \bibinfo {pages} {045024}},\ \bibinfo {note} {publisher:
  American Institute of Physics}\BibitemShut {NoStop}%
\bibitem [{\citenamefont {Krinner}\ \emph {et~al.}()\citenamefont {Krinner},
  \citenamefont {Storz}, \citenamefont {Kurpiers}, \citenamefont {Magnard},
  \citenamefont {Heinsoo}, \citenamefont {Keller}, \citenamefont {Lütolf},
  \citenamefont {Eichler},\ and\ \citenamefont {Wallraff}}]{krinner2019}%
  \BibitemOpen
  \bibfield  {author} {\bibinfo {author} {\bibfnamefont {S.}~\bibnamefont
  {Krinner}}, \bibinfo {author} {\bibfnamefont {S.}~\bibnamefont {Storz}},
  \bibinfo {author} {\bibfnamefont {P.}~\bibnamefont {Kurpiers}}, \bibinfo
  {author} {\bibfnamefont {P.}~\bibnamefont {Magnard}}, \bibinfo {author}
  {\bibfnamefont {J.}~\bibnamefont {Heinsoo}}, \bibinfo {author} {\bibfnamefont
  {R.}~\bibnamefont {Keller}}, \bibinfo {author} {\bibfnamefont
  {J.}~\bibnamefont {Lütolf}}, \bibinfo {author} {\bibfnamefont
  {C.}~\bibnamefont {Eichler}},\ and\ \bibinfo {author} {\bibfnamefont
  {A.}~\bibnamefont {Wallraff}},\ }\bibfield  {title} {\bibinfo {title}
  {Engineering cryogenic setups for 100-qubit scale superconducting circuit
  systems},\ }\href {https://doi.org/10.1140/epjqt/s40507-019-0072-0} {\
  \textbf {\bibinfo {volume} {6}},\ \bibinfo {pages} {2}}\BibitemShut {NoStop}%
\bibitem [{\citenamefont {Rol}\ \emph {et~al.}()\citenamefont {Rol},
  \citenamefont {Ciorciaro}, \citenamefont {Malinowski}, \citenamefont
  {Tarasinski}, \citenamefont {Sagastizabal}, \citenamefont {Bultink},
  \citenamefont {Salathe}, \citenamefont {Haandbaek}, \citenamefont {Sedivy},\
  and\ \citenamefont {{DiCarlo}}}]{rol2020}%
  \BibitemOpen
  \bibfield  {author} {\bibinfo {author} {\bibfnamefont {M.~A.}\ \bibnamefont
  {Rol}}, \bibinfo {author} {\bibfnamefont {L.}~\bibnamefont {Ciorciaro}},
  \bibinfo {author} {\bibfnamefont {F.~K.}\ \bibnamefont {Malinowski}},
  \bibinfo {author} {\bibfnamefont {B.~M.}\ \bibnamefont {Tarasinski}},
  \bibinfo {author} {\bibfnamefont {R.~E.}\ \bibnamefont {Sagastizabal}},
  \bibinfo {author} {\bibfnamefont {C.~C.}\ \bibnamefont {Bultink}}, \bibinfo
  {author} {\bibfnamefont {Y.}~\bibnamefont {Salathe}}, \bibinfo {author}
  {\bibfnamefont {N.}~\bibnamefont {Haandbaek}}, \bibinfo {author}
  {\bibfnamefont {J.}~\bibnamefont {Sedivy}},\ and\ \bibinfo {author}
  {\bibfnamefont {L.}~\bibnamefont {{DiCarlo}}},\ }\bibfield  {title} {\bibinfo
  {title} {Time-domain characterization and correction of on-chip distortion of
  control pulses in a quantum processor},\ }\href
  {https://doi.org/10.1063/1.5133894} {\ \textbf {\bibinfo {volume} {116}},\
  \bibinfo {pages} {054001}},\ \bibinfo {note} {publisher: American Institute
  of Physics}\BibitemShut {NoStop}%
\bibitem [{\citenamefont {Simbierowicz}\ \emph {et~al.}()\citenamefont
  {Simbierowicz}, \citenamefont {Monarkha}, \citenamefont {Singh},
  \citenamefont {Messaoudi}, \citenamefont {Krantz},\ and\ \citenamefont
  {Lake}}]{simbierowicz2022}%
  \BibitemOpen
  \bibfield  {author} {\bibinfo {author} {\bibfnamefont {S.}~\bibnamefont
  {Simbierowicz}}, \bibinfo {author} {\bibfnamefont {V.~Y.}\ \bibnamefont
  {Monarkha}}, \bibinfo {author} {\bibfnamefont {S.}~\bibnamefont {Singh}},
  \bibinfo {author} {\bibfnamefont {N.}~\bibnamefont {Messaoudi}}, \bibinfo
  {author} {\bibfnamefont {P.}~\bibnamefont {Krantz}},\ and\ \bibinfo {author}
  {\bibfnamefont {R.~E.}\ \bibnamefont {Lake}},\ }\bibfield  {title} {\bibinfo
  {title} {Microwave calibration of qubit drive line components at millikelvin
  temperatures},\ }\href {https://doi.org/10.1063/5.0081861} {\ \textbf
  {\bibinfo {volume} {120}},\ \bibinfo {pages} {054004}},\ \bibinfo {note}
  {publisher: American Institute of Physics}\BibitemShut {NoStop}%
\bibitem [{\citenamefont {Koch}\ \emph {et~al.}(2007)\citenamefont {Koch},
  \citenamefont {Yu}, \citenamefont {Gambetta}, \citenamefont {Houck},
  \citenamefont {Schuster}, \citenamefont {Majer}, \citenamefont {Blais},
  \citenamefont {Devoret}, \citenamefont {Girvin},\ and\ \citenamefont
  {Schoelkopf}}]{Koch2007}%
  \BibitemOpen
  \bibfield  {author} {\bibinfo {author} {\bibfnamefont {J.}~\bibnamefont
  {Koch}}, \bibinfo {author} {\bibfnamefont {T.~M.}\ \bibnamefont {Yu}},
  \bibinfo {author} {\bibfnamefont {J.}~\bibnamefont {Gambetta}}, \bibinfo
  {author} {\bibfnamefont {A.~A.}\ \bibnamefont {Houck}}, \bibinfo {author}
  {\bibfnamefont {D.~I.}\ \bibnamefont {Schuster}}, \bibinfo {author}
  {\bibfnamefont {J.}~\bibnamefont {Majer}}, \bibinfo {author} {\bibfnamefont
  {A.}~\bibnamefont {Blais}}, \bibinfo {author} {\bibfnamefont {M.~H.}\
  \bibnamefont {Devoret}}, \bibinfo {author} {\bibfnamefont {S.~M.}\
  \bibnamefont {Girvin}},\ and\ \bibinfo {author} {\bibfnamefont {R.~J.}\
  \bibnamefont {Schoelkopf}},\ }\bibfield  {title} {\bibinfo {title}
  {Charge-insensitive qubit design derived from the cooper pair box},\ }\href
  {https://doi.org/10.1103/PhysRevA.76.042319} {\bibfield  {journal} {\bibinfo
  {journal} {Phys. Rev. A}\ }\textbf {\bibinfo {volume} {76}},\ \bibinfo
  {pages} {042319} (\bibinfo {year} {2007})}\BibitemShut {NoStop}%
\bibitem [{\citenamefont {Chow}\ \emph {et~al.}(2011)\citenamefont {Chow},
  \citenamefont {C\'orcoles}, \citenamefont {Gambetta}, \citenamefont
  {Rigetti}, \citenamefont {Johnson}, \citenamefont {Smolin}, \citenamefont
  {Rozen}, \citenamefont {Keefe}, \citenamefont {Rothwell}, \citenamefont
  {Ketchen},\ and\ \citenamefont {Steffen}}]{Chow2011}%
  \BibitemOpen
  \bibfield  {author} {\bibinfo {author} {\bibfnamefont {J.~M.}\ \bibnamefont
  {Chow}}, \bibinfo {author} {\bibfnamefont {A.~D.}\ \bibnamefont
  {C\'orcoles}}, \bibinfo {author} {\bibfnamefont {J.~M.}\ \bibnamefont
  {Gambetta}}, \bibinfo {author} {\bibfnamefont {C.}~\bibnamefont {Rigetti}},
  \bibinfo {author} {\bibfnamefont {B.~R.}\ \bibnamefont {Johnson}}, \bibinfo
  {author} {\bibfnamefont {J.~A.}\ \bibnamefont {Smolin}}, \bibinfo {author}
  {\bibfnamefont {J.~R.}\ \bibnamefont {Rozen}}, \bibinfo {author}
  {\bibfnamefont {G.~A.}\ \bibnamefont {Keefe}}, \bibinfo {author}
  {\bibfnamefont {M.~B.}\ \bibnamefont {Rothwell}}, \bibinfo {author}
  {\bibfnamefont {M.~B.}\ \bibnamefont {Ketchen}},\ and\ \bibinfo {author}
  {\bibfnamefont {M.}~\bibnamefont {Steffen}},\ }\bibfield  {title} {\bibinfo
  {title} {Simple all-microwave entangling gate for fixed-frequency
  superconducting qubits},\ }\href
  {https://doi.org/10.1103/PhysRevLett.107.080502} {\bibfield  {journal}
  {\bibinfo  {journal} {Phys. Rev. Lett.}\ }\textbf {\bibinfo {volume} {107}},\
  \bibinfo {pages} {080502} (\bibinfo {year} {2011})}\BibitemShut {NoStop}%
\bibitem [{\citenamefont {Shor}(1995)}]{Shore1995}%
  \BibitemOpen
  \bibfield  {author} {\bibinfo {author} {\bibfnamefont {P.~W.}\ \bibnamefont
  {Shor}},\ }\bibfield  {title} {\bibinfo {title} {Scheme for reducing
  decoherence in quantum computer memory},\ }\href
  {https://doi.org/10.1103/PhysRevA.52.R2493} {\bibfield  {journal} {\bibinfo
  {journal} {Phys. Rev. A}\ }\textbf {\bibinfo {volume} {52}},\ \bibinfo
  {pages} {R2493} (\bibinfo {year} {1995})}\BibitemShut {NoStop}%
\bibitem [{\citenamefont {Bacon}(2006)}]{Bacon2006}%
  \BibitemOpen
  \bibfield  {author} {\bibinfo {author} {\bibfnamefont {D.}~\bibnamefont
  {Bacon}},\ }\bibfield  {title} {\bibinfo {title} {Operator quantum
  error-correcting subsystems for self-correcting quantum memories},\ }\href
  {https://doi.org/10.1103/PhysRevA.73.012340} {\bibfield  {journal} {\bibinfo
  {journal} {Phys. Rev. A}\ }\textbf {\bibinfo {volume} {73}},\ \bibinfo
  {pages} {012340} (\bibinfo {year} {2006})}\BibitemShut {NoStop}%
\bibitem [{\citenamefont {Fowler}\ \emph {et~al.}(2012)\citenamefont {Fowler},
  \citenamefont {Mariantoni}, \citenamefont {Martinis},\ and\ \citenamefont
  {Cleland}}]{Fowler2012}%
  \BibitemOpen
  \bibfield  {author} {\bibinfo {author} {\bibfnamefont {A.~G.}\ \bibnamefont
  {Fowler}}, \bibinfo {author} {\bibfnamefont {M.}~\bibnamefont {Mariantoni}},
  \bibinfo {author} {\bibfnamefont {J.~M.}\ \bibnamefont {Martinis}},\ and\
  \bibinfo {author} {\bibfnamefont {A.~N.}\ \bibnamefont {Cleland}},\
  }\bibfield  {title} {\bibinfo {title} {Surface codes: Towards practical
  large-scale quantum computation},\ }\href
  {https://doi.org/10.1103/PhysRevA.86.032324} {\bibfield  {journal} {\bibinfo
  {journal} {Phys. Rev. A}\ }\textbf {\bibinfo {volume} {86}},\ \bibinfo
  {pages} {032324} (\bibinfo {year} {2012})}\BibitemShut {NoStop}%
\bibitem [{\citenamefont {Aliferis}\ and\ \citenamefont
  {Cross}(2007)}]{Cross2007}%
  \BibitemOpen
  \bibfield  {author} {\bibinfo {author} {\bibfnamefont {P.}~\bibnamefont
  {Aliferis}}\ and\ \bibinfo {author} {\bibfnamefont {A.~W.}\ \bibnamefont
  {Cross}},\ }\bibfield  {title} {\bibinfo {title} {Subsystem fault tolerance
  with the bacon-shor code},\ }\href
  {https://doi.org/10.1103/PhysRevLett.98.220502} {\bibfield  {journal}
  {\bibinfo  {journal} {Phys. Rev. Lett.}\ }\textbf {\bibinfo {volume} {98}},\
  \bibinfo {pages} {220502} (\bibinfo {year} {2007})}\BibitemShut {NoStop}%
\bibitem [{\citenamefont {Kandala}\ \emph {et~al.}()\citenamefont {Kandala},
  \citenamefont {Temme}, \citenamefont {Córcoles}, \citenamefont {Mezzacapo},
  \citenamefont {Chow},\ and\ \citenamefont {Gambetta}}]{Kandala_2019}%
  \BibitemOpen
  \bibfield  {author} {\bibinfo {author} {\bibfnamefont {A.}~\bibnamefont
  {Kandala}}, \bibinfo {author} {\bibfnamefont {K.}~\bibnamefont {Temme}},
  \bibinfo {author} {\bibfnamefont {A.~D.}\ \bibnamefont {Córcoles}}, \bibinfo
  {author} {\bibfnamefont {A.}~\bibnamefont {Mezzacapo}}, \bibinfo {author}
  {\bibfnamefont {J.~M.}\ \bibnamefont {Chow}},\ and\ \bibinfo {author}
  {\bibfnamefont {J.~M.}\ \bibnamefont {Gambetta}},\ }\bibfield  {title}
  {\bibinfo {title} {Error mitigation extends the computational reach of a
  noisy quantum processor},\ }\href {https://doi.org/10.1038/s41586-019-1040-7}
  {\ \textbf {\bibinfo {volume} {567}},\ \bibinfo {pages} {491}}\BibitemShut
  {NoStop}%
\bibitem [{\citenamefont {Havlíček}\ \emph {et~al.}()\citenamefont
  {Havlíček}, \citenamefont {Córcoles}, \citenamefont {Temme}, \citenamefont
  {Harrow}, \citenamefont {Kandala}, \citenamefont {Chow},\ and\ \citenamefont
  {Gambetta}}]{Havlicek_2019}%
  \BibitemOpen
  \bibfield  {author} {\bibinfo {author} {\bibfnamefont {V.}~\bibnamefont
  {Havlíček}}, \bibinfo {author} {\bibfnamefont {A.~D.}\ \bibnamefont
  {Córcoles}}, \bibinfo {author} {\bibfnamefont {K.}~\bibnamefont {Temme}},
  \bibinfo {author} {\bibfnamefont {A.~W.}\ \bibnamefont {Harrow}}, \bibinfo
  {author} {\bibfnamefont {A.}~\bibnamefont {Kandala}}, \bibinfo {author}
  {\bibfnamefont {J.~M.}\ \bibnamefont {Chow}},\ and\ \bibinfo {author}
  {\bibfnamefont {J.~M.}\ \bibnamefont {Gambetta}},\ }\bibfield  {title}
  {\bibinfo {title} {Supervised learning with quantum-enhanced feature
  spaces},\ }\href {https://doi.org/10.1038/s41586-019-0980-2} {\ \textbf
  {\bibinfo {volume} {567}},\ \bibinfo {pages} {209}}\BibitemShut {NoStop}%
\bibitem [{\citenamefont {Suzuki}\ \emph {et~al.}(2022)\citenamefont {Suzuki},
  \citenamefont {Endo}, \citenamefont {Fujii},\ and\ \citenamefont
  {Tokunaga}}]{Suzuki2022}%
  \BibitemOpen
  \bibfield  {author} {\bibinfo {author} {\bibfnamefont {Y.}~\bibnamefont
  {Suzuki}}, \bibinfo {author} {\bibfnamefont {S.}~\bibnamefont {Endo}},
  \bibinfo {author} {\bibfnamefont {K.}~\bibnamefont {Fujii}},\ and\ \bibinfo
  {author} {\bibfnamefont {Y.}~\bibnamefont {Tokunaga}},\ }\bibfield  {title}
  {\bibinfo {title} {Quantum error mitigation as a universal error reduction
  technique: Applications from the nisq to the fault-tolerant quantum computing
  eras},\ }\href {https://doi.org/10.1103/PRXQuantum.3.010345} {\bibfield
  {journal} {\bibinfo  {journal} {PRX Quantum}\ }\textbf {\bibinfo {volume}
  {3}},\ \bibinfo {pages} {010345} (\bibinfo {year} {2022})}\BibitemShut
  {NoStop}%
\bibitem [{\citenamefont {Blume-Kohout}\ \emph {et~al.}()\citenamefont
  {Blume-Kohout}, \citenamefont {Gamble}, \citenamefont {Nielsen},
  \citenamefont {Rudinger}, \citenamefont {Mizrahi}, \citenamefont {Fortier},\
  and\ \citenamefont {Maunz}}]{blumekohout2017}%
  \BibitemOpen
  \bibfield  {author} {\bibinfo {author} {\bibfnamefont {R.}~\bibnamefont
  {Blume-Kohout}}, \bibinfo {author} {\bibfnamefont {J.~K.}\ \bibnamefont
  {Gamble}}, \bibinfo {author} {\bibfnamefont {E.}~\bibnamefont {Nielsen}},
  \bibinfo {author} {\bibfnamefont {K.}~\bibnamefont {Rudinger}}, \bibinfo
  {author} {\bibfnamefont {J.}~\bibnamefont {Mizrahi}}, \bibinfo {author}
  {\bibfnamefont {K.}~\bibnamefont {Fortier}},\ and\ \bibinfo {author}
  {\bibfnamefont {P.}~\bibnamefont {Maunz}},\ }\bibfield  {title} {\bibinfo
  {title} {Demonstration of qubit operations below a rigorous fault tolerance
  threshold with gate set tomography},\ }\href
  {https://doi.org/10.1038/ncomms14485} {\ \textbf {\bibinfo {volume} {8}},\
  \bibinfo {pages} {14485}}\BibitemShut {NoStop}%
\bibitem [{\citenamefont {Cross}\ \emph {et~al.}(2019)\citenamefont {Cross},
  \citenamefont {Bishop}, \citenamefont {Sheldon}, \citenamefont {Nation},\
  and\ \citenamefont {Gambetta}}]{cross2019}%
  \BibitemOpen
  \bibfield  {author} {\bibinfo {author} {\bibfnamefont {A.~W.}\ \bibnamefont
  {Cross}}, \bibinfo {author} {\bibfnamefont {L.~S.}\ \bibnamefont {Bishop}},
  \bibinfo {author} {\bibfnamefont {S.}~\bibnamefont {Sheldon}}, \bibinfo
  {author} {\bibfnamefont {P.~D.}\ \bibnamefont {Nation}},\ and\ \bibinfo
  {author} {\bibfnamefont {J.~M.}\ \bibnamefont {Gambetta}},\ }\bibfield
  {title} {\bibinfo {title} {Validating quantum computers using randomized
  model circuits},\ }\href {https://doi.org/10.1103/PhysRevA.100.032328}
  {\bibfield  {journal} {\bibinfo  {journal} {Phys. Rev. A}\ }\textbf {\bibinfo
  {volume} {100}},\ \bibinfo {pages} {032328} (\bibinfo {year}
  {2019})}\BibitemShut {NoStop}%
\bibitem [{\citenamefont {McKay}\ \emph {et~al.}(2019)\citenamefont {McKay},
  \citenamefont {Sheldon}, \citenamefont {Smolin}, \citenamefont {Chow},\ and\
  \citenamefont {Gambetta}}]{mckay2019}%
  \BibitemOpen
  \bibfield  {author} {\bibinfo {author} {\bibfnamefont {D.~C.}\ \bibnamefont
  {McKay}}, \bibinfo {author} {\bibfnamefont {S.}~\bibnamefont {Sheldon}},
  \bibinfo {author} {\bibfnamefont {J.~A.}\ \bibnamefont {Smolin}}, \bibinfo
  {author} {\bibfnamefont {J.~M.}\ \bibnamefont {Chow}},\ and\ \bibinfo
  {author} {\bibfnamefont {J.~M.}\ \bibnamefont {Gambetta}},\ }\bibfield
  {title} {\bibinfo {title} {Three-qubit randomized benchmarking},\ }\href@noop
  {} {\bibfield  {journal} {\bibinfo  {journal} {Phys. Rev. Lett.}\ }\textbf
  {\bibinfo {volume} {122}},\ \bibinfo {pages} {200502} (\bibinfo {year}
  {2019})}\BibitemShut {NoStop}%
\bibitem [{\citenamefont {Magesan}\ \emph {et~al.}(2012)\citenamefont
  {Magesan}, \citenamefont {Gambetta}, \citenamefont {Johnson}, \citenamefont
  {Ryan}, \citenamefont {Chow}, \citenamefont {Merkel}, \citenamefont
  {da~Silva}, \citenamefont {Keefe}, \citenamefont {Rothwell}, \citenamefont
  {Ohki}, \citenamefont {Ketchen},\ and\ \citenamefont
  {Steffen}}]{Magesan2012}%
  \BibitemOpen
  \bibfield  {author} {\bibinfo {author} {\bibfnamefont {E.}~\bibnamefont
  {Magesan}}, \bibinfo {author} {\bibfnamefont {J.~M.}\ \bibnamefont
  {Gambetta}}, \bibinfo {author} {\bibfnamefont {B.~R.}\ \bibnamefont
  {Johnson}}, \bibinfo {author} {\bibfnamefont {C.~A.}\ \bibnamefont {Ryan}},
  \bibinfo {author} {\bibfnamefont {J.~M.}\ \bibnamefont {Chow}}, \bibinfo
  {author} {\bibfnamefont {S.~T.}\ \bibnamefont {Merkel}}, \bibinfo {author}
  {\bibfnamefont {M.~P.}\ \bibnamefont {da~Silva}}, \bibinfo {author}
  {\bibfnamefont {G.~A.}\ \bibnamefont {Keefe}}, \bibinfo {author}
  {\bibfnamefont {M.~B.}\ \bibnamefont {Rothwell}}, \bibinfo {author}
  {\bibfnamefont {T.~A.}\ \bibnamefont {Ohki}}, \bibinfo {author}
  {\bibfnamefont {M.~B.}\ \bibnamefont {Ketchen}},\ and\ \bibinfo {author}
  {\bibfnamefont {M.}~\bibnamefont {Steffen}},\ }\bibfield  {title} {\bibinfo
  {title} {Efficient measurement of quantum gate error by interleaved
  randomized benchmarking},\ }\href
  {https://doi.org/10.1103/PhysRevLett.109.080505} {\bibfield  {journal}
  {\bibinfo  {journal} {Phys. Rev. Lett.}\ }\textbf {\bibinfo {volume} {109}},\
  \bibinfo {pages} {080505} (\bibinfo {year} {2012})}\BibitemShut {NoStop}%
\bibitem [{\citenamefont {Eisert}\ \emph {et~al.}()\citenamefont {Eisert},
  \citenamefont {Hangleiter}, \citenamefont {Walk}, \citenamefont {Roth},
  \citenamefont {Markham}, \citenamefont {Parekh}, \citenamefont {Chabaud},\
  and\ \citenamefont {Kashefi}}]{eisert2020}%
  \BibitemOpen
  \bibfield  {author} {\bibinfo {author} {\bibfnamefont {J.}~\bibnamefont
  {Eisert}}, \bibinfo {author} {\bibfnamefont {D.}~\bibnamefont {Hangleiter}},
  \bibinfo {author} {\bibfnamefont {N.}~\bibnamefont {Walk}}, \bibinfo {author}
  {\bibfnamefont {I.}~\bibnamefont {Roth}}, \bibinfo {author} {\bibfnamefont
  {D.}~\bibnamefont {Markham}}, \bibinfo {author} {\bibfnamefont
  {R.}~\bibnamefont {Parekh}}, \bibinfo {author} {\bibfnamefont
  {U.}~\bibnamefont {Chabaud}},\ and\ \bibinfo {author} {\bibfnamefont
  {E.}~\bibnamefont {Kashefi}},\ }\bibfield  {title} {\bibinfo {title} {Quantum
  certification and benchmarking},\ }\href
  {https://doi.org/10.1038/s42254-020-0186-4} {\bibfield  {journal} {\bibinfo
  {journal} {Nature Reviews Physics}\ }\textbf {\bibinfo {volume} {2}},\
  \bibinfo {pages} {382}}\BibitemShut {NoStop}%
\bibitem [{\citenamefont {C\'orcoles}\ \emph {et~al.}(2021)\citenamefont
  {C\'orcoles}, \citenamefont {Takita}, \citenamefont {Inoue}, \citenamefont
  {Lekuch}, \citenamefont {Minev}, \citenamefont {Chow},\ and\ \citenamefont
  {Gambetta}}]{Corcoles2021}%
  \BibitemOpen
  \bibfield  {author} {\bibinfo {author} {\bibfnamefont {A.~D.}\ \bibnamefont
  {C\'orcoles}}, \bibinfo {author} {\bibfnamefont {M.}~\bibnamefont {Takita}},
  \bibinfo {author} {\bibfnamefont {K.}~\bibnamefont {Inoue}}, \bibinfo
  {author} {\bibfnamefont {S.}~\bibnamefont {Lekuch}}, \bibinfo {author}
  {\bibfnamefont {Z.~K.}\ \bibnamefont {Minev}}, \bibinfo {author}
  {\bibfnamefont {J.~M.}\ \bibnamefont {Chow}},\ and\ \bibinfo {author}
  {\bibfnamefont {J.~M.}\ \bibnamefont {Gambetta}},\ }\bibfield  {title}
  {\bibinfo {title} {Exploiting dynamic quantum circuits in a quantum algorithm
  with superconducting qubits},\ }\href
  {https://doi.org/10.1103/PhysRevLett.127.100501} {\bibfield  {journal}
  {\bibinfo  {journal} {Phys. Rev. Lett.}\ }\textbf {\bibinfo {volume} {127}},\
  \bibinfo {pages} {100501} (\bibinfo {year} {2021})}\BibitemShut {NoStop}%
\bibitem [{\citenamefont {Overwater}\ \emph {et~al.}(2022)\citenamefont
  {Overwater}, \citenamefont {Babaie},\ and\ \citenamefont
  {Sebastiano}}]{Overwater_2022}%
  \BibitemOpen
  \bibfield  {author} {\bibinfo {author} {\bibfnamefont {R.~J.}\ \bibnamefont
  {Overwater}}, \bibinfo {author} {\bibfnamefont {M.}~\bibnamefont {Babaie}},\
  and\ \bibinfo {author} {\bibfnamefont {F.}~\bibnamefont {Sebastiano}},\
  }\bibfield  {title} {\bibinfo {title} {Neural-network decoders for quantum
  error correction using surface codes: A space exploration of the hardware
  cost-performance tradeoffs},\ }\href
  {https://doi.org/10.1109/TQE.2022.3174017} {\bibfield  {journal} {\bibinfo
  {journal} {IEEE Transactions on Quantum Engineering}\ }\textbf {\bibinfo
  {volume} {3}},\ \bibinfo {pages} {1} (\bibinfo {year} {2022})}\BibitemShut
  {NoStop}%
\bibitem [{\citenamefont {Chen}\ \emph {et~al.}(2022)\citenamefont {Chen},
  \citenamefont {Yoder}, \citenamefont {Kim}, \citenamefont {Sundaresan},
  \citenamefont {Srinivasan}, \citenamefont {Li}, \citenamefont {C\'orcoles},
  \citenamefont {Cross},\ and\ \citenamefont {Takita}}]{Chen_2022}%
  \BibitemOpen
  \bibfield  {author} {\bibinfo {author} {\bibfnamefont {E.~H.}\ \bibnamefont
  {Chen}}, \bibinfo {author} {\bibfnamefont {T.~J.}\ \bibnamefont {Yoder}},
  \bibinfo {author} {\bibfnamefont {Y.}~\bibnamefont {Kim}}, \bibinfo {author}
  {\bibfnamefont {N.}~\bibnamefont {Sundaresan}}, \bibinfo {author}
  {\bibfnamefont {S.}~\bibnamefont {Srinivasan}}, \bibinfo {author}
  {\bibfnamefont {M.}~\bibnamefont {Li}}, \bibinfo {author} {\bibfnamefont
  {A.~D.}\ \bibnamefont {C\'orcoles}}, \bibinfo {author} {\bibfnamefont
  {A.~W.}\ \bibnamefont {Cross}},\ and\ \bibinfo {author} {\bibfnamefont
  {M.}~\bibnamefont {Takita}},\ }\bibfield  {title} {\bibinfo {title}
  {Calibrated decoders for experimental quantum error correction},\ }\href
  {https://doi.org/10.1103/PhysRevLett.128.110504} {\bibfield  {journal}
  {\bibinfo  {journal} {Phys. Rev. Lett.}\ }\textbf {\bibinfo {volume} {128}},\
  \bibinfo {pages} {110504} (\bibinfo {year} {2022})}\BibitemShut {NoStop}%
\bibitem [{\citenamefont {Wack}\ \emph {et~al.}(2021)\citenamefont {Wack},
  \citenamefont {Paik}, \citenamefont {Javadi-Abhari}, \citenamefont
  {Jurcevic}, \citenamefont {Faro}, \citenamefont {Gambetta},\ and\
  \citenamefont {Johnson}}]{Wack2021}%
  \BibitemOpen
  \bibfield  {author} {\bibinfo {author} {\bibfnamefont {A.}~\bibnamefont
  {Wack}}, \bibinfo {author} {\bibfnamefont {H.}~\bibnamefont {Paik}}, \bibinfo
  {author} {\bibfnamefont {A.}~\bibnamefont {Javadi-Abhari}}, \bibinfo {author}
  {\bibfnamefont {P.}~\bibnamefont {Jurcevic}}, \bibinfo {author}
  {\bibfnamefont {I.}~\bibnamefont {Faro}}, \bibinfo {author} {\bibfnamefont
  {J.~M.}\ \bibnamefont {Gambetta}},\ and\ \bibinfo {author} {\bibfnamefont
  {B.~R.}\ \bibnamefont {Johnson}},\ }\bibfield  {title} {\bibinfo {title}
  {Quality, speed, and scale: three key attributes to measure the performance
  of near-term quantum computers}\ }\href
  {https://doi.org/10.48550/ARXIV.2110.14108} {10.48550/ARXIV.2110.14108}
  (\bibinfo {year} {2021})\BibitemShut {NoStop}%
\bibitem [{\citenamefont {Bhattacharya}\ and\ \citenamefont
  {Jha}(2014)}]{Bhattacharya2014}%
  \BibitemOpen
  \bibfield  {author} {\bibinfo {author} {\bibfnamefont {D.}~\bibnamefont
  {Bhattacharya}}\ and\ \bibinfo {author} {\bibfnamefont {N.~K.}\ \bibnamefont
  {Jha}},\ }\bibfield  {title} {\bibinfo {title} {{FinFETs}: from devices to
  architectures},\ }\href {https://doi.org/10.1155/2014/365689} {\bibfield
  {journal} {\bibinfo  {journal} {Advances in Electronics}\ }\textbf {\bibinfo
  {volume} {2014}},\ \bibinfo {pages} {365689} (\bibinfo {year}
  {2014})}\BibitemShut {NoStop}%
\bibitem [{\citenamefont {Park}\ \emph {et~al.}(2021)\citenamefont {Park},
  \citenamefont {Subramanian}, \citenamefont {Lampert}, \citenamefont
  {Mladenov}, \citenamefont {Klotchkov}, \citenamefont {Kurian}, \citenamefont
  {Ju{\'a}rez-Hern{\'a}ndez}, \citenamefont {Esparza}, \citenamefont {Kale},
  \citenamefont {Beevi}, \citenamefont {Premaratne}, \citenamefont {Watson},
  \citenamefont {Suzuki}, \citenamefont {Rahman}, \citenamefont {Timbadiya},
  \citenamefont {Soni},\ and\ \citenamefont {Pellerano}}]{Park2021AFI}%
  \BibitemOpen
  \bibfield  {author} {\bibinfo {author} {\bibfnamefont {J.-S.}\ \bibnamefont
  {Park}}, \bibinfo {author} {\bibfnamefont {S.}~\bibnamefont {Subramanian}},
  \bibinfo {author} {\bibfnamefont {L.}~\bibnamefont {Lampert}}, \bibinfo
  {author} {\bibfnamefont {T.}~\bibnamefont {Mladenov}}, \bibinfo {author}
  {\bibfnamefont {I.~V.}\ \bibnamefont {Klotchkov}}, \bibinfo {author}
  {\bibfnamefont {D.}~\bibnamefont {Kurian}}, \bibinfo {author} {\bibfnamefont
  {E.}~\bibnamefont {Ju{\'a}rez-Hern{\'a}ndez}}, \bibinfo {author}
  {\bibfnamefont {B.~P.}\ \bibnamefont {Esparza}}, \bibinfo {author}
  {\bibfnamefont {S.~R.}\ \bibnamefont {Kale}}, \bibinfo {author}
  {\bibfnamefont {K.~T.~A.}\ \bibnamefont {Beevi}}, \bibinfo {author}
  {\bibfnamefont {S.~P.}\ \bibnamefont {Premaratne}}, \bibinfo {author}
  {\bibfnamefont {T.}~\bibnamefont {Watson}}, \bibinfo {author} {\bibfnamefont
  {S.}~\bibnamefont {Suzuki}}, \bibinfo {author} {\bibfnamefont
  {M.}~\bibnamefont {Rahman}}, \bibinfo {author} {\bibfnamefont
  {J.}~\bibnamefont {Timbadiya}}, \bibinfo {author} {\bibfnamefont
  {S.}~\bibnamefont {Soni}},\ and\ \bibinfo {author} {\bibfnamefont
  {S.}~\bibnamefont {Pellerano}},\ }\bibfield  {title} {\bibinfo {title} {A
  fully integrated cryo-{CMOS} {SoC} for qubit control in quantum computers
  capable of state manipulation, readout and high-speed gate pulsing of spin
  qubits in {Intel} 22nm {FFL} {FinFET} technology},\ }\href@noop {} {\bibfield
   {journal} {\bibinfo  {journal} {2021 IEEE International Solid-State Circuits
  Conference (ISSCC)}\ }\textbf {\bibinfo {volume} {64}},\ \bibinfo {pages}
  {208} (\bibinfo {year} {2021})}\BibitemShut {NoStop}%
\bibitem [{\citenamefont {Patra}\ \emph {et~al.}(2020)\citenamefont {Patra},
  \citenamefont {van Dijk}, \citenamefont {Subramanian}, \citenamefont {Corna},
  \citenamefont {Xue}, \citenamefont {Jeon}, \citenamefont {Sheikh},
  \citenamefont {Juarez-Hernandez}, \citenamefont {Esparza}, \citenamefont
  {Rampurawala}, \citenamefont {Carlton}, \citenamefont {Samkharadze},
  \citenamefont {Ravikumar}, \citenamefont {Nieva}, \citenamefont {Kim},
  \citenamefont {Lee}, \citenamefont {Sammak}, \citenamefont {Scappucci},
  \citenamefont {Veldhorst}, \citenamefont {Vandersypen}, \citenamefont
  {Babaie}, \citenamefont {Sebastiano}, \citenamefont {Charbon},\ and\
  \citenamefont {Pellerano}}]{Patra2020}%
  \BibitemOpen
  \bibfield  {author} {\bibinfo {author} {\bibfnamefont {B.}~\bibnamefont
  {Patra}}, \bibinfo {author} {\bibfnamefont {J.~P.~G.}\ \bibnamefont {van
  Dijk}}, \bibinfo {author} {\bibfnamefont {S.}~\bibnamefont {Subramanian}},
  \bibinfo {author} {\bibfnamefont {A.}~\bibnamefont {Corna}}, \bibinfo
  {author} {\bibfnamefont {X.}~\bibnamefont {Xue}}, \bibinfo {author}
  {\bibfnamefont {C.}~\bibnamefont {Jeon}}, \bibinfo {author} {\bibfnamefont
  {F.}~\bibnamefont {Sheikh}}, \bibinfo {author} {\bibfnamefont
  {E.}~\bibnamefont {Juarez-Hernandez}}, \bibinfo {author} {\bibfnamefont
  {B.~P.}\ \bibnamefont {Esparza}}, \bibinfo {author} {\bibfnamefont
  {H.}~\bibnamefont {Rampurawala}}, \bibinfo {author} {\bibfnamefont
  {B.}~\bibnamefont {Carlton}}, \bibinfo {author} {\bibfnamefont
  {N.}~\bibnamefont {Samkharadze}}, \bibinfo {author} {\bibfnamefont
  {S.}~\bibnamefont {Ravikumar}}, \bibinfo {author} {\bibfnamefont
  {C.}~\bibnamefont {Nieva}}, \bibinfo {author} {\bibfnamefont
  {S.}~\bibnamefont {Kim}}, \bibinfo {author} {\bibfnamefont {H.-J.}\
  \bibnamefont {Lee}}, \bibinfo {author} {\bibfnamefont {A.}~\bibnamefont
  {Sammak}}, \bibinfo {author} {\bibfnamefont {G.}~\bibnamefont {Scappucci}},
  \bibinfo {author} {\bibfnamefont {M.}~\bibnamefont {Veldhorst}}, \bibinfo
  {author} {\bibfnamefont {L.~M.~K.}\ \bibnamefont {Vandersypen}}, \bibinfo
  {author} {\bibfnamefont {M.}~\bibnamefont {Babaie}}, \bibinfo {author}
  {\bibfnamefont {F.}~\bibnamefont {Sebastiano}}, \bibinfo {author}
  {\bibfnamefont {E.}~\bibnamefont {Charbon}},\ and\ \bibinfo {author}
  {\bibfnamefont {S.}~\bibnamefont {Pellerano}},\ }\bibfield  {title} {\bibinfo
  {title} {A scalable cryo-{CMOS} 2-to-{20GHz} digitally intensive controller
  for 4×32 frequency multiplexed spin qubits/transmons in 22nm {FinFET}
  technology for quantum computers},\ }in\ \href
  {https://doi.org/10.1109/ISSCC19947.2020.9063109} {\emph {\bibinfo
  {booktitle} {2020 IEEE International Solid-State Circuits Conference -
  (ISSCC)}}}\ (\bibinfo {year} {2020})\ pp.\ \bibinfo {pages}
  {304--305}\BibitemShut {NoStop}%
\bibitem [{\citenamefont {McKay}\ \emph {et~al.}(2017)\citenamefont {McKay},
  \citenamefont {Wood}, \citenamefont {Sheldon}, \citenamefont {Chow},\ and\
  \citenamefont {Gambetta}}]{McKay_2017}%
  \BibitemOpen
  \bibfield  {author} {\bibinfo {author} {\bibfnamefont {D.~C.}\ \bibnamefont
  {McKay}}, \bibinfo {author} {\bibfnamefont {C.~J.}\ \bibnamefont {Wood}},
  \bibinfo {author} {\bibfnamefont {S.}~\bibnamefont {Sheldon}}, \bibinfo
  {author} {\bibfnamefont {J.~M.}\ \bibnamefont {Chow}},\ and\ \bibinfo
  {author} {\bibfnamefont {J.~M.}\ \bibnamefont {Gambetta}},\ }\bibfield
  {title} {\bibinfo {title} {Efficient z gates for quantum computing},\
  }\bibfield  {journal} {\bibinfo  {journal} {Physical Review A}\ }\textbf
  {\bibinfo {volume} {96}},\ \href {https://doi.org/10.1103/physreva.96.022330}
  {10.1103/physreva.96.022330} (\bibinfo {year} {2017})\BibitemShut {NoStop}%
\bibitem [{\citenamefont {Sheldon}\ \emph
  {et~al.}(2016{\natexlab{a}})\citenamefont {Sheldon}, \citenamefont {Magesan},
  \citenamefont {Chow},\ and\ \citenamefont {Gambetta}}]{Sheldon2016}%
  \BibitemOpen
  \bibfield  {author} {\bibinfo {author} {\bibfnamefont {S.}~\bibnamefont
  {Sheldon}}, \bibinfo {author} {\bibfnamefont {E.}~\bibnamefont {Magesan}},
  \bibinfo {author} {\bibfnamefont {J.~M.}\ \bibnamefont {Chow}},\ and\
  \bibinfo {author} {\bibfnamefont {J.~M.}\ \bibnamefont {Gambetta}},\
  }\bibfield  {title} {\bibinfo {title} {Procedure for systematically tuning up
  cross-talk in the cross-resonance gate},\ }\href
  {https://doi.org/10.1103/PhysRevA.93.060302} {\bibfield  {journal} {\bibinfo
  {journal} {Phys. Rev. A}\ }\textbf {\bibinfo {volume} {93}},\ \bibinfo
  {pages} {060302} (\bibinfo {year} {2016}{\natexlab{a}})}\BibitemShut
  {NoStop}%
\bibitem [{\citenamefont {Moro}\ \emph {et~al.}()\citenamefont {Moro},
  \citenamefont {Paris}, \citenamefont {Restelli},\ and\ \citenamefont
  {Prati}}]{moro_2021}%
  \BibitemOpen
  \bibfield  {author} {\bibinfo {author} {\bibfnamefont {L.}~\bibnamefont
  {Moro}}, \bibinfo {author} {\bibfnamefont {M.~G.~A.}\ \bibnamefont {Paris}},
  \bibinfo {author} {\bibfnamefont {M.}~\bibnamefont {Restelli}},\ and\
  \bibinfo {author} {\bibfnamefont {E.}~\bibnamefont {Prati}},\ }\bibfield
  {title} {\bibinfo {title} {Quantum compiling by deep reinforcement
  learning},\ }\href {https://doi.org/10.1038/s42005-021-00684-3} {\ \textbf
  {\bibinfo {volume} {4}},\ \bibinfo {pages} {178}}\BibitemShut {NoStop}%
\bibitem [{\citenamefont {Pedersen}\ \emph {et~al.}(2007)\citenamefont
  {Pedersen}, \citenamefont {Møller},\ and\ \citenamefont
  {Mølmer}}]{Pedersen_2007}%
  \BibitemOpen
  \bibfield  {author} {\bibinfo {author} {\bibfnamefont {L.~H.}\ \bibnamefont
  {Pedersen}}, \bibinfo {author} {\bibfnamefont {N.~M.}\ \bibnamefont
  {Møller}},\ and\ \bibinfo {author} {\bibfnamefont {K.}~\bibnamefont
  {Mølmer}},\ }\bibfield  {title} {\bibinfo {title} {Fidelity of quantum
  operations},\ }\href
  {https://doi.org/https://doi.org/10.1016/j.physleta.2007.02.069} {\bibfield
  {journal} {\bibinfo  {journal} {Physics Letters A}\ }\textbf {\bibinfo
  {volume} {367}},\ \bibinfo {pages} {47} (\bibinfo {year} {2007})}\BibitemShut
  {NoStop}%
\bibitem [{\citenamefont {O'Malley}\ \emph {et~al.}(2015)\citenamefont
  {O'Malley}, \citenamefont {Kelly}, \citenamefont {Barends}, \citenamefont
  {Campbell}, \citenamefont {Chen}, \citenamefont {Chen}, \citenamefont
  {Chiaro}, \citenamefont {Dunsworth}, \citenamefont {Fowler}, \citenamefont
  {Hoi}, \citenamefont {Jeffrey}, \citenamefont {Megrant}, \citenamefont
  {Mutus}, \citenamefont {Neill}, \citenamefont {Quintana}, \citenamefont
  {Roushan}, \citenamefont {Sank}, \citenamefont {Vainsencher}, \citenamefont
  {Wenner}, \citenamefont {White}, \citenamefont {Korotkov}, \citenamefont
  {Cleland},\ and\ \citenamefont {Martinis}}]{Omalley_2015}%
  \BibitemOpen
  \bibfield  {author} {\bibinfo {author} {\bibfnamefont {P.~J.~J.}\
  \bibnamefont {O'Malley}}, \bibinfo {author} {\bibfnamefont {J.}~\bibnamefont
  {Kelly}}, \bibinfo {author} {\bibfnamefont {R.}~\bibnamefont {Barends}},
  \bibinfo {author} {\bibfnamefont {B.}~\bibnamefont {Campbell}}, \bibinfo
  {author} {\bibfnamefont {Y.}~\bibnamefont {Chen}}, \bibinfo {author}
  {\bibfnamefont {Z.}~\bibnamefont {Chen}}, \bibinfo {author} {\bibfnamefont
  {B.}~\bibnamefont {Chiaro}}, \bibinfo {author} {\bibfnamefont
  {A.}~\bibnamefont {Dunsworth}}, \bibinfo {author} {\bibfnamefont {A.~G.}\
  \bibnamefont {Fowler}}, \bibinfo {author} {\bibfnamefont {I.-C.}\
  \bibnamefont {Hoi}}, \bibinfo {author} {\bibfnamefont {E.}~\bibnamefont
  {Jeffrey}}, \bibinfo {author} {\bibfnamefont {A.}~\bibnamefont {Megrant}},
  \bibinfo {author} {\bibfnamefont {J.}~\bibnamefont {Mutus}}, \bibinfo
  {author} {\bibfnamefont {C.}~\bibnamefont {Neill}}, \bibinfo {author}
  {\bibfnamefont {C.}~\bibnamefont {Quintana}}, \bibinfo {author}
  {\bibfnamefont {P.}~\bibnamefont {Roushan}}, \bibinfo {author} {\bibfnamefont
  {D.}~\bibnamefont {Sank}}, \bibinfo {author} {\bibfnamefont {A.}~\bibnamefont
  {Vainsencher}}, \bibinfo {author} {\bibfnamefont {J.}~\bibnamefont {Wenner}},
  \bibinfo {author} {\bibfnamefont {T.~C.}\ \bibnamefont {White}}, \bibinfo
  {author} {\bibfnamefont {A.~N.}\ \bibnamefont {Korotkov}}, \bibinfo {author}
  {\bibfnamefont {A.~N.}\ \bibnamefont {Cleland}},\ and\ \bibinfo {author}
  {\bibfnamefont {J.~M.}\ \bibnamefont {Martinis}},\ }\bibfield  {title}
  {\bibinfo {title} {Qubit metrology of ultralow phase noise using randomized
  benchmarking},\ }\href {https://doi.org/10.1103/PhysRevApplied.3.044009}
  {\bibfield  {journal} {\bibinfo  {journal} {Phys. Rev. Applied}\ }\textbf
  {\bibinfo {volume} {3}},\ \bibinfo {pages} {044009} (\bibinfo {year}
  {2015})}\BibitemShut {NoStop}%
\bibitem [{\citenamefont {Willsch}\ \emph {et~al.}(2017)\citenamefont
  {Willsch}, \citenamefont {Nocon}, \citenamefont {Jin}, \citenamefont
  {De~Raedt},\ and\ \citenamefont {Michielsen}}]{Willsch_2017}%
  \BibitemOpen
  \bibfield  {author} {\bibinfo {author} {\bibfnamefont {D.}~\bibnamefont
  {Willsch}}, \bibinfo {author} {\bibfnamefont {M.}~\bibnamefont {Nocon}},
  \bibinfo {author} {\bibfnamefont {F.}~\bibnamefont {Jin}}, \bibinfo {author}
  {\bibfnamefont {H.}~\bibnamefont {De~Raedt}},\ and\ \bibinfo {author}
  {\bibfnamefont {K.}~\bibnamefont {Michielsen}},\ }\bibfield  {title}
  {\bibinfo {title} {Gate-error analysis in simulations of quantum computers
  with transmon qubits},\ }\href {https://doi.org/10.1103/PhysRevA.96.062302}
  {\bibfield  {journal} {\bibinfo  {journal} {Phys. Rev. A}\ }\textbf {\bibinfo
  {volume} {96}},\ \bibinfo {pages} {062302} (\bibinfo {year}
  {2017})}\BibitemShut {NoStop}%
\bibitem [{\citenamefont {Abad}\ \emph {et~al.}(2022)\citenamefont {Abad},
  \citenamefont {Fern\'andez-Pend\'as}, \citenamefont {Frisk~Kockum},\ and\
  \citenamefont {Johansson}}]{Tahereh_2022}%
  \BibitemOpen
  \bibfield  {author} {\bibinfo {author} {\bibfnamefont {T.}~\bibnamefont
  {Abad}}, \bibinfo {author} {\bibfnamefont {J.}~\bibnamefont
  {Fern\'andez-Pend\'as}}, \bibinfo {author} {\bibfnamefont {A.}~\bibnamefont
  {Frisk~Kockum}},\ and\ \bibinfo {author} {\bibfnamefont {G.}~\bibnamefont
  {Johansson}},\ }\bibfield  {title} {\bibinfo {title} {Universal fidelity
  reduction of quantum operations from weak dissipation},\ }\href
  {https://doi.org/10.1103/PhysRevLett.129.150504} {\bibfield  {journal}
  {\bibinfo  {journal} {Phys. Rev. Lett.}\ }\textbf {\bibinfo {volume} {129}},\
  \bibinfo {pages} {150504} (\bibinfo {year} {2022})}\BibitemShut {NoStop}%
\bibitem [{\citenamefont {Kelly}\ \emph {et~al.}(2014)\citenamefont {Kelly},
  \citenamefont {Barends}, \citenamefont {Campbell}, \citenamefont {Chen},
  \citenamefont {Chen}, \citenamefont {Chiaro}, \citenamefont {Dunsworth},
  \citenamefont {Fowler}, \citenamefont {Hoi}, \citenamefont {Jeffrey},
  \citenamefont {Megrant}, \citenamefont {Mutus}, \citenamefont {Neill},
  \citenamefont {O'Malley}, \citenamefont {Quintana}, \citenamefont {Roushan},
  \citenamefont {Sank}, \citenamefont {Vainsencher}, \citenamefont {Wenner},
  \citenamefont {White}, \citenamefont {Cleland},\ and\ \citenamefont
  {Martinis}}]{Kelly_2014}%
  \BibitemOpen
  \bibfield  {author} {\bibinfo {author} {\bibfnamefont {J.}~\bibnamefont
  {Kelly}}, \bibinfo {author} {\bibfnamefont {R.}~\bibnamefont {Barends}},
  \bibinfo {author} {\bibfnamefont {B.}~\bibnamefont {Campbell}}, \bibinfo
  {author} {\bibfnamefont {Y.}~\bibnamefont {Chen}}, \bibinfo {author}
  {\bibfnamefont {Z.}~\bibnamefont {Chen}}, \bibinfo {author} {\bibfnamefont
  {B.}~\bibnamefont {Chiaro}}, \bibinfo {author} {\bibfnamefont
  {A.}~\bibnamefont {Dunsworth}}, \bibinfo {author} {\bibfnamefont {A.~G.}\
  \bibnamefont {Fowler}}, \bibinfo {author} {\bibfnamefont {I.-C.}\
  \bibnamefont {Hoi}}, \bibinfo {author} {\bibfnamefont {E.}~\bibnamefont
  {Jeffrey}}, \bibinfo {author} {\bibfnamefont {A.}~\bibnamefont {Megrant}},
  \bibinfo {author} {\bibfnamefont {J.}~\bibnamefont {Mutus}}, \bibinfo
  {author} {\bibfnamefont {C.}~\bibnamefont {Neill}}, \bibinfo {author}
  {\bibfnamefont {P.~J.~J.}\ \bibnamefont {O'Malley}}, \bibinfo {author}
  {\bibfnamefont {C.}~\bibnamefont {Quintana}}, \bibinfo {author}
  {\bibfnamefont {P.}~\bibnamefont {Roushan}}, \bibinfo {author} {\bibfnamefont
  {D.}~\bibnamefont {Sank}}, \bibinfo {author} {\bibfnamefont {A.}~\bibnamefont
  {Vainsencher}}, \bibinfo {author} {\bibfnamefont {J.}~\bibnamefont {Wenner}},
  \bibinfo {author} {\bibfnamefont {T.~C.}\ \bibnamefont {White}}, \bibinfo
  {author} {\bibfnamefont {A.~N.}\ \bibnamefont {Cleland}},\ and\ \bibinfo
  {author} {\bibfnamefont {J.~M.}\ \bibnamefont {Martinis}},\ }\bibfield
  {title} {\bibinfo {title} {Optimal quantum control using randomized
  benchmarking},\ }\href {https://doi.org/10.1103/PhysRevLett.112.240504}
  {\bibfield  {journal} {\bibinfo  {journal} {Phys. Rev. Lett.}\ }\textbf
  {\bibinfo {volume} {112}},\ \bibinfo {pages} {240504} (\bibinfo {year}
  {2014})}\BibitemShut {NoStop}%
\bibitem [{\citenamefont {Rodionov}\ \emph {et~al.}(2014)\citenamefont
  {Rodionov}, \citenamefont {Veitia}, \citenamefont {Barends}, \citenamefont
  {Kelly}, \citenamefont {Sank}, \citenamefont {Wenner}, \citenamefont
  {Martinis}, \citenamefont {Kosut},\ and\ \citenamefont
  {Korotkov}}]{Rodionov_2014}%
  \BibitemOpen
  \bibfield  {author} {\bibinfo {author} {\bibfnamefont {A.~V.}\ \bibnamefont
  {Rodionov}}, \bibinfo {author} {\bibfnamefont {A.}~\bibnamefont {Veitia}},
  \bibinfo {author} {\bibfnamefont {R.}~\bibnamefont {Barends}}, \bibinfo
  {author} {\bibfnamefont {J.}~\bibnamefont {Kelly}}, \bibinfo {author}
  {\bibfnamefont {D.}~\bibnamefont {Sank}}, \bibinfo {author} {\bibfnamefont
  {J.}~\bibnamefont {Wenner}}, \bibinfo {author} {\bibfnamefont {J.~M.}\
  \bibnamefont {Martinis}}, \bibinfo {author} {\bibfnamefont {R.~L.}\
  \bibnamefont {Kosut}},\ and\ \bibinfo {author} {\bibfnamefont {A.~N.}\
  \bibnamefont {Korotkov}},\ }\bibfield  {title} {\bibinfo {title} {Compressed
  sensing quantum process tomography for superconducting quantum gates},\
  }\href {https://doi.org/10.1103/PhysRevB.90.144504} {\bibfield  {journal}
  {\bibinfo  {journal} {Phys. Rev. B}\ }\textbf {\bibinfo {volume} {90}},\
  \bibinfo {pages} {144504} (\bibinfo {year} {2014})}\BibitemShut {NoStop}%
\bibitem [{\citenamefont {Gaikwad}\ \emph {et~al.}()\citenamefont {Gaikwad},
  \citenamefont {Shende}, \citenamefont {{Arvind}},\ and\ \citenamefont
  {Dorai}}]{gaikwad_2022}%
  \BibitemOpen
  \bibfield  {author} {\bibinfo {author} {\bibfnamefont {A.}~\bibnamefont
  {Gaikwad}}, \bibinfo {author} {\bibfnamefont {K.}~\bibnamefont {Shende}},
  \bibinfo {author} {\bibnamefont {{Arvind}}},\ and\ \bibinfo {author}
  {\bibfnamefont {K.}~\bibnamefont {Dorai}},\ }\bibfield  {title} {\bibinfo
  {title} {Implementing efficient selective quantum process tomography of
  superconducting quantum gates on {IBM} quantum experience},\ }\href
  {https://doi.org/10.1038/s41598-022-07721-3} {\ \textbf {\bibinfo {volume}
  {12}},\ \bibinfo {pages} {3688}}\BibitemShut {NoStop}%
\bibitem [{\citenamefont {Quantum}()}]{qiskit_text}%
  \BibitemOpen
  \bibfield  {author} {\bibinfo {author} {\bibfnamefont {I.}~\bibnamefont
  {Quantum}},\ }\bibfield  {title} {\bibinfo {title} {Calibrating qubits with
  qiskit pulse},\ }\bibfield  {booktitle} {\emph {\bibinfo {booktitle} {Learn
  Quantum Computation using Qiskit}},\ }\href
  {https://qiskit.org/textbook/ch-quantum-hardware/index-pulses.html} {\
  }\BibitemShut {NoStop}%
\bibitem [{\citenamefont {Patterson}\ \emph {et~al.}(2019)\citenamefont
  {Patterson}, \citenamefont {Rahamim}, \citenamefont {Tsunoda}, \citenamefont
  {Spring}, \citenamefont {Jebari}, \citenamefont {Ratter}, \citenamefont
  {Mergenthaler}, \citenamefont {Tancredi}, \citenamefont {Vlastakis},
  \citenamefont {Esposito},\ and\ \citenamefont {Leek}}]{Patterson2019}%
  \BibitemOpen
  \bibfield  {author} {\bibinfo {author} {\bibfnamefont {A.}~\bibnamefont
  {Patterson}}, \bibinfo {author} {\bibfnamefont {J.}~\bibnamefont {Rahamim}},
  \bibinfo {author} {\bibfnamefont {T.}~\bibnamefont {Tsunoda}}, \bibinfo
  {author} {\bibfnamefont {P.}~\bibnamefont {Spring}}, \bibinfo {author}
  {\bibfnamefont {S.}~\bibnamefont {Jebari}}, \bibinfo {author} {\bibfnamefont
  {K.}~\bibnamefont {Ratter}}, \bibinfo {author} {\bibfnamefont
  {M.}~\bibnamefont {Mergenthaler}}, \bibinfo {author} {\bibfnamefont
  {G.}~\bibnamefont {Tancredi}}, \bibinfo {author} {\bibfnamefont
  {B.}~\bibnamefont {Vlastakis}}, \bibinfo {author} {\bibfnamefont
  {M.}~\bibnamefont {Esposito}},\ and\ \bibinfo {author} {\bibfnamefont
  {P.}~\bibnamefont {Leek}},\ }\bibfield  {title} {\bibinfo {title}
  {Calibration of a cross-resonance two-qubit gate between directly coupled
  transmons},\ }\href {https://doi.org/10.1103/PhysRevApplied.12.064013}
  {\bibfield  {journal} {\bibinfo  {journal} {Phys. Rev. Applied}\ }\textbf
  {\bibinfo {volume} {12}},\ \bibinfo {pages} {064013} (\bibinfo {year}
  {2019})}\BibitemShut {NoStop}%
\bibitem [{\citenamefont {Gambetta}\ \emph {et~al.}(2011)\citenamefont
  {Gambetta}, \citenamefont {Motzoi}, \citenamefont {Merkel},\ and\
  \citenamefont {Wilhelm}}]{Gambetta2011}%
  \BibitemOpen
  \bibfield  {author} {\bibinfo {author} {\bibfnamefont {J.~M.}\ \bibnamefont
  {Gambetta}}, \bibinfo {author} {\bibfnamefont {F.}~\bibnamefont {Motzoi}},
  \bibinfo {author} {\bibfnamefont {S.~T.}\ \bibnamefont {Merkel}},\ and\
  \bibinfo {author} {\bibfnamefont {F.~K.}\ \bibnamefont {Wilhelm}},\
  }\bibfield  {title} {\bibinfo {title} {Analytic control methods for
  high-fidelity unitary operations in a weakly nonlinear oscillator},\ }\href
  {https://doi.org/10.1103/PhysRevA.83.012308} {\bibfield  {journal} {\bibinfo
  {journal} {Phys. Rev. A}\ }\textbf {\bibinfo {volume} {83}},\ \bibinfo
  {pages} {012308} (\bibinfo {year} {2011})}\BibitemShut {NoStop}%
\bibitem [{\citenamefont {Motzoi}\ \emph {et~al.}(2009)\citenamefont {Motzoi},
  \citenamefont {Gambetta}, \citenamefont {Rebentrost},\ and\ \citenamefont
  {Wilhelm}}]{Motzoi2009}%
  \BibitemOpen
  \bibfield  {author} {\bibinfo {author} {\bibfnamefont {F.}~\bibnamefont
  {Motzoi}}, \bibinfo {author} {\bibfnamefont {J.~M.}\ \bibnamefont
  {Gambetta}}, \bibinfo {author} {\bibfnamefont {P.}~\bibnamefont
  {Rebentrost}},\ and\ \bibinfo {author} {\bibfnamefont {F.~K.}\ \bibnamefont
  {Wilhelm}},\ }\bibfield  {title} {\bibinfo {title} {Simple pulses for
  elimination of leakage in weakly nonlinear qubits},\ }\href
  {https://doi.org/10.1103/PhysRevLett.103.110501} {\bibfield  {journal}
  {\bibinfo  {journal} {Phys. Rev. Lett.}\ }\textbf {\bibinfo {volume} {103}},\
  \bibinfo {pages} {110501} (\bibinfo {year} {2009})}\BibitemShut {NoStop}%
\bibitem [{\citenamefont {Chow}\ \emph {et~al.}(2010)\citenamefont {Chow},
  \citenamefont {DiCarlo}, \citenamefont {Gambetta}, \citenamefont {Motzoi},
  \citenamefont {Frunzio}, \citenamefont {Girvin},\ and\ \citenamefont
  {Schoelkopf}}]{Chow2010}%
  \BibitemOpen
  \bibfield  {author} {\bibinfo {author} {\bibfnamefont {J.~M.}\ \bibnamefont
  {Chow}}, \bibinfo {author} {\bibfnamefont {L.}~\bibnamefont {DiCarlo}},
  \bibinfo {author} {\bibfnamefont {J.~M.}\ \bibnamefont {Gambetta}}, \bibinfo
  {author} {\bibfnamefont {F.}~\bibnamefont {Motzoi}}, \bibinfo {author}
  {\bibfnamefont {L.}~\bibnamefont {Frunzio}}, \bibinfo {author} {\bibfnamefont
  {S.~M.}\ \bibnamefont {Girvin}},\ and\ \bibinfo {author} {\bibfnamefont
  {R.~J.}\ \bibnamefont {Schoelkopf}},\ }\bibfield  {title} {\bibinfo {title}
  {Optimized driving of superconducting artificial atoms for improved
  single-qubit gates},\ }\href {https://doi.org/10.1103/PhysRevA.82.040305}
  {\bibfield  {journal} {\bibinfo  {journal} {Phys. Rev. A}\ }\textbf {\bibinfo
  {volume} {82}},\ \bibinfo {pages} {040305} (\bibinfo {year}
  {2010})}\BibitemShut {NoStop}%
\bibitem [{\citenamefont {Sheldon}\ \emph
  {et~al.}(2016{\natexlab{b}})\citenamefont {Sheldon}, \citenamefont {Bishop},
  \citenamefont {Magesan}, \citenamefont {Filipp}, \citenamefont {Chow},\ and\
  \citenamefont {Gambetta}}]{Sheldon2016_v2}%
  \BibitemOpen
  \bibfield  {author} {\bibinfo {author} {\bibfnamefont {S.}~\bibnamefont
  {Sheldon}}, \bibinfo {author} {\bibfnamefont {L.~S.}\ \bibnamefont {Bishop}},
  \bibinfo {author} {\bibfnamefont {E.}~\bibnamefont {Magesan}}, \bibinfo
  {author} {\bibfnamefont {S.}~\bibnamefont {Filipp}}, \bibinfo {author}
  {\bibfnamefont {J.~M.}\ \bibnamefont {Chow}},\ and\ \bibinfo {author}
  {\bibfnamefont {J.~M.}\ \bibnamefont {Gambetta}},\ }\bibfield  {title}
  {\bibinfo {title} {Characterizing errors on qubit operations via iterative
  randomized benchmarking},\ }\href
  {https://doi.org/10.1103/PhysRevA.93.012301} {\bibfield  {journal} {\bibinfo
  {journal} {Phys. Rev. A}\ }\textbf {\bibinfo {volume} {93}},\ \bibinfo
  {pages} {012301} (\bibinfo {year} {2016}{\natexlab{b}})}\BibitemShut
  {NoStop}%
\bibitem [{\citenamefont {Vitanov}(2020)}]{Vitanov_2020}%
  \BibitemOpen
  \bibfield  {author} {\bibinfo {author} {\bibfnamefont {N.~V.}\ \bibnamefont
  {Vitanov}},\ }\bibfield  {title} {\bibinfo {title} {Relations between single
  and repeated qubit gates: coherent error amplification for high-fidelity
  quantum-gate tomography},\ }\href {https://doi.org/10.1088/1367-2630/ab6a38}
  {\bibfield  {journal} {\bibinfo  {journal} {New Journal of Physics}\ }\textbf
  {\bibinfo {volume} {22}},\ \bibinfo {pages} {023015} (\bibinfo {year}
  {2020})}\BibitemShut {NoStop}%
\bibitem [{\citenamefont {Malekakhlagh}\ and\ \citenamefont
  {Magesan}(2022)}]{Malekakhlagh_2022}%
  \BibitemOpen
  \bibfield  {author} {\bibinfo {author} {\bibfnamefont {M.}~\bibnamefont
  {Malekakhlagh}}\ and\ \bibinfo {author} {\bibfnamefont {E.}~\bibnamefont
  {Magesan}},\ }\bibfield  {title} {\bibinfo {title} {Mitigating off-resonant
  error in the cross-resonance gate},\ }\href
  {https://doi.org/10.1103/PhysRevA.105.012602} {\bibfield  {journal} {\bibinfo
   {journal} {Phys. Rev. A}\ }\textbf {\bibinfo {volume} {105}},\ \bibinfo
  {pages} {012602} (\bibinfo {year} {2022})}\BibitemShut {NoStop}%
\bibitem [{\citenamefont {Magesan}\ and\ \citenamefont
  {Gambetta}(2020)}]{Magesan_2020}%
  \BibitemOpen
  \bibfield  {author} {\bibinfo {author} {\bibfnamefont {E.}~\bibnamefont
  {Magesan}}\ and\ \bibinfo {author} {\bibfnamefont {J.~M.}\ \bibnamefont
  {Gambetta}},\ }\bibfield  {title} {\bibinfo {title} {Effective {Hamiltonian}
  models of the cross-resonance gate},\ }\href
  {https://doi.org/10.1103/PhysRevA.101.052308} {\bibfield  {journal} {\bibinfo
   {journal} {Phys. Rev. A}\ }\textbf {\bibinfo {volume} {101}},\ \bibinfo
  {pages} {052308} (\bibinfo {year} {2020})}\BibitemShut {NoStop}%
\bibitem [{\citenamefont {Malekakhlagh}\ \emph {et~al.}(2020)\citenamefont
  {Malekakhlagh}, \citenamefont {Magesan},\ and\ \citenamefont
  {McKay}}]{Malekakhlagh_2020}%
  \BibitemOpen
  \bibfield  {author} {\bibinfo {author} {\bibfnamefont {M.}~\bibnamefont
  {Malekakhlagh}}, \bibinfo {author} {\bibfnamefont {E.}~\bibnamefont
  {Magesan}},\ and\ \bibinfo {author} {\bibfnamefont {D.~C.}\ \bibnamefont
  {McKay}},\ }\bibfield  {title} {\bibinfo {title} {First-principles analysis
  of cross-resonance gate operation},\ }\href
  {https://doi.org/10.1103/PhysRevA.102.042605} {\bibfield  {journal} {\bibinfo
   {journal} {Phys. Rev. A}\ }\textbf {\bibinfo {volume} {102}},\ \bibinfo
  {pages} {042605} (\bibinfo {year} {2020})}\BibitemShut {NoStop}%
\bibitem [{\citenamefont {Sundaresan}\ \emph {et~al.}(2020)\citenamefont
  {Sundaresan}, \citenamefont {Lauer}, \citenamefont {Pritchett}, \citenamefont
  {Magesan}, \citenamefont {Jurcevic},\ and\ \citenamefont
  {Gambetta}}]{Sundaresan2020}%
  \BibitemOpen
  \bibfield  {author} {\bibinfo {author} {\bibfnamefont {N.}~\bibnamefont
  {Sundaresan}}, \bibinfo {author} {\bibfnamefont {I.}~\bibnamefont {Lauer}},
  \bibinfo {author} {\bibfnamefont {E.}~\bibnamefont {Pritchett}}, \bibinfo
  {author} {\bibfnamefont {E.}~\bibnamefont {Magesan}}, \bibinfo {author}
  {\bibfnamefont {P.}~\bibnamefont {Jurcevic}},\ and\ \bibinfo {author}
  {\bibfnamefont {J.~M.}\ \bibnamefont {Gambetta}},\ }\bibfield  {title}
  {\bibinfo {title} {Reducing unitary and spectator errors in cross resonance
  with optimized rotary echoes},\ }\href
  {https://doi.org/10.1103/PRXQuantum.1.020318} {\bibfield  {journal} {\bibinfo
   {journal} {PRX Quantum}\ }\textbf {\bibinfo {volume} {1}},\ \bibinfo {pages}
  {020318} (\bibinfo {year} {2020})}\BibitemShut {NoStop}%
\bibitem [{\citenamefont {Kandala}\ \emph {et~al.}(2021)\citenamefont
  {Kandala}, \citenamefont {Wei}, \citenamefont {Srinivasan}, \citenamefont
  {Magesan}, \citenamefont {Carnevale}, \citenamefont {Keefe}, \citenamefont
  {Klaus}, \citenamefont {Dial},\ and\ \citenamefont {McKay}}]{Kandala2021}%
  \BibitemOpen
  \bibfield  {author} {\bibinfo {author} {\bibfnamefont {A.}~\bibnamefont
  {Kandala}}, \bibinfo {author} {\bibfnamefont {K.~X.}\ \bibnamefont {Wei}},
  \bibinfo {author} {\bibfnamefont {S.}~\bibnamefont {Srinivasan}}, \bibinfo
  {author} {\bibfnamefont {E.}~\bibnamefont {Magesan}}, \bibinfo {author}
  {\bibfnamefont {S.}~\bibnamefont {Carnevale}}, \bibinfo {author}
  {\bibfnamefont {G.~A.}\ \bibnamefont {Keefe}}, \bibinfo {author}
  {\bibfnamefont {D.}~\bibnamefont {Klaus}}, \bibinfo {author} {\bibfnamefont
  {O.}~\bibnamefont {Dial}},\ and\ \bibinfo {author} {\bibfnamefont {D.~C.}\
  \bibnamefont {McKay}},\ }\bibfield  {title} {\bibinfo {title} {Demonstration
  of a high-fidelity {CNOT} gate for fixed-frequency transmons with engineered
  $zz$ suppression},\ }\href {https://doi.org/10.1103/PhysRevLett.127.130501}
  {\bibfield  {journal} {\bibinfo  {journal} {Phys. Rev. Lett.}\ }\textbf
  {\bibinfo {volume} {127}},\ \bibinfo {pages} {130501} (\bibinfo {year}
  {2021})}\BibitemShut {NoStop}%
\bibitem [{\citenamefont {Wei}\ \emph {et~al.}(2023)\citenamefont {Wei},
  \citenamefont {Pritchett}, \citenamefont {Zajac}, \citenamefont {McKay},\
  and\ \citenamefont {Merkel}}]{Pritchett_2023}%
  \BibitemOpen
  \bibfield  {author} {\bibinfo {author} {\bibfnamefont {K.~X.}\ \bibnamefont
  {Wei}}, \bibinfo {author} {\bibfnamefont {E.}~\bibnamefont {Pritchett}},
  \bibinfo {author} {\bibfnamefont {D.}~\bibnamefont {Zajac}}, \bibinfo
  {author} {\bibfnamefont {D.}~\bibnamefont {McKay}},\ and\ \bibinfo {author}
  {\bibfnamefont {S.}~\bibnamefont {Merkel}},\ }\bibfield  {title} {\bibinfo
  {title} {Characterizing non-markovian off-resonant errors in quantum gates},\
  }\href@noop {} {\bibfield  {journal} {\bibinfo  {journal} {arxiv.2302.10881}\
  } (\bibinfo {year} {2023})}\BibitemShut {NoStop}%
\bibitem [{\citenamefont {Ball}\ \emph {et~al.}()\citenamefont {Ball},
  \citenamefont {Oliver},\ and\ \citenamefont {Biercuk}}]{Ball2016}%
  \BibitemOpen
  \bibfield  {author} {\bibinfo {author} {\bibfnamefont {H.}~\bibnamefont
  {Ball}}, \bibinfo {author} {\bibfnamefont {W.~D.}\ \bibnamefont {Oliver}},\
  and\ \bibinfo {author} {\bibfnamefont {M.~J.}\ \bibnamefont {Biercuk}},\
  }\bibfield  {title} {\bibinfo {title} {The role of master clock stability in
  quantum information processing},\ }\href
  {https://doi.org/10.1038/npjqi.2016.33} {\ \textbf {\bibinfo {volume} {2}},\
  \bibinfo {pages} {16033}}\BibitemShut {NoStop}%
\bibitem [{\citenamefont {Yan}\ \emph {et~al.}()\citenamefont {Yan},
  \citenamefont {Gustavsson}, \citenamefont {Bylander}, \citenamefont {Jin},
  \citenamefont {Yoshihara}, \citenamefont {Cory}, \citenamefont {Nakamura},
  \citenamefont {Orlando},\ and\ \citenamefont {Oliver}}]{Yan2013}%
  \BibitemOpen
  \bibfield  {author} {\bibinfo {author} {\bibfnamefont {F.}~\bibnamefont
  {Yan}}, \bibinfo {author} {\bibfnamefont {S.}~\bibnamefont {Gustavsson}},
  \bibinfo {author} {\bibfnamefont {J.}~\bibnamefont {Bylander}}, \bibinfo
  {author} {\bibfnamefont {X.}~\bibnamefont {Jin}}, \bibinfo {author}
  {\bibfnamefont {F.}~\bibnamefont {Yoshihara}}, \bibinfo {author}
  {\bibfnamefont {D.~G.}\ \bibnamefont {Cory}}, \bibinfo {author}
  {\bibfnamefont {Y.}~\bibnamefont {Nakamura}}, \bibinfo {author}
  {\bibfnamefont {T.~P.}\ \bibnamefont {Orlando}},\ and\ \bibinfo {author}
  {\bibfnamefont {W.~D.}\ \bibnamefont {Oliver}},\ }\bibfield  {title}
  {\bibinfo {title} {Rotating-frame relaxation as a noise spectrum analyser of
  a superconducting qubit undergoing driven evolution},\ }\href
  {https://doi.org/10.1038/ncomms3337} {\ \textbf {\bibinfo {volume} {4}},\
  \bibinfo {pages} {2337}}\BibitemShut {NoStop}%
\bibitem [{\citenamefont {Bylander}\ \emph {et~al.}()\citenamefont {Bylander},
  \citenamefont {Gustavsson}, \citenamefont {Yan}, \citenamefont {Yoshihara},
  \citenamefont {Harrabi}, \citenamefont {Fitch}, \citenamefont {Cory},
  \citenamefont {Nakamura}, \citenamefont {Tsai},\ and\ \citenamefont
  {Oliver}}]{Bylander2011}%
  \BibitemOpen
  \bibfield  {author} {\bibinfo {author} {\bibfnamefont {J.}~\bibnamefont
  {Bylander}}, \bibinfo {author} {\bibfnamefont {S.}~\bibnamefont
  {Gustavsson}}, \bibinfo {author} {\bibfnamefont {F.}~\bibnamefont {Yan}},
  \bibinfo {author} {\bibfnamefont {F.}~\bibnamefont {Yoshihara}}, \bibinfo
  {author} {\bibfnamefont {K.}~\bibnamefont {Harrabi}}, \bibinfo {author}
  {\bibfnamefont {G.}~\bibnamefont {Fitch}}, \bibinfo {author} {\bibfnamefont
  {D.~G.}\ \bibnamefont {Cory}}, \bibinfo {author} {\bibfnamefont
  {Y.}~\bibnamefont {Nakamura}}, \bibinfo {author} {\bibfnamefont {J.-S.}\
  \bibnamefont {Tsai}},\ and\ \bibinfo {author} {\bibfnamefont {W.~D.}\
  \bibnamefont {Oliver}},\ }\bibfield  {title} {\bibinfo {title} {Noise
  spectroscopy through dynamical decoupling with a superconducting flux
  qubit},\ }\href {https://doi.org/10.1038/nphys1994} {\ \textbf {\bibinfo
  {volume} {7}},\ \bibinfo {pages} {565}}\BibitemShut {NoStop}%
\bibitem [{\citenamefont {Ku}\ \emph {et~al.}(2020)\citenamefont {Ku},
  \citenamefont {Xu}, \citenamefont {Brink}, \citenamefont {McKay},
  \citenamefont {Hertzberg}, \citenamefont {Ansari},\ and\ \citenamefont
  {Plourde}}]{Ku2020}%
  \BibitemOpen
  \bibfield  {author} {\bibinfo {author} {\bibfnamefont {J.}~\bibnamefont
  {Ku}}, \bibinfo {author} {\bibfnamefont {X.}~\bibnamefont {Xu}}, \bibinfo
  {author} {\bibfnamefont {M.}~\bibnamefont {Brink}}, \bibinfo {author}
  {\bibfnamefont {D.~C.}\ \bibnamefont {McKay}}, \bibinfo {author}
  {\bibfnamefont {J.~B.}\ \bibnamefont {Hertzberg}}, \bibinfo {author}
  {\bibfnamefont {M.~H.}\ \bibnamefont {Ansari}},\ and\ \bibinfo {author}
  {\bibfnamefont {B.~L.~T.}\ \bibnamefont {Plourde}},\ }\bibfield  {title}
  {\bibinfo {title} {Suppression of unwanted $zz$ interactions in a hybrid
  two-qubit system},\ }\href {https://doi.org/10.1103/PhysRevLett.125.200504}
  {\bibfield  {journal} {\bibinfo  {journal} {Phys. Rev. Lett.}\ }\textbf
  {\bibinfo {volume} {125}},\ \bibinfo {pages} {200504} (\bibinfo {year}
  {2020})}\BibitemShut {NoStop}%
\bibitem [{\citenamefont {Ding}\ \emph {et~al.}(2020)\citenamefont {Ding},
  \citenamefont {Gokhale}, \citenamefont {Lin}, \citenamefont {Rines},
  \citenamefont {Propson},\ and\ \citenamefont {Chong}}]{Ding2020}%
  \BibitemOpen
  \bibfield  {author} {\bibinfo {author} {\bibfnamefont {Y.}~\bibnamefont
  {Ding}}, \bibinfo {author} {\bibfnamefont {P.}~\bibnamefont {Gokhale}},
  \bibinfo {author} {\bibfnamefont {S.}~\bibnamefont {Lin}}, \bibinfo {author}
  {\bibfnamefont {R.}~\bibnamefont {Rines}}, \bibinfo {author} {\bibfnamefont
  {T.}~\bibnamefont {Propson}},\ and\ \bibinfo {author} {\bibfnamefont {F.~T.}\
  \bibnamefont {Chong}},\ }\bibfield  {title} {\bibinfo {title} {Systematic
  crosstalk mitigation for superconducting qubits via frequency-aware
  compilation},\ }in\ \href {https://doi.org/10.1109/MICRO50266.2020.00028}
  {\emph {\bibinfo {booktitle} {2020 53rd Annual IEEE/ACM International
  Symposium on Microarchitecture (MICRO)}}}\ (\bibinfo  {publisher} {IEEE
  Computer Society},\ \bibinfo {address} {Los Alamitos, CA, USA},\ \bibinfo
  {year} {2020})\ pp.\ \bibinfo {pages} {201--214}\BibitemShut {NoStop}%
\bibitem [{\citenamefont {Wanga}\ \emph {et~al.}()\citenamefont {Wanga},
  \citenamefont {Zhaob}, \citenamefont {Jin},\ and\ \citenamefont
  {Yu}}]{Wanga2022}%
  \BibitemOpen
  \bibfield  {author} {\bibinfo {author} {\bibfnamefont {R.}~\bibnamefont
  {Wanga}}, \bibinfo {author} {\bibfnamefont {P.}~\bibnamefont {Zhaob}},
  \bibinfo {author} {\bibfnamefont {Y.}~\bibnamefont {Jin}},\ and\ \bibinfo
  {author} {\bibfnamefont {H.}~\bibnamefont {Yu}},\ }\bibfield  {title}
  {\bibinfo {title} {Control and mitigation of microwave crosstalk effect with
  superconducting qubits},\ }\href {https://doi.org/10.1063/5.0115393} {\
  \textbf {\bibinfo {volume} {121}}}\BibitemShut {NoStop}%
\bibitem [{\citenamefont {Werninghaus}\ \emph {et~al.}()\citenamefont
  {Werninghaus}, \citenamefont {Egger}, \citenamefont {Roy}, \citenamefont
  {Machnes}, \citenamefont {Wilhelm},\ and\ \citenamefont
  {Filipp}}]{Werninghaus2021}%
  \BibitemOpen
  \bibfield  {author} {\bibinfo {author} {\bibfnamefont {M.}~\bibnamefont
  {Werninghaus}}, \bibinfo {author} {\bibfnamefont {D.~J.}\ \bibnamefont
  {Egger}}, \bibinfo {author} {\bibfnamefont {F.}~\bibnamefont {Roy}}, \bibinfo
  {author} {\bibfnamefont {S.}~\bibnamefont {Machnes}}, \bibinfo {author}
  {\bibfnamefont {F.~K.}\ \bibnamefont {Wilhelm}},\ and\ \bibinfo {author}
  {\bibfnamefont {S.}~\bibnamefont {Filipp}},\ }\bibfield  {title} {\bibinfo
  {title} {Leakage reduction in fast superconducting qubit gates via optimal
  control},\ }\href {https://doi.org/10.1038/s41534-020-00346-2} {\ \textbf
  {\bibinfo {volume} {7}},\ \bibinfo {pages} {14}}\BibitemShut {NoStop}%
\bibitem [{\citenamefont {Schuster}\ \emph {et~al.}(2005)\citenamefont
  {Schuster}, \citenamefont {Wallraff}, \citenamefont {Blais}, \citenamefont
  {Frunzio}, \citenamefont {Huang}, \citenamefont {Majer}, \citenamefont
  {Girvin},\ and\ \citenamefont {Schoelkopf}}]{Schuster2005}%
  \BibitemOpen
  \bibfield  {author} {\bibinfo {author} {\bibfnamefont {D.~I.}\ \bibnamefont
  {Schuster}}, \bibinfo {author} {\bibfnamefont {A.}~\bibnamefont {Wallraff}},
  \bibinfo {author} {\bibfnamefont {A.}~\bibnamefont {Blais}}, \bibinfo
  {author} {\bibfnamefont {L.}~\bibnamefont {Frunzio}}, \bibinfo {author}
  {\bibfnamefont {R.-S.}\ \bibnamefont {Huang}}, \bibinfo {author}
  {\bibfnamefont {J.}~\bibnamefont {Majer}}, \bibinfo {author} {\bibfnamefont
  {S.~M.}\ \bibnamefont {Girvin}},\ and\ \bibinfo {author} {\bibfnamefont
  {R.~J.}\ \bibnamefont {Schoelkopf}},\ }\bibfield  {title} {\bibinfo {title}
  {ac stark shift and dephasing of a superconducting qubit strongly coupled to
  a cavity field},\ }\href {https://doi.org/10.1103/PhysRevLett.94.123602}
  {\bibfield  {journal} {\bibinfo  {journal} {Phys. Rev. Lett.}\ }\textbf
  {\bibinfo {volume} {94}},\ \bibinfo {pages} {123602} (\bibinfo {year}
  {2005})}\BibitemShut {NoStop}%
\bibitem [{\citenamefont {Oka}\ \emph {et~al.}(2020)\citenamefont {Oka},
  \citenamefont {Matsukawa}, \citenamefont {Kato}, \citenamefont {Iizuka},
  \citenamefont {Mizubayashi}, \citenamefont {Endo}, \citenamefont {Yasuda},\
  and\ \citenamefont {Mori}}]{Oka_2020}%
  \BibitemOpen
  \bibfield  {author} {\bibinfo {author} {\bibfnamefont {H.}~\bibnamefont
  {Oka}}, \bibinfo {author} {\bibfnamefont {T.}~\bibnamefont {Matsukawa}},
  \bibinfo {author} {\bibfnamefont {K.}~\bibnamefont {Kato}}, \bibinfo {author}
  {\bibfnamefont {S.}~\bibnamefont {Iizuka}}, \bibinfo {author} {\bibfnamefont
  {W.}~\bibnamefont {Mizubayashi}}, \bibinfo {author} {\bibfnamefont
  {K.}~\bibnamefont {Endo}}, \bibinfo {author} {\bibfnamefont {T.}~\bibnamefont
  {Yasuda}},\ and\ \bibinfo {author} {\bibfnamefont {T.}~\bibnamefont {Mori}},\
  }\bibfield  {title} {\bibinfo {title} {Toward long-coherence-time si spin
  qubit: The origin of low-frequency noise in cryo-cmos},\ }in\ \href
  {https://doi.org/10.1109/VLSITechnology18217.2020.9265013} {\emph {\bibinfo
  {booktitle} {2020 IEEE Symposium on VLSI Technology}}}\ (\bibinfo {year}
  {2020})\ pp.\ \bibinfo {pages} {1--2}\BibitemShut {NoStop}%
\bibitem [{\citenamefont {Gustavsson}\ \emph {et~al.}(2012)\citenamefont
  {Gustavsson}, \citenamefont {Bylander}, \citenamefont {Yan}, \citenamefont
  {Forn-D\'{\i}az}, \citenamefont {Bolkhovsky}, \citenamefont {Braje},
  \citenamefont {Fitch}, \citenamefont {Harrabi}, \citenamefont {Lennon},
  \citenamefont {Miloshi}, \citenamefont {Murphy}, \citenamefont {Slattery},
  \citenamefont {Spector}, \citenamefont {Turek}, \citenamefont {Weir},
  \citenamefont {Welander}, \citenamefont {Yoshihara}, \citenamefont {Cory},
  \citenamefont {Nakamura}, \citenamefont {Orlando},\ and\ \citenamefont
  {Oliver}}]{Gustavsson2012}%
  \BibitemOpen
  \bibfield  {author} {\bibinfo {author} {\bibfnamefont {S.}~\bibnamefont
  {Gustavsson}}, \bibinfo {author} {\bibfnamefont {J.}~\bibnamefont
  {Bylander}}, \bibinfo {author} {\bibfnamefont {F.}~\bibnamefont {Yan}},
  \bibinfo {author} {\bibfnamefont {P.}~\bibnamefont {Forn-D\'{\i}az}},
  \bibinfo {author} {\bibfnamefont {V.}~\bibnamefont {Bolkhovsky}}, \bibinfo
  {author} {\bibfnamefont {D.}~\bibnamefont {Braje}}, \bibinfo {author}
  {\bibfnamefont {G.}~\bibnamefont {Fitch}}, \bibinfo {author} {\bibfnamefont
  {K.}~\bibnamefont {Harrabi}}, \bibinfo {author} {\bibfnamefont
  {D.}~\bibnamefont {Lennon}}, \bibinfo {author} {\bibfnamefont
  {J.}~\bibnamefont {Miloshi}}, \bibinfo {author} {\bibfnamefont
  {P.}~\bibnamefont {Murphy}}, \bibinfo {author} {\bibfnamefont
  {R.}~\bibnamefont {Slattery}}, \bibinfo {author} {\bibfnamefont
  {S.}~\bibnamefont {Spector}}, \bibinfo {author} {\bibfnamefont
  {B.}~\bibnamefont {Turek}}, \bibinfo {author} {\bibfnamefont
  {T.}~\bibnamefont {Weir}}, \bibinfo {author} {\bibfnamefont {P.~B.}\
  \bibnamefont {Welander}}, \bibinfo {author} {\bibfnamefont {F.}~\bibnamefont
  {Yoshihara}}, \bibinfo {author} {\bibfnamefont {D.~G.}\ \bibnamefont {Cory}},
  \bibinfo {author} {\bibfnamefont {Y.}~\bibnamefont {Nakamura}}, \bibinfo
  {author} {\bibfnamefont {T.~P.}\ \bibnamefont {Orlando}},\ and\ \bibinfo
  {author} {\bibfnamefont {W.~D.}\ \bibnamefont {Oliver}},\ }\bibfield  {title}
  {\bibinfo {title} {Driven dynamics and rotary echo of a qubit tunably coupled
  to a harmonic oscillator},\ }\href
  {https://doi.org/10.1103/PhysRevLett.108.170503} {\bibfield  {journal}
  {\bibinfo  {journal} {Phys. Rev. Lett.}\ }\textbf {\bibinfo {volume} {108}},\
  \bibinfo {pages} {170503} (\bibinfo {year} {2012})}\BibitemShut {NoStop}%
\end{thebibliography}%

\end{document}